\documentclass{jfm}
\usepackage{graphicx}
\usepackage{epstopdf, epsfig}
\usepackage{amsmath}
\usepackage{enumitem}
\usepackage{natbib}
\usepackage{mathtools}
\usepackage{amsfonts}
\usepackage{tikz}
\usepackage[english]{babel}
\usepackage{enumitem}
\usepackage{setspace}
\usepackage{verbatim}
\usepackage{caption}
\usepackage{subcaption}
\usepackage{wrapfig}
\usepackage{nextpage}
\usepackage[titletoc]{appendix}
\usepackage{titlesec}
\usepackage{standalone}
\usepackage{bm}
\usepackage{bibentry}
\usepackage{pdflscape}
\usepackage{rotating}
\usepackage{xcolor}
\usepackage{soul}
\usepackage{titlesec}
\usepackage{floatrow}
\usepackage{sidecap}
\usepackage{color}
\usepackage{tikz}
\usepackage{rotating}
\usetikzlibrary{shapes.misc}
\usepackage[colorlinks=true, linkcolor=blue, urlcolor=blue, citecolor=blue]{hyperref}

\allowdisplaybreaks
\tikzset{cross/.style={cross out, draw=black, minimum size=2*(#1-\pgflinewidth), inner sep=0pt, outer sep=0pt}, cross/.default={1pt}}

\shorttitle{Predicting burst events in a forced 2D flow: A wavelet-based analysis}
\shortauthor{A. Madhusudanan, R. R. Kerswell}

\title{Predicting burst events in a forced 2D flow: A wavelet-based analysis}

\author{Anagha Madhusudanan\aff{1,2}
  \corresp{\email{anaghamadhu91@gmail.com}}
  \and Rich R. Kerswell\aff{3}}

\affiliation{
$^1$ Isaac Newton Institute for Mathematical Sciences, CB3 0EH, Cambridge, UK \\
$^2$ Department of Aerospace Engineering, Indian Institute of Science, Bangalore, India \\
$^3$ Department of Applied Mathematics and Theoretical Physics, Centre for Mathematical Sciences, Wilberforce Road, Cambridge CB3 0WA, UK 
}

\begin{document}

\maketitle

\begin{abstract}
Predicting and perhaps mitigating against rare, extreme events in fluid flows is an important challenge. 
Due to the time-localised nature of these events, Fourier-based methods prove inefficient in capturing them. 
Instead, this paper uses wavelet-based methods to understand the underlying patterns in a forced flow over a 2-torus which has intermittent high-energy burst events interrupting an ambient low energy `quiet’ flow. 
Two wavelet-based methods are examined to predict burst events: (1) a wavelet proper orthogonal decomposition (WPOD) based method which uncovers and utilises the key flow patterns seen in the quiet regions and the bursting episodes; and (2) a wavelet resolvent analysis (WRA) based method that relies on the forcing structures which amplify the underlying flow patterns. 
These methods are compared to a straightforward energy tracking approach which acts as a benchmark. 
Both the wavelet-based approaches succeed in producing better predictions than a simple energy criterion, i.e.\ earlier prediction times and/or fewer false positives and the WRA-based technique always performs better than WPOD. 
However, the improvement of WRA over WPOD is not as substantial as anticipated.  
We conjecture that this is because the  mechanism for the bursts in the flow studied is found to be largely modal, associated with the unstable eigenfunction of the Navier-Stokes operator linearized around the mean flow.  
The WRA approach should deliver much better improvement over the WPOD approach for generically non-modal bursting mechanisms where there is a lag between the imposed forcing and the final response pattern.
\end{abstract}

\section{Introduction}
\label{sec:introduction}

Intermittent extreme events, characterised by a sudden increase in observables like energy or dissipation, are frequently encountered in both natural and engineering fluid flow systems. 
Some examples include extreme weather events \citep[e.g.][]{neelin1998enso}, rogue waves in the ocean \citep[e.g.][]{dysthe2008oceanic} and high dissipation events in turbulent flows \citep[e.g.][]{yeung2015extreme, chandler2013invariant}. 
These events can have significant impacts. 
Understanding the mechanisms that generate such events, and predicting them, are therefore active areas of research \citep{farazmand2019extreme, sapsis2021statistics}. 
For instance, \citet{donzis2010short} analysed direct numerical simulation (DNS) data to identify the observables that act as precursors to the intermittent events. 
In another work, \citet{babaee2016minimization} used data 
to identify time-dependent orthonormal bases that characterise the transient instabilities in a system, which were then used to forecast extreme events \citep{farazmand2016dynamical}. 
However, working with time-dependent modes can become computationally expensive, and therefore a variational framework that identified static structures responsible for the extreme events was formulated by \citet{farazmand2017variational, farazmand2019extreme}. 
By tracking these identified structures, methods to predict and control the burst events were introduced \citep{farazmand2019closed, blonigan2019extreme}. 
A graph theoretic approach that uses clustering algorithms to find a hierarchy of coherent structures in intermittent flows has also been used \citep{schmid2018description}. 
More recently, the use of machine learning to identify and control extreme events has also gained popularity \citep[e.g.][]{wan2018data, guth2019machine, pyragas2020using, qi2020using, doan2021short, racca2022data, rudy2022prediction, fox2023predicting}. 
Another method, and one that is of more direct relevance to the current work, is the use of a variation of proper orthogonal decomposition (POD), called conditional-POD, to identify the dominant structures that are responsible for intermittent events \citep{schmidt2019conditional}. 

Due to the time-localised nature of intermittent events, the Fourier basis is generally inefficient at characterising them. 
Wavelets are better adapted for this purpose, which suggests that wavelet-based methods could potentially provide another class of techniques for understanding and predicting intermittent events.   
Wavelets have found application in the analysis of turbulent signals since the early 1990s \citep[e.g.][]{farge1988transformee,meneveau1991analysis,farge1992wavelet}.  
Since then, they have found varied applications in turbulence such as, for instance, extraction of the coherent and incoherent parts of a turbulent flow-field \citep[e.g.][]{farge2001coherent, farge2003coherent, farge2006extraction} and development of simulation methods using wavelet-based numerical algorithms and turbulence models (see review by \citet{schneider2010wavelet}). 
Wavelet coefficients have also been used for detecting pipe bursts in water distribution systems \citep{srirangarajan2013wavelet} and predicting rogue waves \citep{bayindir2016early}.  
Data-driven decomposition techniques based on wavelets have been introduced by \citet{floryan2021discovering} to generate a hierarchical orthogonal basis for a flow and by \citet{ren2021image} to extract the different dominant scales in a flow. 
Recent studies by \citet{gupta2022impact}, \citet{krah2022wavelet} and \citet{barthel2023harnessing} have used wavelet analysis along with POD for identifying coherent structures in intermittent flows. 
In a parallel line of work, \citet{ballouz2023transient} and \citet{ballouz2023wavelet} extended the extensively used resolvent analysis technique to incorporate a temporal wavelet basis, thereby broadening the scope of the analysis. 

In the current work, we aim to contribute towards understanding and predicting intermittent high-energy events in fluid flows by employing wavelet-based methods.  
A two-dimensional (2D) Kolmogorov flow at a Reynolds number of 40, forced by a sinusoidal body force with wavenumber 4, is considered.
This flow is temporally characterised by a persistent quiet region that is intermittently
interrupted by high-energy burst events \citep[e.g.][]{chandler2013invariant, page2021revealing}. 
To understand the temporal characteristics of the flow, we first project it onto a wavelet basis.
Two wavelet-based methods will then be used to analyse the flow: (1) Wavelet-based Proper Orthogonal Decomposition (WPOD) to distinguish the dominant flow patterns of the quiet region and burst events and (2) Wavelet-based Resolvent Analysis (WRA) to identify forcing structures that generate the quiet region and burst events. 
The identification of these flow patterns poses a question: can these patterns be utilised to predict oncoming burst events?
We will therefore explore whether tracking the flow patterns obtained from WPOD (wavelet-based prediction method 1) and from WRA (wavelet-based prediction method 2) enables prediction of oncoming burst events. 
The predictions will be compared to those obtained from the more straightforward approach of tracking the energy of the flow. 

We find that both the WPOD-based and the WRA-based methods give better predictions when compared to the energy-based method, i.e.\ earlier prediction times and/or fewer false positives. 
Additionally, the WRA-based method outperforms the WPOD-based method. 
This improvement, however, is not as significant as initially expected.  
Energy-amplification in the flow considered here is likely caused by normal-mode mechanisms, where the amplification arises due to the system being excited at a frequency close to its eigenvalue \citep{trefethen1993hydrodynamic}. 
This could potentially explain the prediction performance observed here (see \S\ref{sec:a discussion of the predictions obtained}). 
There is therefore a future need to assess how the prediction performance changes for flows governed by non-normal energy amplification mechanisms, where we can expect a lag between the forcing and the final response pattern. 

The outline of the rest of this manuscript is as follows. 
We first introduce the 2D Kolmogorov flow in \S\ref{sec:2d kolmogorov flow introduction}.  
The aim thereafter is to decompose the flow into the quiet region and the burst events. 
The efficiency of the wavelet basis in achieving such a decomposition is shown in \S\ref{sec:decomposing the flow: wavelet bases}, and this is compared to the efficiency of the Fourier basis in \S\ref{sec:decomposing the flow: fourier bases}. 
Thereafter, in \S\ref{sec:coherent structures of the flow: a wavelet-pod analysis}, we probe the underlying flow patterns in this intermittent flow using WPOD. 
Predictions of the burst events obtained by tracking these WPOD modes (WPOD-based prediction method) are then studied in \S\ref{sec:predicting burst events using wpod modes - method 2}. 
These predictions are compared to those obtained from a straightforward energy-tracking approach. 
Following this, in \S\ref{sec:forcing to the coherent structures} we use WRA to distinguish the structures that force the quiet region and burst events. 
Predictions obtained from tracking these forcing structures (WRA-based prediction method) are then compared to those obtained from the WPOD-based and energy-based methods in \S\ref{sec:predicting burst events using resolvent forcing structures - method 3}. 
Following this, in \S\ref{sec:a discussion of the predictions obtained} we will look at a more detailed comparison of the performances of the prediction methods and also implications for future work. 
A final discussion follows in \S\ref{sec:conclusions}. 

\section{2D Kolmogorov flow}
\label{sec:2d kolmogorov flow introduction}

In this section, we consider the 2D Kolmogorov flow. 
We will first discuss the linearized Navier--Stokes equations that describe the flow (\S\ref{sec:2d kolmogorov flow}), and thereafter consider the data obtained from the DNS of this flow (\S\ref{sec:direct numerical simulations} and \S\ref{sec:symmetrisation of the mean profile}).

\subsection{Linearized Navier--Stokes equations}
\label{sec:2d kolmogorov flow}

%
%
\begin{figure}[t]
\captionsetup[subfigure]{labelformat=empty,skip=-60pt}
\begin{subfigure}[b]{\textwidth}
\centering
\includegraphics[width=\textwidth]{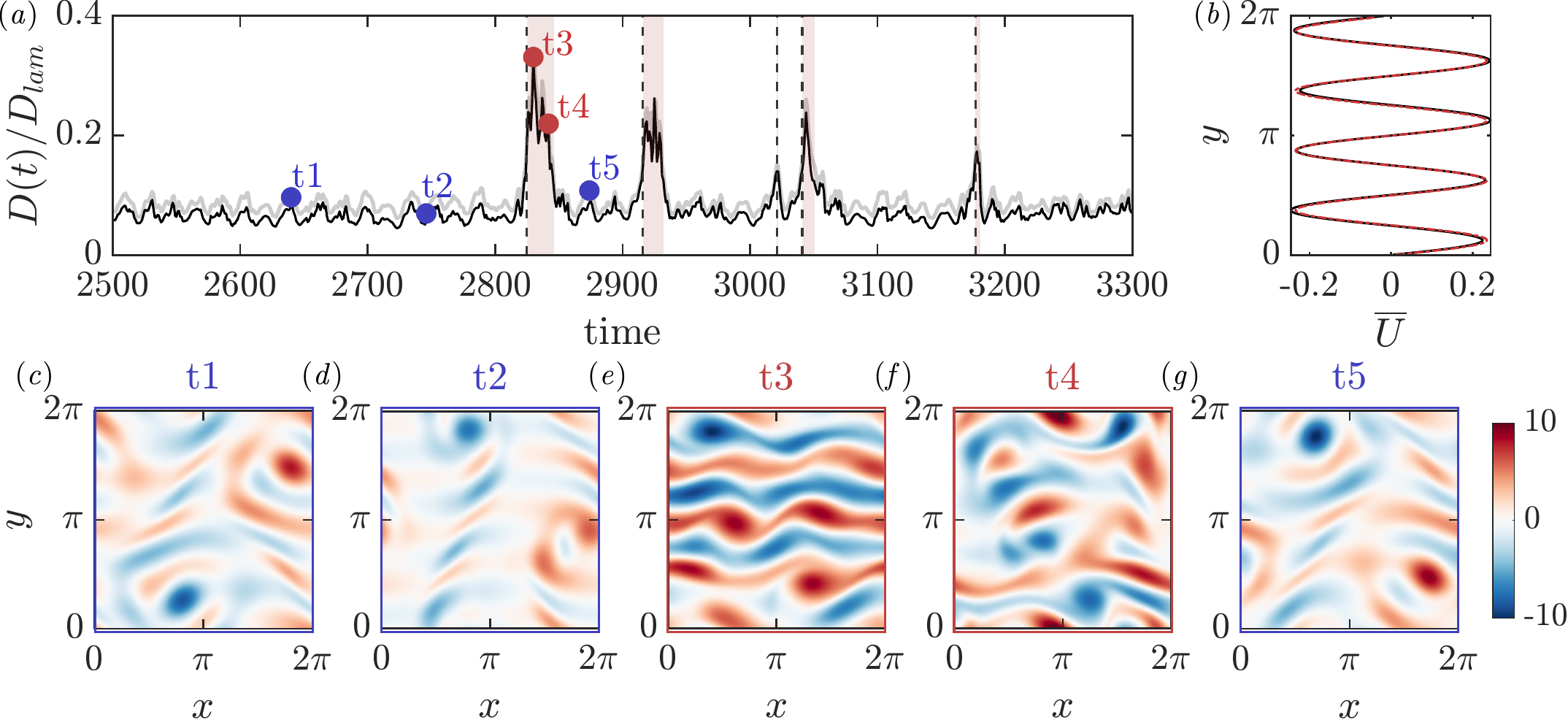}
\caption{}
\label{fig:Intro_TimeSeries}
\end{subfigure}%
\begin{subfigure}[b]{\textwidth}
\centering
\caption{}
\label{fig:Intro_Mean}
\end{subfigure}
\begin{subfigure}[b]{\textwidth}
\centering
\caption{}
\label{fig:Intro_field_1}
\end{subfigure}%
\begin{subfigure}[b]{\textwidth}
\centering
\caption{}
\label{fig:Intro_field_2}
\end{subfigure}
\begin{subfigure}[b]{\textwidth}
\centering
\caption{}
\label{fig:Intro_field_3}
\end{subfigure}
\begin{subfigure}[b]{\textwidth}
\centering
\caption{}
\label{fig:Intro_field_4}
\end{subfigure}
\begin{subfigure}[b]{\textwidth}
\centering
\caption{}
\label{fig:Intro_field_5}
\end{subfigure}
\caption{(\textit{a}) The time-series of $D(t)/D_{lam}$ from the full data is shown in grey for a sample time interval. 
Additionally, the time series obtained from just the streamwise wavenumbers of $|k_x| \leq 3$ is shown in black. 
Burst events, defined as times when $D(t)/D_{lam} >0.15$, are shown as the red-shaded regions, with dashed black vertical lines indicating the beginnings of the burst events. 
(\textit{c-g}) The vorticity field at the five time instances $t1$-$t5$  indicated on (\textit{a}) are also shown, where the contours indicate negative (red) and positive (blue) vorticity, respectively. 
The colour limits are kept the same across the five time snapshots.  
In (\textit{b}) the mean profile obtained from the data (black line) is compared to the symmetrised mean profile (red line).   }
\label{fig:Intro}
\end{figure}

The intermittent flow that we use as our example in this work is the incompressible 2D Kolmogorov flow. 
The two dimensions are denoted by $x$ and $y$, respectively. 
The domain is doubly periodic, with size $L_x$ in the $x$-direction and $L_y$ in the $y$-direction. 
The flow is forced in the $x$-direction by a sinusoidal body forcing of the form $\zeta \sin(2\pi n\grave{y}/L_y)$, where $\zeta$ is the amplitude of the forcing per unit mass of fluid, $n$ is the wavenumber of the forcing and $\grave{y}$ indicates the dimensional $y$-coordinate. 
The mean-velocity is $U(y)\widehat{\bm{i}}$, where $(\widehat{\bm{i}},\widehat{\bm{j}})$ are unit vectors in the $(x,y)$ directions. 
The mean is here defined over the $x$-direction and time. 
Additionally, it is symmetrised in the $y$-direction (why such a symmetrisation is required is discussed in \S\ref{sec:symmetrisation of the mean profile}). 
The fluctuations of velocities around this mean are denoted by $u$ and $v$ in $x$ and $y$ directions, respectively. 
Pressure is denoted by $p$ and time by $t$. 
The length-scale $L_y/2\pi$ and timescale $\sqrt{L_y/2\pi\zeta}$ is used to non-dimensionalise the system. 
The non-dimensional number that characterises this system is the Reynolds number, defined as $Re:=(\sqrt{\zeta}/\nu)(L_y/2\pi)^{(3/2)}$. 
Here we consider a flow with $n=4$, $Re=40$ and $L_x=L_y=2\pi$, which is a regime characterised by intermittent burst events as will be discussed in \S\ref{sec:direct numerical simulations}. 
 
The non-dimensional equations linearized around the mean state $(U(y),0)$ are:
\begin{equation}
\begin{split}
\frac{\partial \bm{u}}{\partial t} + U\frac{\partial \bm{u}}{\partial x} + v\frac{\partial U}{\partial y} \widehat{\bm{i}} + \nabla p - \frac{1}{Re} \nabla^2 \bm{u} = \underbrace{ \overline{\bm{u}\cdot \nabla \bm{u}} - \bm{u}\cdot \nabla \bm{u}}_{\bm{f}}, \quad \nabla \cdot \bm{u} = 0, 
\label{eqn:NS}
\end{split}
\end{equation}
where $\bm{u}=(u,v)$. 
The non-linear terms of the equation $\overline{\bm{u}\cdot \nabla \bm{u}} - \bm{u}\cdot \nabla \bm{u}$ are hereafter represented by a forcing term $\bm{f}=(f_x,f_y)$, where $f_x$ and $f_y$ represent the forcing to the $x$ and $y$ momentum equations, respectively. 

Using the incompressibility condition and taking the curl of \eqref{eqn:NS}, we obtain the vorticity ($\omega$) form of the equation as:
\begin{equation}
\begin{split}
\frac{\partial \omega}{\partial t} + \left[U\frac{\partial }{\partial x} 
-\frac{1}{Re} \Delta -  U'' \Delta^{-1} \frac{\partial }{\partial x} \right] \omega 
&
= F,  
\label{eqn:NS_vort}
\end{split}
\end{equation}
where $\omega := \partial v/\partial x - \partial u/\partial y$ and $F:= \partial f_y/\partial x - \partial f_x/\partial y$. 
Let us consider a Fourier transform in the homogeneous $x$-direction. 
The one-dimensional (1D) discrete Fourier transform of $\omega$ gives
\begin{equation}
\omega(x,y,t) = \sum_{k_x=-(N_x/2)}^{(N_x/2)} \widetilde{\omega}(y,t;k_x)e^{ik_x x}, 
\end{equation}
where $\widetilde{\cdot}$ represents the 1D Fourier transform, $k_x$ is the streamwise wavenumber non-dimensionalised by $L_y/2\pi$ and $N_x$ is the number of grid-points used to discretise the $x$-direction. 
Similarly, if $\widetilde{F}$ is the 1D Fourier transform of $F$, we obtain the following equation: 
\begin{equation}
\begin{split}
\dot{\widetilde{\omega}}(y,t;k_x) + \bm{A}\widetilde{\omega}(y,t;k_x) = \widetilde{F}(y,t;k_x),  
\label{eqn:state_space}
\end{split}
\end{equation}
where $(\dot{\mbox{ }})$ denotes the derivative in time. 
The matrix $\bm{A}$ contains the finite-dimensional discrete approximations of the linearized momentum equation from \eqref{eqn:NS_vort} in terms of the 1-D Fourier transforms with $\partial/\partial x$ replaced by $ik_x$. 

\subsection{Direct Numerical Simulation data}
\label{sec:direct numerical simulations}

To obtain the DNS data for the 2D Kolmogorov flow, we used the pseudo-spectral code as used in \citet{chandler2013invariant}. 
We consider the flow at a Reynolds number $Re=40$ and forcing frequency $n=4$.
The DNS was run on a grid with $N_x=256$ points in the $x$-direction and $N_y=257$ points in the $y$-direction. 
Figure \ref{fig:Intro}(\subref{fig:Intro_TimeSeries}) shows the time series of total dissipation $D(t):=\langle \omega, \omega \rangle_{x,y}$ normalised by the laminar dissipation rate $D_{lam}:=Re/(2n^2)$. 
Here $\langle a,b \rangle_{x,y}$ denotes the inner product $\int_{0}^{L_x} \int_{0}^{L_y} b^*(x,y,t) a(x,y,t) dx dy$. 
The time series represented by the grey line is obtained from the full dataset. 
On the other hand, the time series in black is obtained by retaining only the smallest streamwise wavenumbers with $|k_x|\leq 3$. 
The black line follows the grey line reasonably well, thereby suggesting that the $|k_x|\leq 3$ modes are responsible for the dominant dynamics in the flow. 
Example snapshots of the flow at the five times denoted in figure \ref{fig:Intro}(\subref{fig:Intro_TimeSeries}) are shown in figures \ref{fig:Intro}(\subref{fig:Intro_field_1})-\ref{fig:Intro}(\subref{fig:Intro_field_5}), where the red and blue contours represent the positive and negative vorticity fluctuations, respectively. 
Here again we observe the presence of large-scale structures in the flow, further confirming that the smallest wavenumbers dominate the flow dynamics. 
This is consistent with the observations in \citet{page2021revealing}. 
For the rest of this manuscript, we therefore project the full DNS data onto the $|k_x|\leq 3$ modes, and use the truncated data for the POD and resolvent-based analyses. 
This truncation using a Fourier transform is not essential for the work that follows, but just allows us to expedite the computations. 

Considering the burst events, if we define them as times when $D(t)/D_{lam} \geq 0.15$ \citep{page2021revealing}, then they are indicated by the red-shaded regions in figure \ref{fig:Intro}(\subref{fig:Intro_TimeSeries}) 
(the vertical black dashed lines therefore indicate the beginnings of these burst events). 
Throughout the manuscript, we use this criterion of $D(t)/D_{lam} \geq 0.15$ to define burst events, and also $D(t)/D_{lam} \leq 0.1$ to identify regions that are assured to be quiet. 
Changing these numbers for $D(t)/D_{lam}$ will impact the quantitative values obtained. 
However, we are here interested only in the relative differences between the prediction methods, and we can expect these trends to remain insensitive to this change. 
Appendix \ref{sec:insensitivity of prediction methods to parameter choices} confirms that these relative trends are not impacted by this definition of the burst event. 

\subsection{Symmetrisation of the mean profile}
\label{sec:symmetrisation of the mean profile}

Let us briefly consider the mean velocity profiles $U(y)$ obtained from this flow, where the mean is defined over the $x$-direction and time. 
The converged mean profile is expected to follow two symmetries: (i) a shift-and-reflect symmetry
$\mathcal{S}:U(y) \to - U(y+\pi/4)$
and (ii) a rotational symmetry 
$\mathcal{R}:U(y) \to - U(-y)$. 
\citet{chandler2013invariant} investigated the mean profiles obtained from $10^5$ time units of the flow, and observed that even this was not long enough to obtain a converged mean profile that shows the expected symmetries. 
The black line (which is asymmetric when investigated closely) in figure \ref{fig:Intro}(\subref{fig:Intro_Mean}) represents the mean computed from $16800$ snapshots of data. 
This obtained mean does not follow the symmetries. 
Therefore, as done in \citet{chandler2013invariant}, here we use a symmetrised mean profile $U(y)$ which is extracted from the asymmetric mean obtained from the DNS $U_a(y)$ as
\begin{equation}
    U(y) := \frac{1}{2n} \sum_{m=0}^{2n-1} \mathcal{S}^{-m}U^\dagger(\mathcal{S}^my), \mbox{ where } U^\dagger:=\frac{1}{2} \left[ U_a(y) + \mathcal{R}^{-1}U_a(\mathcal{R}y) \right]. 
\end{equation}
Here $n=4$. 
The definitions of the symmetry operations follows from before as $\mathcal{R}^{-1}:U(y) \to - U(-y)$, $\mathcal{S}^m:U(y) \to (-1)^m U(y+m\pi/4)$ and $\mathcal{S}^{-m}:U(y) \to (-1)^m U(y-m\pi/4)$, where $0\leq m \leq (2n-1)$. 
The red line in figure \ref{fig:Intro}(\subref{fig:Intro_Mean}) shows $U(y)$, and this symmetrised mean will be used for the rest of this work. 

\section{Decomposing the flow into quiet regions and burst events}
\label{sec:Decomposing the flow into quiet regions and burst events}

Both the WPOD-based and WRA-based prediction methods that will be introduced later depend on the identification of flow structures that exist in the quiet region and the burst events. 
To identify such structures, it is necessary to first decompose the flow into the quiet region and the burst events. 

\subsection{Decomposing the flow: Fourier bases}
\label{sec:decomposing the flow: fourier bases}

In this section, we use the Fourier basis for decomposing the flow into the quiet region and the burst events. 
Looking ahead, isolating the time-localised burst events using the global Fourier basis may prove to be inefficient. 
Therefore, the aim of this section is to provide motivation for moving towards using a wavelet basis. 

%
%
\begin{figure}[t]
\captionsetup[subfigure]{labelformat=empty,skip=-25pt}
\begin{subfigure}[b]{0.8\textwidth}
\centering
\includegraphics[width=\textwidth]{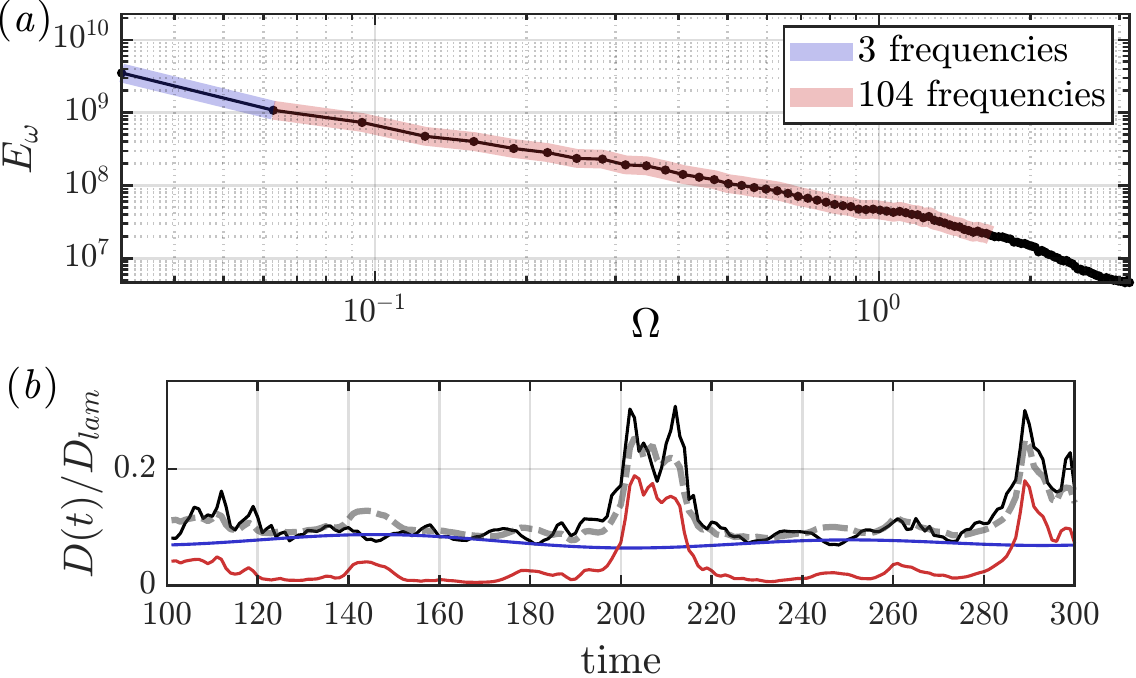}
\caption{}
\label{fig:DNSSplit_Fourier_spec}
\end{subfigure}
\begin{subfigure}[b]{0.75\textwidth}
\centering
\caption{}
\label{fig:DNSSplit_Fourier_timeseries}
\end{subfigure}
\caption{(\textit{a}) The ensemble-averaged Fourier spectrum is shown in black, and the frequencies used to reconstruct the quiet region and the burst event, separately, are shaded in blue and red, respectively. 
(\textit{b}) The full data (for $k_x={0,1,2,3}$) (black) is compared to the reconstructions of the quiet region (blue) and the burst event (red) using the shaded frequencies in (\textit{a}). 
For comparison, the grey dashed-dot line shows the sum of the blue and the red lines. 
}
\label{fig:DNSSplit_fourier}
\end{figure}

\begin{figure}[t]
\captionsetup[subfigure]{labelformat=empty,skip=-30pt}
\begin{subfigure}[b]{\textwidth}
\centering
\includegraphics[width=\textwidth]{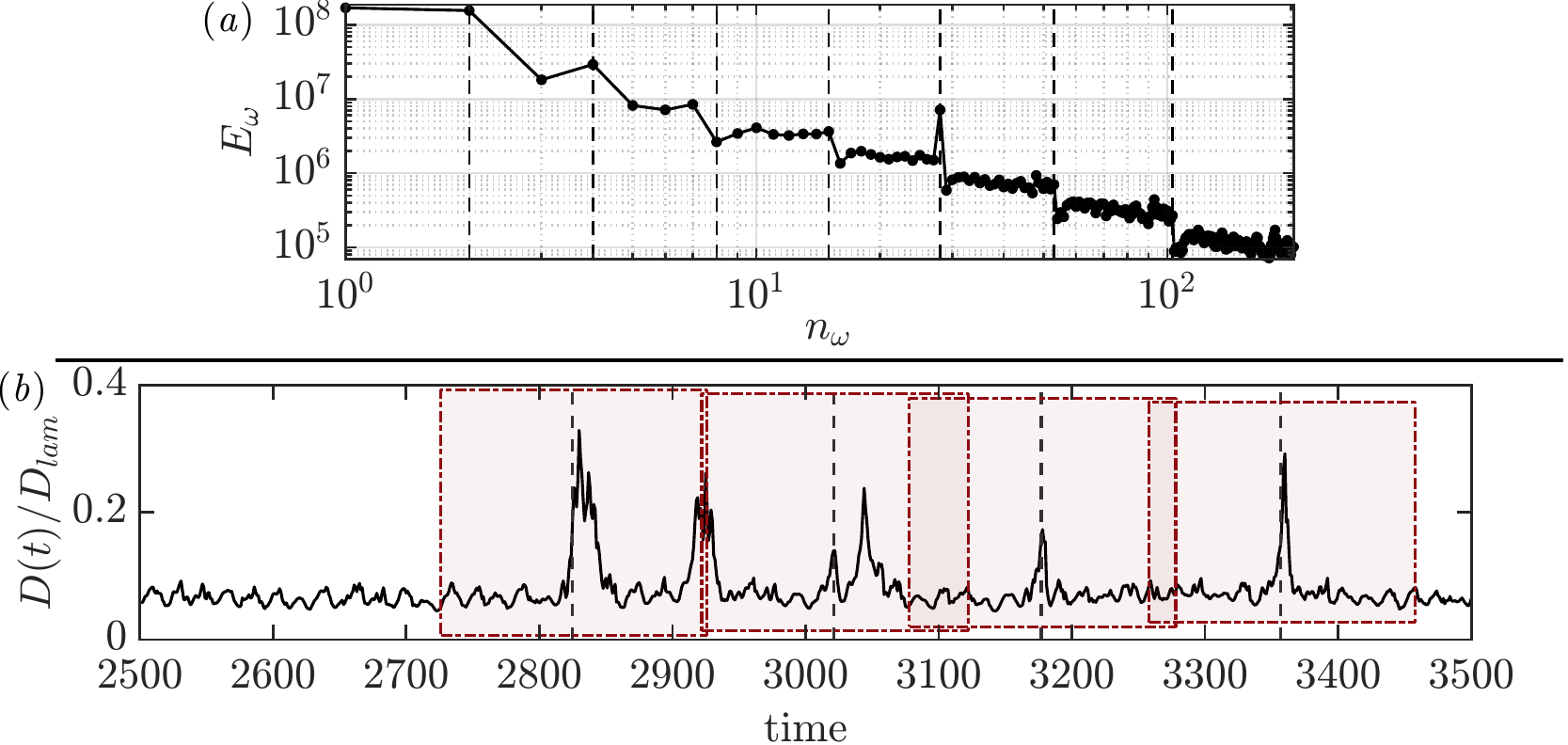}
\caption{}
\label{fig:windowing_NoCentering_spec}
\end{subfigure}
\begin{subfigure}[b]{\textwidth}
\centering
\caption{}
\label{fig:windowing_windows}
\end{subfigure}
\caption{(\textit{a}) The ensemble-averaged energy of the wavelet coefficients is shown for the case when ensembles are chosen as consecutive time-blocks.
(\textit{b}) A different strategy for choosing ensembles is also illustrated, where the window is chosen such that a burst lies at the centre of it.  
}
\label{fig:windowing}
\end{figure}

We are interested in analysing a burst event, and require multiple realisations of such events. 
Consider $N_e$ realisations of the burst event, each containing $N_t$ time-snapshots of vorticity $\omega(x,y,t)$ spaced apart by time $dt$. 
Each of the $N_e$ realisations is therefore $T=N_t dt$ long. 
Here, to obtain one such realisation of the burst event, the start of a burst event $T_b$ is first identified using the condition $D(t)/D_{lam}\geq0.15$. 
The realisation of the burst event is then taken to be $T_b-100\leq t \leq T_b+100$. 
(The reasons for choosing such time-windows, instead of the more generally used consecutive time-blocks, will become apparent when considering the energy of the wavelet coefficients in figure \ref{fig:windowing} of the next section.) 
For the DNS dataset considered here, we split the data to obtain $N_e=50$ realisations of the burst event, with each realisation having $N_t=200$ and $dt=1$, and therefore $T=200$. 
It is ensured that no two realisations of the burst event have more than $50\%$ overlap. 

Let us now consider the Fourier transform of the vorticity $\omega(x,y,t)$ in time,   
\begin{equation}
\omega(x,y,t) = \sum_{\Omega=0}^{N_t-1} \breve{\omega}(x,y;\Omega)e^{i(2\pi/T)\Omega t},  
\end{equation}
where $(\breve{\cdot})$ here represents the temporal Fourier transform. 
The corresponding Fourier spectrum is 
$
E_{\omega}(\Omega) = \left[ \int_0^{2\pi} \int_0^{2\pi} \breve{\omega} \breve{\omega}^* \mbox{ } dx dy \right]_{N_e}, 
$
with $(\cdot)^*$ representing complex conjugate. 
Here $\left[\vphantom{l}\cdot\vphantom{l}\right]_{N_e}$ represents an averaging across the $N_e$ different realisations of the burst event. 
Figure \ref{fig:DNSSplit_fourier}(\subref{fig:DNSSplit_Fourier_spec}) shows the obtained Fourier spectrum. 
To isolate the quiet region and the burst events, we use distinct sets of frequencies. 
The blue-shaded frequencies in figure \ref{fig:DNSSplit_fourier}(\subref{fig:DNSSplit_Fourier_spec}) are used for the quiet region, and the red shaded frequencies for the burst event. 
Let us denote the reconstruction of the data using this truncated Fourier basis as $\omega_{r}(x,y,t)$. 
Figure \ref{fig:DNSSplit_fourier}(\subref{fig:DNSSplit_Fourier_timeseries}) shows the time-series of the reconstructions $\langle \omega_r, \omega_r \rangle_{x,y}/D_{lam}$ of the quiet region in blue and the burst event in red. 
These reconstructions are compared with the full data $\langle \omega, \omega \rangle_{x,y}/D_{lam}$ in black. 

The number of modes required for isolating the different regions is identified by defining an error $\epsilon(t)$ in the projection of $\omega_{r}(x,y,t)$ onto $\omega(x,y,t)$, where $\epsilon(t) := 1 - \langle \omega_r, \omega \rangle_{x,y}/\langle \omega, \omega \rangle_{x,y}$. 
Let $\left[\vphantom{l}\cdot\vphantom{l}\right]_{t}$ denote averaging in time. 
The blue-shaded frequencies in figure \ref{fig:DNSSplit_fourier}(\subref{fig:DNSSplit_Fourier_spec}) ensure that $\left[ \epsilon(t_q)\right]_t < 0.2$ where $t_q$ is defined as the times $t$ when $D(t)/D_{lam} \leq 0.1$, i.e.\ the quiet region.
Similarly, the red-shaded frequencies in figure \ref{fig:DNSSplit_fourier}(\subref{fig:DNSSplit_Fourier_spec}) ensure that $\left[ \epsilon(t_b)\right]_t < 0.4$ where $t_b$ is defined as the times $t$ when $D(t)/D_{lam} \geq 0.15$, i.e.\ the burst event. 
The numerical values of $\left[ \epsilon(t_q)\right]_t < 0.2$ and $\left[ \epsilon(t_b)\right]_t < 0.4$ simply ensure that the dominant frequencies (or wavelets, as in the next section) are included, and changing the values does not significantly impact any of the discussions in this work (see Appendix \ref{sec:insensitivity of prediction methods to parameter choices}). 

From figure \ref{fig:DNSSplit_fourier}(\subref{fig:DNSSplit_Fourier_timeseries}), we find that the Fourier basis is not able to efficiently isolate the burst event, because the red line does not go to zero in the quiet regions of the flow.  
Additionally, we require a significant number of frequencies to reconstruct the burst event. 
This is expected since we are trying to represent a localised event using the Fourier basis that is global. 
In order to remedy these problems, in the next section we will explore using a localised basis -- wavelets -- for this purpose.  

\subsection{Daubechies 1 wavelet basis and wavelet transform}
\label{sec:daubechies 1 wavelet basis and wavelet transform}

%
%
\begin{figure}[t]
\captionsetup[subfigure]{labelformat=empty,skip=-10pt}
\begin{subfigure}[b]{\textwidth}
\centering
\includegraphics[width=\textwidth]{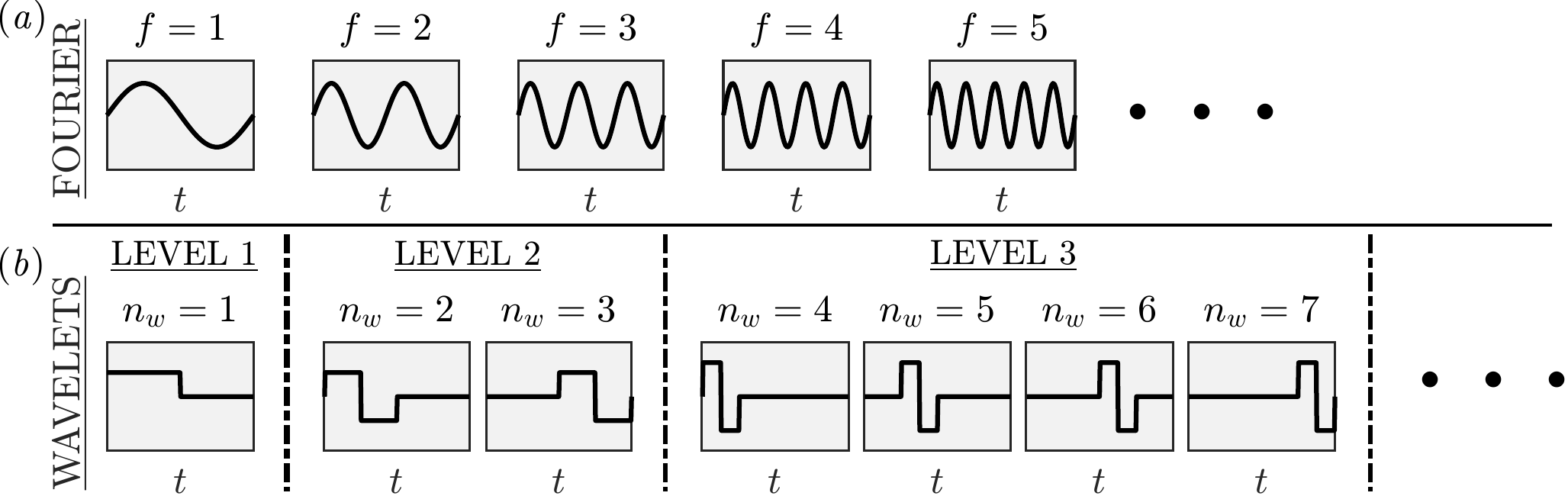}
\caption{}
\label{fig:wavelet_schematic_fourier}
\end{subfigure}%
\begin{subfigure}[b]{\textwidth}
\centering
\caption{}
\label{fig:wavelet_schematic_wavelet}
\end{subfigure}
\caption{Schematic showing (\textit{a}) the Fourier basis and (\textit{b}) the discrete wavelet basis of Daubechies 1. 
The vertical dashed-dot lines in (\textit{b}) demarcate the different levels of the wavelet basis, and will be used again in figure \ref{fig:DNSSplit_wavelet}(\subref{fig:DNSSplit_Wavelet_spec}) when plotting the energy of the wavelet coefficients. }
\label{fig:wavelet_schematic}
\end{figure}

%
%

%
%
\begin{figure}[t]
\captionsetup[subfigure]{labelformat=empty,skip=-25pt}
\begin{subfigure}[b]{0.75\textwidth}
\centering
\includegraphics[width=\textwidth]{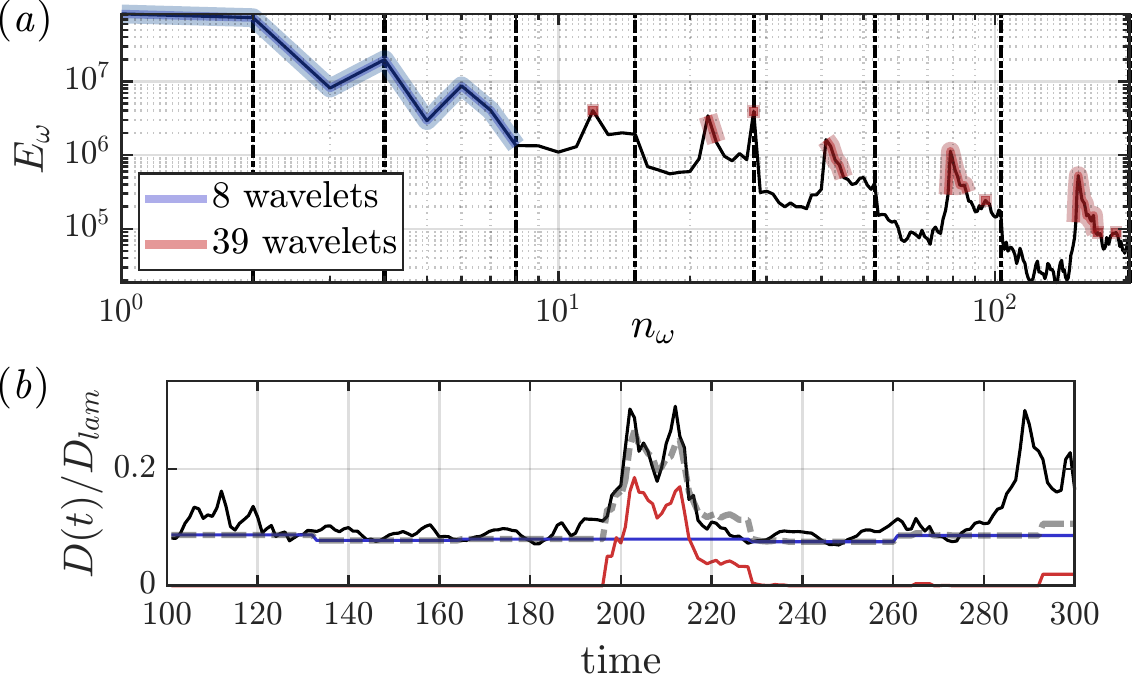}
\caption{}
\label{fig:DNSSplit_Wavelet_spec}
\end{subfigure}
\begin{subfigure}[b]{0.75\textwidth}
\centering
\caption{}
\label{fig:DNSSplit_Wavelet_timeseries}
\end{subfigure}
\caption{(\textit{a}) The ensemble-averaged energy of the wavelet coefficients is shown in black, and the wavelets used to reconstruct the separate regions in (\textit{b}) are shaded in different colours: the $8$ wavelets for the quiet region are shaded in blue and the $39$ wavelets for the burst event are in red. 
The vertical dashed lines indicate the different levels of the wavelet transform (see figure \ref{fig:wavelet_schematic}). 
(\textit{b}) The full data for $k_x={0,1,2,3}$ (black) is compared to the reconstructions (using the shaded frequencies in (\textit{a})) of the quiet region (blue) and the burst event (red). 
For comparison, the grey dashed-dot line shows the sum of the blue and the red lines. 
}
\label{fig:DNSSplit_wavelet}
\end{figure}

We now employ a wavelet basis to capture the burst events. 
The discrete Daubechies 1 (DB1) wavelet, which is also referred to as the Haar wavelet, will be used for a majority of this work. 
The only exception is Appendix \ref{sec:prediction using daubechies 2 wavelets}, where for a different choice of wavelets, that of Daubechies 2, we see that the discussions in this work remain similar. 

To obtain a wavelet basis $\Theta_{n_w}(t)$, we first make a choice of the wavelet corresponding to $n_w=1$, i.e.\ $\Theta_{1}(t)$, which is the mother wavelet. 
For the Daubechies 1 wavelet basis used in this work, the mother wavelet is a step function, as shown in the first panel of figure \ref{fig:wavelet_schematic}(\subref{fig:wavelet_schematic_wavelet}). 
The mother wavelet covers the entire time domain, and here we denote this mother wavelet as belonging to level 1 of the wavelet basis. 
To get level 2 of this wavelet basis, the mother wavelet is compressed by half. 
Now we need two wavelets to cover the entire domain, and these are shown as $n_w=2$ and $n_w=3$ in figure \ref{fig:wavelet_schematic}(\subref{fig:wavelet_schematic_wavelet}). 
Level 3 consists of the mother wavelet compressed by a factor of four, and then repeated four times to cover the domain. 
This level is as shown by $n_w=4-7$ in figure \ref{fig:wavelet_schematic}(\subref{fig:wavelet_schematic_wavelet}). 
The vertical black dashed-dot lines in figure \ref{fig:wavelet_schematic}(\subref{fig:wavelet_schematic_wavelet}) demarcate the different levels of the wavelet transform. 
Throughout this manuscript, we use such vertical black dashed-dot lines to demarcate the levels. 
(It should be noted that, in practise, there is a level $0$ corresponding to $n_w=0$ that contains the lowest frequencies of the data that are not included in the other levels. 
In more technical terms, the wavelet transform for $N=200$ includes the `detail coefficients' from $7$ levels of the transform as well as the `approximation coefficient' from level $1$ \citep[see][]{daubechies1988orthonormal}.)
We see that, when considering wavelets, there are two factors that are important: (i) the compression of the wavelet, which is related to the concept of frequency in the Fourier domain and (ii) the location of the wavelet in time. 

Similar to the Fourier transform in \S\ref{sec:decomposing the flow: fourier bases}, we obtain a wavelet transform of $\omega(x,y,t)$ as
\begin{equation}
\omega(x,y,t) = \sum_{n_w=0}^{N_w} \widehat{\omega}(x,y;n_w)\Theta_{n_w}(t), 
\end{equation}
where $(\widehat{\cdot})$ represents the transform onto the basis $\Theta_{n_w}$, which is here the wavelet basis. 
The energy of the wavelet coefficients can then be obtained as 
$
E_{\omega}(n_\omega) = \left[ \int_0^{2\pi} \int_0^{2\pi} \widehat{\omega} \widehat{\omega}^* \mbox{ } dx dy \right]_{N_e}
$. 
It should be noted that the $\Theta_{n_w}(t)$ used in this work is real-valued. 
However, since it is a complete basis, it provides a basis for both real-valued and complex-valued data. 
This is relevant when we later consider the wavelet transform of the 1-D Fourier transformed (in the $x$-direction) data. 

A wavelet transform requires a choice of boundary conditions and throughout this manuscript, periodic boundary conditions have been used. 
The WPOD described later gives similar results for other physically relevant boundary conditions, such as symmetric and reflecting boundary conditions. 

\subsection{Burst centred windowing}
\label{sec:Burst centred windowing}

Before looking at the energy of the wavelet coefficients, let us briefly discuss the choice of the $N_e$ realisations of the burst event. 
Generally, to obtain the different ensembles for ensemble averaging, we split the available data into several consecutive data-blocks of equal length, and compute the spectrum for each of these ensembles. 
The spectra are then averaged across the ensembles. 
The energy of the wavelet coefficients obtained from such consecutive ensembles is shown in figure \ref{fig:windowing}(\subref{fig:windowing_NoCentering_spec}). 
Within each of the levels (indicated by the vertical dashed-dot lines), we see that there are no noticeable trends, and peaks are notably absent. 
This absence of peaks can be explained by recalling that the location in time is important when considering wavelets. 
In this case, for each ensemble, the burst events occur at different locations in time and therefore appear at different $n_w$. 

To remedy this, we here adopt the alternative windowing strategy that was briefly mentioned in \S\ref{sec:decomposing the flow: fourier bases} and is depicted in figure \ref{fig:windowing}(\subref{fig:windowing_windows}). 
A realisation of the burst event is defined using the start of a burst event $T_b$ as $T_b-100\leq t \leq T_b+100$, where $T_b$ is identified using the condition $D(t)/D_{lam}\geq0.15$ (note, time is here non-dimensionalised by $\sqrt{L_y/2\pi\zeta}$).  
As in \S\ref{sec:decomposing the flow: fourier bases}, here we consider $N_e=50$ realisations of the burst event, each containing $N_t=200$ time-snapshots of vorticity $\omega(x,y,t)$ spaced apart by time $dt=1$. 
The spectrum obtained from this new windowing strategy is shown in figure \ref{fig:DNSSplit_wavelet}(\subref{fig:DNSSplit_Wavelet_spec}). 
From the red shaded regions of the spectrum in figure \ref{fig:DNSSplit_wavelet}(\subref{fig:DNSSplit_Wavelet_spec}), we observe clear peaks that are present because of the burst events. 
In the next section, we will more clearly show that these peaks correspond to the burst events. 

\subsection{Decomposing the flow: Wavelet bases}
\label{sec:decomposing the flow: wavelet bases}

The number of levels present in the spectrum is determined by the length of the time window (here $T=200$) and the type of wavelet used (here DB1). 
In figure \ref{fig:DNSSplit_wavelet}(\subref{fig:DNSSplit_Wavelet_spec}) there are $8$ levels that are demarcated by the vertical dashed-dot lines. 
We here use distinct sets of wavelets to isolate the quiet region and the burst events. 
The first $3$ levels will be used to reconstruct the quiet region, and these are shaded in blue. 
Next, $20$\% of the most energetic modes in levels $4$-$8$ will be used to reconstruct the burst event, and these are shaded in red (Appendix \ref{sec:insensitivity of prediction methods to parameter choices} shows the insensitivity of the results presented to this specific choice of $20\%$ of the wavelets). 
As in \S\ref{sec:decomposing the flow: fourier bases}, the number of wavelets used to reconstruct the flow is chosen such that the error $\left[ \epsilon(t_q)\right]_t < 0.2$ and $\left[ \epsilon(t_b)\right]_t < 0.4$ (see \S\ref{sec:decomposing the flow: fourier bases} for the definition of the error metric). 
Let us denote the data reconstructed using a truncated wavelet basis as $\omega_{r}(x,y,t)$. 
Figure \ref{fig:DNSSplit_wavelet}(\subref{fig:DNSSplit_Wavelet_timeseries}) shows the time-series of the reconstructions $\langle  \omega_{r},\omega_{r} \rangle_{x,y}/D_{lam}$ of the quiet region in blue and the burst event in red. 
The reconstructions are compared with the full data $\langle  \omega,\omega \rangle_{x,y}/D_{lam}$ in black. 
We observe that isolating the quiet region and the burst events is approximately possible using wavelets. 

In comparing the wavelet-based reconstruction in figure \ref{fig:DNSSplit_wavelet}(\subref{fig:DNSSplit_Wavelet_timeseries}) to the Fourier-based reconstruction in figure \ref{fig:DNSSplit_fourier}(\subref{fig:DNSSplit_Fourier_timeseries}), three observations stand out.
Firstly, and most importantly, wavelets are able to isolate the burst events better than the Fourier basis. 
To understand why, let us turn our attention to the relatively small magnitude (relative to the burst events) oscillations in the quiet region of the flow. 
When considering the Fourier basis, the same Fourier frequencies contribute both to these quiet oscillations and the burst events. 
However, since these oscillations occur at different locations in time relative to the burst events, the wavelet basis is able to produce reconstructions of the burst event uncontaminated by the quiet oscillations. 
Secondly, to reconstruct the burst event, fewer wavelets are required in comparison to Fourier frequencies: $39$ wavelets as opposed to $108$ ($54$ positive) Fourier frequencies. 
Finally, from figure \ref{fig:DNSSplit_wavelet}(\subref{fig:DNSSplit_Wavelet_timeseries}) we see that wavelets are able to isolate a single burst event, in contrast to the Fourier-based reconstruction in figure \ref{fig:DNSSplit_fourier}(\subref{fig:DNSSplit_Fourier_timeseries}) where both the burst events are captured equally. 
Therefore, isolating single burst events, when there are multiple similar events present in a time window, is only possible using the wavelets bases. 

We can therefore conclude that wavelets are better at isolating burst events. 
Using wavelets, we are able to obtain a signal of the burst event that is uncontaminated by the oscillations in the quiet region. 
Additionally, wavelets enable a lower-order representation of the burst event by requiring fewer wavelets (than Fourier frequencies) to reconstruct the burst event. 
In the next section, we will therefore concentrate on using the wavelet basis to probe the flow patterns active in the quiet region and the burst events. 

\section{Prediction method 1: Wavelet-POD (WPOD) based}
\label{sec:coherent structures of the flow: a wavelet-pod analysis}

Proper Orthogonal Decomposition (POD), introduced by \citet{lumley1967structure, lumley1970stochastic}, is a common technique employed to find the flow patterns that are energetically dominant. 
In this section, we consider two POD-based methods that use a wavelet basis: (i) WPOD (\S\ref{sec:a description of wpod} and \S\ref{sec:wpod spectrum and modes for the Kolmogorov flow}) and (ii) its variation composite-WPOD (\S\ref{sec:coherent structures in quiet and burst regions - a composite-wpod Analysis}). 
The aim of these methods is the identification of the coherent structures in an intermittent flow. 
Here, a coherent structure is defined as a flow pattern that maintains a significant degree of correlation with itself over a range of space and time \citep{robinson1991coherent}. 
After identifying the flow patterns, we use them to predict the burst events (\S\ref{sec:tracking composite-wpod modes} and \S\ref{sec:predicting burst events using wpod modes - method 2}). 

\subsection{A description of WPOD}
\label{sec:a description of wpod}

Many adaptations of POD, tailored for various classes of problems, are found in the literature (for instance, see reviews by \citet{taira2017modal} and \citet{rowley2017model} and references therein). 
We are here interested in finding structures that are coherent in space and time, and a recently introduced POD-based technique for this purpose is spectral-proper-orthogonal-decomposition (SPOD) \citep{towne2018spectral}. 
In this method, the flow is first projected onto a Fourier basis.
Thereafter, a proper orthogonal decomposition (POD) is performed at each Fourier frequency that gives coherent structures at that particular frequency. 
These structures are coherent in space due to the properties of POD and coherent in time since they are Fourier modes. 
However, in \S\ref{sec:decomposing the flow: wavelet bases}, we saw that a wavelet basis is better at characterising intermittent flows. 
Therefore, instead of a SPOD, here we use a wavelet-proper orthogonal decomposition (WPOD). 
Using this method we obtain structures that are coherent in space, and both coherent and localised in time since they are wavelet coefficients. 

Before describing WPOD, to provide context, let us briefly consider the regular POD. 
In this case, we have velocity fields $\bm{q}(x,y,t)=(u(x,y,t),v(x,y,t))$ for a range of time. 
(For POD, we are required to choose a norm that is maximised to obtain the POD modes. 
Kinetic energy is a physically relevant norm, and choosing velocity as data, instead of vorticity, ensures that the kinetic energy is used as the POD norm.)
The POD modes $\bm{\phi}_i(x,y)$ give an orthogonal basis for $\bm{q}(x,y,t)$, such that the $1^{st}$ POD mode captures the largest variance of the data, the $i^{th}$ POD mode captures the $i^{th}$ largest variance, and so on. 
In other words, if $\mathcal{E}^{pod}_i$ is the projection of the data $\bm{q}(x,y,t)$ onto the $i^{th}$ POD mode, $\bm{\phi}_i(x,y)$
\begin{equation}
\mathcal{E}^{pod}_i = \frac{\left[\left| \langle \bm{q}(x,y,t),\bm{\phi}_i(x,y)\rangle_{x,y} \right|^2\right]_t}{\langle \bm{\phi}_i(x,y),\bm{\phi}_i(x,y)\rangle_{x,y}},  
\label{eqn:POD_projection}
\end{equation}
then $\mathcal{E}^{pod}_i>\mathcal{E}^{pod}_{i+1}$. 
Here, as before, $\left[\cdot\right]_t$ denotes averaging in time, $\langle a,b \rangle_{x,y}$ denotes the inner product $\int_{0}^{L_y} \int_{0}^{L_x} b^*(x,y,t) a(x,y,t) dx dy$, and $\left|\cdot\right|$ represents absolute value. 
The problem of finding POD modes reduces to finding the eigenvalues of the correlation matrix $QQ^*$ where $Q$ is a matrix with columns containing $\bm{q}(x,y,t)$, i.e. the $1$st column of $Q$ is $\bm{q}(x,y,t_1)$ and the $i^{th}$ column is $\bm{q}(x,y,t_i)$. 
The eigenvectors of $QQ^*$ are equivalently the left singular vectors of the matrix $Q$, and therefore the problem can be solved by performing a singular value decomposition (SVD) of the data-matrix $Q$.  

Rather than finding the coherent structures in the entire flow using POD, we are interested in finding the coherent structures at particular wavelets.
Let us therefore consider wavelet-POD. 
For this, first consider the $N_e$ different realisations of the burst event (see \S\ref{sec:Burst centred windowing} for how each realisation is chosen from the data). 
From these $N_e$ realisations, consider one realisation $n_e$, where now $1 \leq n_e \leq N_e$. 
At this $n_e$, we have the data $\bm{q}(x,y,t;n_e) = (u(x,y,t;n_e), v(x,y,t;n_e))$. 
A wavelet transform of this $n_e^{th}$ realisation of the burst event will give us  $\widehat{\bm{q}}(x,y;n_w,n_e)=(\widehat{u}(x,y;n_w,n_e),\widehat{v}(x,y;n_w,n_e))$, where $\widehat{u}$ and $\widehat{v}$ represent the wavelet transform of $u$ and $v$. 
Therefore, for each wavelet $n_w$, and from each realisation of the burst event $n_e$, we have the state vector $\widehat{\bm{q}}(x,y;n_w,n_e)$. 
If we now concentrate on just one wavelet $n_w$, we have $N_e$ data vectors $\widehat{\bm{q}}(x,y;n_w,1:N_e)$. 

The objective now is to find basis vectors $\widehat{\bm{\phi}}_i(x,y;n_w)$ such that
\begin{equation}
\widehat{\bm{q}}(x,y;n_w,n_e) = \sum_{i=0}^{N_e} a_i(n_w,n_e) \widehat{\bm{\phi}}_i(x,y;n_w). 
\end{equation}
For each $n_w$, the equivalent of the POD energy in \eqref{eqn:POD_projection} now becomes 
\begin{equation}
\mathcal{E}^{wpod}_i(n_w) = \frac{\left[\left| \langle \widehat{\bm{q}}(x,y;n_w,n_e),\widehat{\bm{\phi}}_i(x,y;n_w)\rangle_{x,y} \right|^2\right]_{n_e}}{\langle \widehat{\bm{\phi}}_i(x,y;n_w),\widehat{\bm{\phi}}_i(x,y;n_w)\rangle_{x,y}},  
\label{eqn:WPOD_projection_nw}
\end{equation}
where $\mathcal{E}^{wpod}_i(n_w)\geq \mathcal{E}^{wpod}_{i+1}(n_w)$. 
It should be noted that, unlike in \eqref{eqn:POD_projection} where the averaging is over time, in \eqref{eqn:WPOD_projection_nw} the averaging is across the $N_e$ different realisations of the burst event (denoted by $\left[\cdot\right]_{n_e}$). 
If we consider the WPOD modes across all $n_w$ and pick the dominant mode, i.e.\ the mode with maximum energy across $n_w$, then the projection that is maximised is
\begin{equation}
\max_{n_w}(\mathcal{E}^{wpod}_1(n_w)) = \frac{\left[\left| \langle \bm{q}(x,y,t;n_e),\bm{\phi}_1(x,y,t)\rangle_{x,y,t} \right|^2\right]_{n_e}}{\langle \bm{\phi}_1(x,y,t),\bm{\phi}_1(x,y,t)\rangle_{x,y,t}},  
\label{eqn:WPOD_projection_full}
\end{equation}
where $\langle a,b \rangle_{x,y,t}$ denotes the inner product $\int_{0}^{T} \int_{0}^{L_y} \int_{0}^{L_x} b^*(x,y,t) a(x,y,t) dx dy dt$. 
Here $\bm{\phi}_i(x,y,t)$ represents the WPOD mode $\widehat{\bm{\phi}}_i(x,y;n_w)$ in time, i.e.\ $\bm{\phi}_i(x,y,t)$ is an inverse wavelet transform of $\widehat{\bm{\phi}}_i(x,y;n_w)$. 
Similar to POD, for WPOD we perform a SVD of the data matrix $\widehat{Q}_{n_w}$, where columns of $\widehat{Q}_{n_w}$ are the data $\widehat{\bm{q}}(x,y;n_w,1:N_e)$. 
The $i^{th}$ left singular vector is the $i^{th}$ WPOD mode, and the square of the $i^{th}$ singular value is the corresponding WPOD energy $\mathcal{E}^{wpod}_i(n_w)$ \eqref{eqn:WPOD_projection_nw}. 

\subsection{WPOD-based analysis of the 2D Kolmogorov flow}
\label{sec:wpod spectrum and modes for the Kolmogorov flow}

%
%
\begin{figure}[t]
\captionsetup[subfigure]{labelformat=empty,skip=-5pt}
\begin{subfigure}[b]{\textwidth}
\centering
\includegraphics[width=0.8\textwidth]{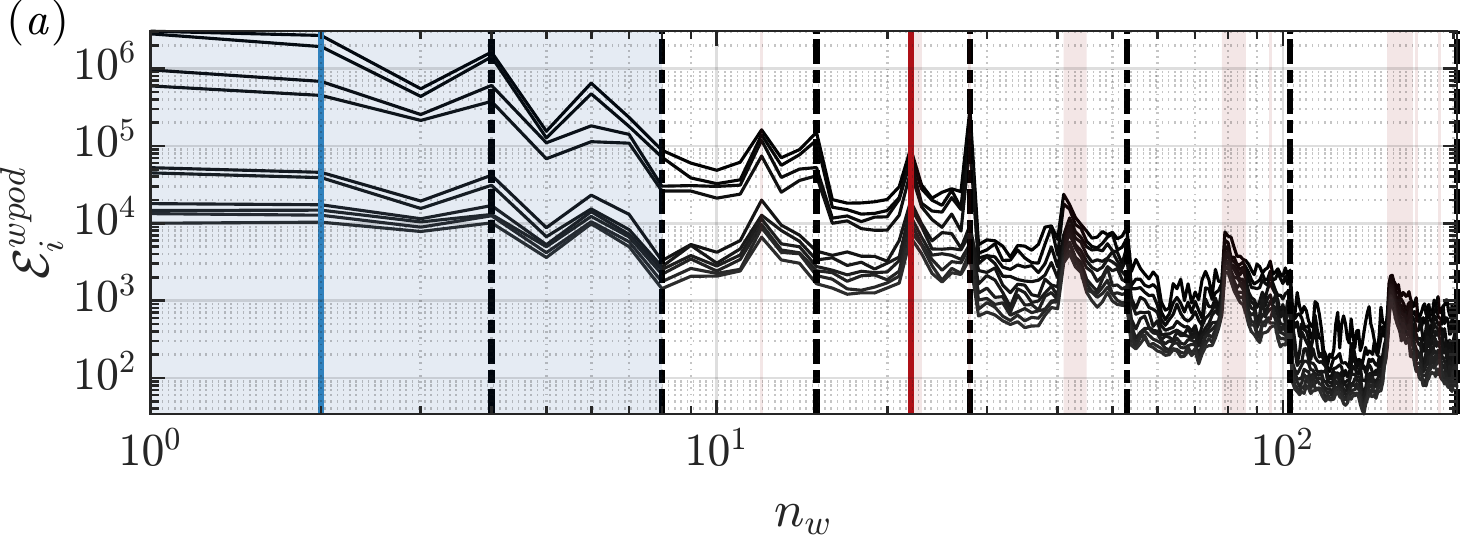}
\caption{}
\label{fig:WPOD_spectrum}
\end{subfigure}
\begin{subfigure}[b]{\textwidth}
\centering
\includegraphics[width=1\textwidth]{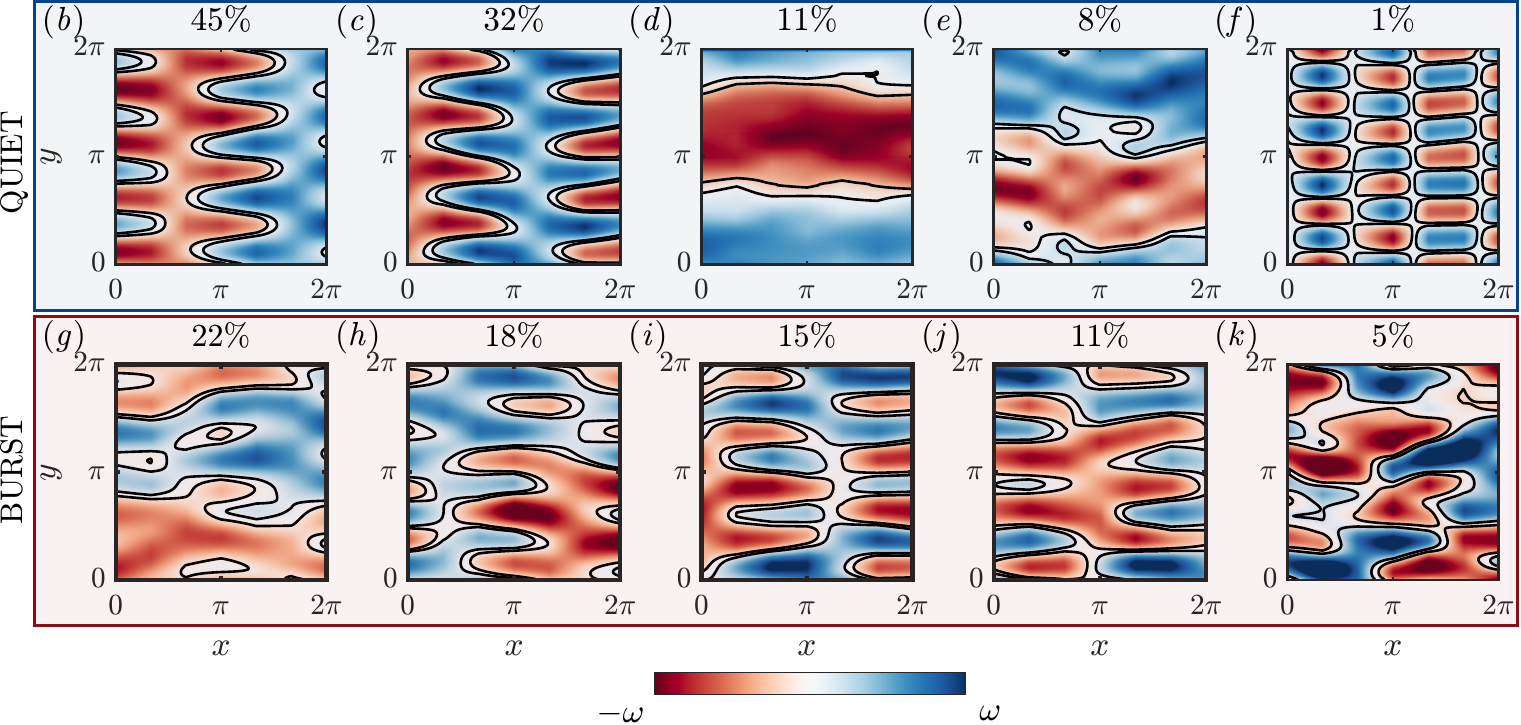}
\caption{}
\label{fig:WPOD_structures_1}
\end{subfigure}%
\begin{subfigure}[b]{\textwidth}
\caption{}
\label{fig:WPOD_structures_2}
\end{subfigure}%
\begin{subfigure}[b]{\textwidth}
\caption{}
\label{fig:WPOD_structures_3}
\end{subfigure}%
\begin{subfigure}[b]{\textwidth}
\caption{}
\label{fig:WPOD_structures_4}
\end{subfigure}%
\begin{subfigure}[b]{\textwidth}
\caption{}
\label{fig:WPOD_structures_5}
\end{subfigure}%
\begin{subfigure}[b]{\textwidth}
\caption{}
\label{fig:WPOD_structures_6}
\end{subfigure}%
\begin{subfigure}[b]{\textwidth}
\caption{}
\label{fig:WPOD_structures_7}
\end{subfigure}%
\begin{subfigure}[b]{\textwidth}
\caption{}
\label{fig:WPOD_structures_8}
\end{subfigure}%
\begin{subfigure}[b]{\textwidth}
\caption{}
\label{fig:WPOD_structures_9}
\end{subfigure}%
\begin{subfigure}[b]{\textwidth}
\caption{}
\label{fig:WPOD_structures_10}
\end{subfigure}%
\caption{(\textit{a}) The WPOD spectrum is shown as a function of wavelet $n_w$. 
The energies of the first $10$ WPOD modes are shown, with the lighter lines representing the higher modes. 
The vertical dashed-dot lines demarcate the levels of the wavelet transform.  
The shaded regions represent the wavelets that are responsible for the quiet region in blue and the burst event in red. 
The first $5$ WPOD modes are shown for two different wavelets: (\textit{b}-\textit{f}) $n_w=2$ responsible for the quiet region marked by the blue vertical line in (\textit{a}) and (\textit{g}-\textit{k}) $n_w=22$ responsible for the burst event marked by the red vertical line in (\textit{a}).
}
\label{fig:WPOD_spectrum_structures}
\end{figure}

Figure \ref{fig:WPOD_spectrum_structures}(\subref{fig:WPOD_spectrum}) shows the WPOD energies $\mathcal{E}^{wpod}_i(n_w)$ \eqref{eqn:WPOD_projection_nw} as a function of wavelet $n_w$. 
In other words, each vertical line in figure \ref{fig:WPOD_spectrum_structures}(\subref{fig:WPOD_spectrum}) corresponds to the WPOD obtained at that particular $n_w$.
Let us first focus on the wavelets responsible for the quiet region shaded in blue, i.e.\ the first three levels of the wavelet transform. 
Notably, only these lower levels show any appreciable low-rank behaviour, i.e.\ within these levels the first few WPOD modes capture significantly more energy than the higher modes. 
Low-rankness in the WPOD spectrum suggests that there is an energetically dominant mechanism that is responsible for the quiet region of the flow. 
To probe this further, let us now look at the WPOD modes at $n_w=2$ in figures \ref{fig:WPOD_spectrum_structures}(\subref{fig:WPOD_structures_1}-\subref{fig:WPOD_structures_5}).  
The WPOD modes $1$ and $2$, as well as modes $3$ and $4$, are shifted versions of the same mode, a result of the inherent symmetries in the flow. 
(While symmetries can be incorporated into the POD modes, the instantaneous fluctuations of the flow do not adhere to these symmetries, and so we present modes without symmetrisation).  
Together, modes $1$ and $2$ account for $77$\% of the energy at this $n_w$.  
Crucially, these first two modes closely resemble the unstable eigenfunction obtained from the Navier-Stokes equations linearized around the mean flow (see appendix \ref{sec:unstable eigenvector of the linearized navier--stokes equations}).  
This indicates that this unstable eigenfunction is mainly responsible for the quiet region of this flow. 
This is consistent with the observations in \citet{farazmand2017variational}, where they employed an alternative strategy to capture this unstable eigenfunction, and used it to predict the burst events. 
Moving on to modes $3$ and $4$, while these modes are energetically significant, they capture less energy compared to modes $1$ and $2$. 
Together they capture $19\%$ of the energy at this $n_w$. 
Structurally, modes $3$ and $4$ capture shearing motions. 

Let us now turn our attention to the wavelets responsible for the burst event. 
These wavelets are shaded in red in figure \ref{fig:WPOD_spectrum_structures}(\subref{fig:WPOD_spectrum}). 
One initial observation is that there is no significant low-rankness at a majority of these wavelets. 
This lack of low-rankness generally suggests that there is no one dominant mechanism that is responsible for the burst event. 
To investigate these modes further, consider the WPOD modes at $n_w=22$ in figures \ref{fig:WPOD_spectrum_structures}(\subref{fig:WPOD_structures_6}-\subref{fig:WPOD_structures_10}). 
Notably, the first $4$ WPOD modes in the burst event are structurally similar to the modes in the quiet region.  
Additionally, we observe that modes $2$-$4$, while being similar to the unstable eigenfunction, are fragmented versions of this eigenfunction. 
It should be noted that, this fragmentation of the modes is more apparent for certain $n_w$ in the burst region (for instance, it is not as clearly apparent for $n_w=12$ in level 4). 
Therefore, for a more conclusive illustration of these fragmented modes, we will consider composite-WPOD modes in the next section. 

From these observations, we can hypothesise that the shearing motions are responsible for disrupting the flow due to the unstable eigenfunction. 
Intermittently, this shearing disrupts the flow enough to cause a burst event.  
While a WPOD analysis can suggest that such a mechanism causes the burst events, to conclusively show this, we need to do a dynamic mode decomposition \citep{schmid2022dynamic} analysis that identifies the relevant instabilities of the flow generated by the unstable eigenfunction. 
However, this falls beyond the scope of the current manuscript. 

\subsection{Coherent Structures in Quiet and Burst Regions - A Composite-WPOD Analysis}
\label{sec:coherent structures in quiet and burst regions - a composite-wpod Analysis}

%
%
\begin{figure}[t]
\captionsetup[subfigure]{labelformat=empty,skip=-50pt}
\begin{subfigure}[b]{\textwidth}
\centering
\includegraphics[width=1\textwidth]{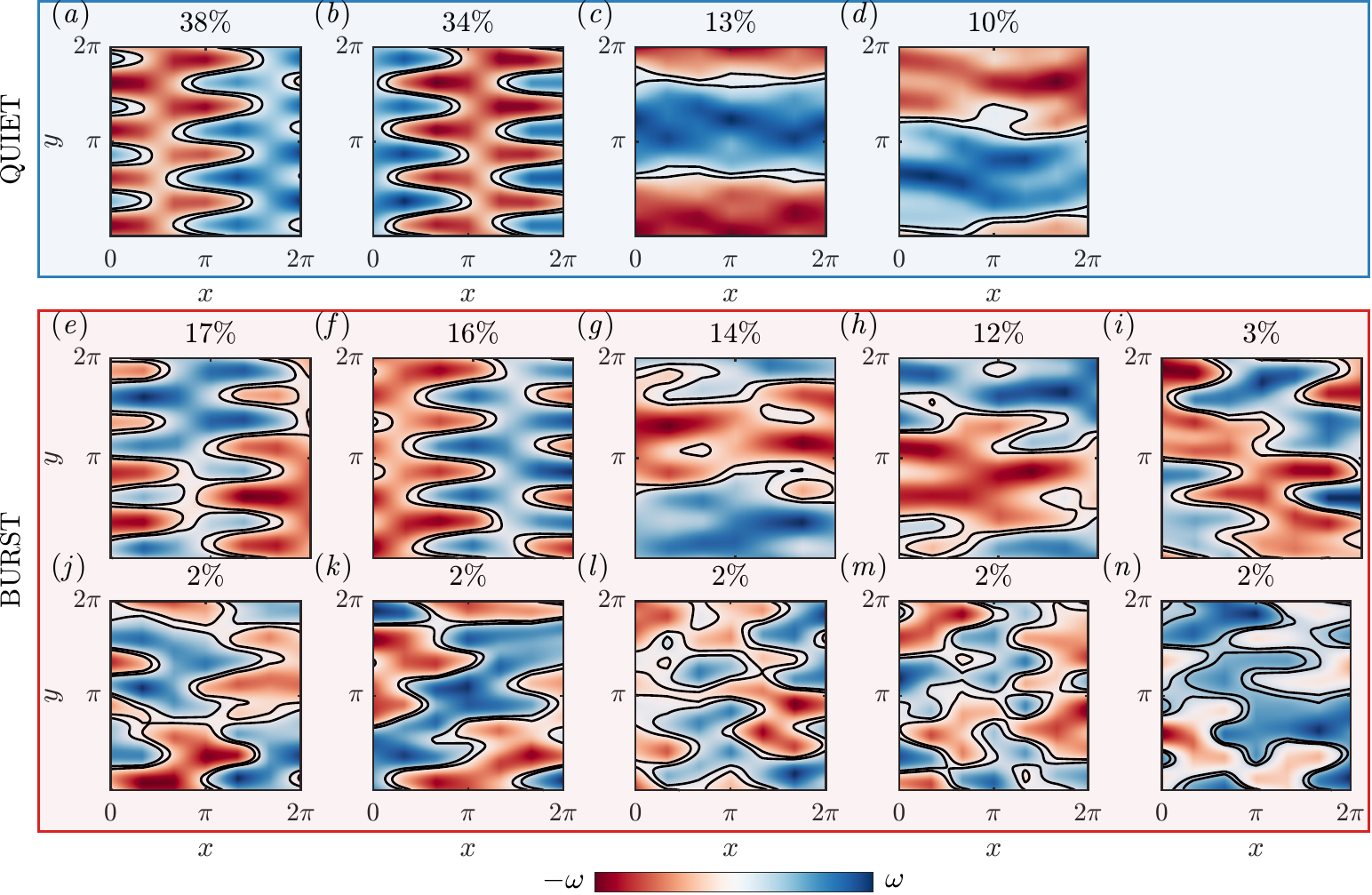}
\caption{}
\label{fig:db1_CoherentStructures_nwcollection_1}
\end{subfigure}%
\begin{subfigure}[b]{\textwidth}
\centering
\caption{}
\label{fig:db1_CoherentStructures_nwcollection_2}
\end{subfigure}%
\begin{subfigure}[b]{\textwidth}
\centering
\caption{}
\label{fig:db1_CoherentStructures_nwcollection_3}
\end{subfigure}%
\begin{subfigure}[b]{\textwidth}
\centering
\caption{}
\label{fig:db1_CoherentStructures_nwcollection_4}
\end{subfigure}%
\begin{subfigure}[b]{\textwidth}
\centering
\caption{}
\label{fig:db1_CoherentStructures_nwcollection_5}
\end{subfigure}%
\begin{subfigure}[b]{\textwidth}
\centering
\caption{}
\label{fig:db1_CoherentStructures_nwcollection_6}
\end{subfigure}%
\begin{subfigure}[b]{\textwidth}
\centering
\caption{}
\label{fig:db1_CoherentStructures_nwcollection_7}
\end{subfigure}%
\begin{subfigure}[b]{\textwidth}
\centering
\caption{}
\label{fig:db1_CoherentStructures_nwcollection_8}
\end{subfigure}%
\begin{subfigure}[b]{\textwidth}
\centering
\caption{}
\label{fig:db1_CoherentStructures_nwcollection_9}
\end{subfigure}%
\begin{subfigure}[b]{\textwidth}
\centering
\caption{}
\label{fig:db1_CoherentStructures_nwcollection_10}
\end{subfigure}
\begin{subfigure}[b]{\textwidth}
\centering
\caption{}
\label{fig:db1_CoherentStructures_nwcollection_11}
\end{subfigure}
\begin{subfigure}[b]{\textwidth}
\centering
\caption{}
\label{fig:db1_CoherentStructures_nwcollection_12}
\end{subfigure}
\begin{subfigure}[b]{\textwidth}
\centering
\caption{}
\label{fig:db1_CoherentStructures_nwcollection_13}
\end{subfigure}
\begin{subfigure}[b]{\textwidth}
\centering
\caption{}
\label{fig:db1_CoherentStructures_nwcollection_14}
\end{subfigure}
\captionsetup[subfigure]{labelformat=empty,skip=-20pt}
\begin{subfigure}[b]{\textwidth}
\centering
\includegraphics[width=0.75\textwidth]{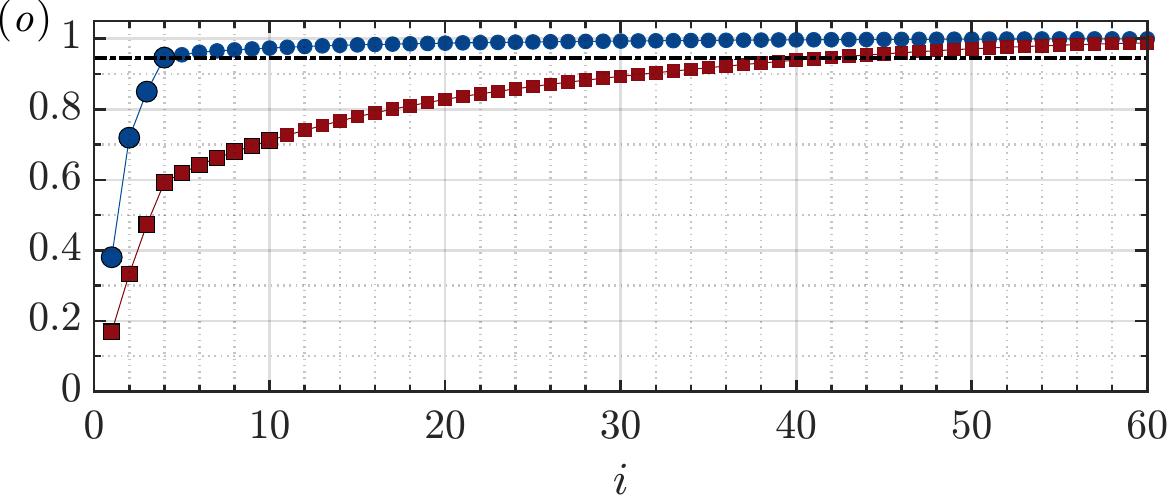}
\caption{}
\label{fig:db1_CoherentStructures_nwcollection_spec}
\end{subfigure}
\caption{The first few composite-WPOD modes are shown for two different sets of wavelets: 
(\textit{a}-\textit{d}) wavelets responsible for the quiet region marked by the blue shaded region in figure \ref{fig:WPOD_spectrum_structures}(\subref{fig:WPOD_spectrum}) and 
(\textit{e}-\textit{n}) wavelets responsible for the burst event marked by the red shaded regions in figure \ref{fig:WPOD_spectrum_structures}(\subref{fig:WPOD_spectrum}).  
The titles of these plots show the percentage energy captured by the mode computed as $100\times\mathcal{E}_i^q/\sum_{j=0}^M \mathcal{E}_j^q$ for the quiet region, and similarly for the burst events. 
(\textit{o}) Also shown is the cumulative contribution of the first $i$ modes to the total energy for the quiet regions (blue line, circle markers) and the burst events (red line, square markers). }
\label{fig:db1_CoherentStructures_nwcollection}
\end{figure}

In this section, we introduce composite-WPOD. 
Instead of identifying structures at specific wavelets $n_w$ as done for WPOD in the previous section, in composite-WPOD we identify the coherent structures for a set of wavelets $\{n_{w1},n_{w2},\dots\}$.  
This is important since, from \S\ref{sec:decomposing the flow: wavelet bases} we know that, rather than individual wavelets, sets of wavelets contribute to the quiet region and the burst events.  
For example, to find modes for the quiet region, we need modes that represent the set of wavelets $\bm{n}^q=\{n^q_{w1},n^q_{w2},\dots n^q_{wN_q}\}$ that are responsible for the quiet region (represented by the blue shaded values of $n_w$ in figure \ref{fig:WPOD_spectrum_structures}(\subref{fig:WPOD_spectrum})). 
We therefore use a composite version of the WPOD used in \S\ref{sec:a description of wpod}. 

In \S\ref{sec:a description of wpod}, to obtain the WPOD modes, we computed the SVD of a data matrix $\widehat{Q}_{n_w}$. 
The columns of $\widehat{Q}_{n_w}$ were taken to be the data $\widehat{\bm{q}}(x,y;n_w,1:N_e)$, such that the $1$st column corresponds to $\widehat{\bm{q}}(x,y;n_w,1)$, the $i^{th}$ column to $\widehat{\bm{q}}(x,y;n_w,i)$, and so on. 
Here we instead consider the data matrix $\widehat{P}$ that contains coefficients at all the $n_w$ that is contained in $\bm{n}^q$ such that: 
\begin{equation}
\widehat{P} = \left[ \widehat{\bm{q}}(x,y;n^q_{w1},1), \cdots, \widehat{\bm{q}}(x,y;n^q_{w1},N_e), \mbox{ } \cdots\widehat{\bm{q}}(x,y;n^q_{wi},1), \cdots, \widehat{\bm{q}}(x,y;n^q_{wi},N_e), \cdots \right]. 
\end{equation}
In effect, to obtain $\widehat{P}$, we horizontally stack all the $N_q$ number of  $\widehat{Q}_{n_w}$ that correspond to the $n_w$ in $\bm{n}^q$, i.e.\ $\widehat{P} = [\widehat{Q}_{n_{w1}^q}, \widehat{Q}_{n_{w2}^q}, \cdots, \widehat{Q}_{n_{wi}^q}, \cdots]$. 
The SVD of $\widehat{P}$ will give us the dominant structures that are responsible for the set of wavelets $\bm{n}^q$. 

We denote the composite-WPOD energies for wavelets $\bm{n}^q$, obtained as the square of the singular values of $\widehat{P}$, as $\mathcal{E}_i^q$. 
In other words, $\mathcal{E}_1^q$ is the energy of the dominant composite-WPOD mode that captures the dynamics across the set of wavelets $\bm{n}^q$. 
(To clarify the notation for the POD norms used in this section, we used $\mathcal{E}^{pod}$ for the POD norm, $\mathcal{E}^{wpod}$ for the WPOD norm and $\mathcal{E}$ for the composite-WPOD norm.)
The first few such composite-WPOD modes for the set of wavelets $\bm{n}^q$ are shown in 
figures \ref{fig:db1_CoherentStructures_nwcollection}(\subref{fig:db1_CoherentStructures_nwcollection_1}-\subref{fig:db1_CoherentStructures_nwcollection_4}).  
The percentage energy captured by the mode is also shown in the figure. 
Similarly, composite-WPOD modes can be obtained for the wavelets $\bm{n}^b$ that contribute to the burst events, and here we denote the  WPOD energies of these modes as $\mathcal{E}_i^b$. 
The first $10$ among these WPOD modes for the burst event are shown in figures \ref{fig:db1_CoherentStructures_nwcollection}(\subref{fig:db1_CoherentStructures_nwcollection_5}-\subref{fig:db1_CoherentStructures_nwcollection_14}). 
Figure \ref{fig:db1_CoherentStructures_nwcollection}(\subref{fig:db1_CoherentStructures_nwcollection_spec}) shows the cumulative contribution of the first $i$ modes to the total energy. 
In other words, the blue line in figure \ref{fig:db1_CoherentStructures_nwcollection}(\subref{fig:db1_CoherentStructures_nwcollection_spec}) shows $(\sum_{j=0}^{i}\mathcal{E}_j^q)/(\sum_{j=0}^{M_q}\mathcal{E}_j^q)$, where $M_q=N_e\times N_q$ is the total number of composite-WPOD modes obtained for the set of wavelets $\bm{n}^q$ for the quiet region.   
The red line shows the same quantity for the $M_b=N_e\times N_b$ composite-WPOD modes, corresponding to the set of wavelets $\bm{n}^b$ for the burst event. 
In this study $N_e=50$, $N_q=8$ and $N_b=39$ (see figure \ref{fig:DNSSplit_wavelet}(\subref{fig:DNSSplit_Wavelet_spec})), and therefore $M_q=400$ and $M_b=1950$. 
The horizontal dashed-dot line in figure \ref{fig:db1_CoherentStructures_nwcollection}(\subref{fig:db1_CoherentStructures_nwcollection_spec}) indicates $95\%$. 

Let us first focus on the quiet region. 
From the blue line in figure \ref{fig:db1_CoherentStructures_nwcollection}(\subref{fig:db1_CoherentStructures_nwcollection_spec}) we see that the curve very quickly approaches the dashed-dot line indicating $95\%$. 
The first four modes capture most of the energy of the flow. 
These four modes are shown in figures \ref{fig:db1_CoherentStructures_nwcollection}(\subref{fig:db1_CoherentStructures_nwcollection_1}-\subref{fig:db1_CoherentStructures_nwcollection_4}), and they can be compared to their counterparts in figures \ref{fig:WPOD_spectrum_structures}(\subref{fig:WPOD_structures_1}-\subref{fig:WPOD_structures_4}). 
We note that the leading modes obtained from both WPOD versions are similar. 
Consistent with the observations from figures \ref{fig:WPOD_spectrum_structures}, composite-WPOD modes $1$ and $2$ correspond to the unstable eigenfunction and modes $3$ and $4$ represent shearing motions in the $y$-direction. 
(The modes in figure \ref{fig:db1_CoherentStructures_nwcollection} exhibit greater convergence compared to those in figure \ref{fig:WPOD_spectrum_structures} due to the inclusion of data from a collection of $n_w$, thereby increasing the input data used.) 

Now consider the burst region. 
From the red line in figure \ref{fig:db1_CoherentStructures_nwcollection}(\subref{fig:db1_CoherentStructures_nwcollection_spec}), we see that the increase in total energy is more gradual. 
The first four modes, shown in figures \ref{fig:db1_CoherentStructures_nwcollection}(\subref{fig:db1_CoherentStructures_nwcollection_5}-\subref{fig:db1_CoherentStructures_nwcollection_8}), still capture a considerable part of the energy. 
Structurally, these four modes resemble the modes in the quiet region. 
This is consistent with the observation in \S\ref{sec:wpod spectrum and modes for the Kolmogorov flow}, and shows that the dominant modes in the quiet region persist during the burst events as well. 
In effect, this flow cannot be precisely divided into the quiet region and the burst events. 
However, going back to figure \ref{fig:db1_CoherentStructures_nwcollection}(\subref{fig:db1_CoherentStructures_nwcollection_spec}), the red line shows a very slow increase, and many suboptimal modes are required to reach $95\%$ energy (the horizontal dashed-dot line). 
Therefore, although these suboptimal modes (i.e.\ mode 5 and onward) each contribute very little energy, large numbers of them together play a significant role in the burst events. 
A few of these modes are shown in figures \ref{fig:db1_CoherentStructures_nwcollection}(\subref{fig:db1_CoherentStructures_nwcollection_9}-\subref{fig:db1_CoherentStructures_nwcollection_14}). 
The leading among these modes appear to be modified, here sheared, versions of the unstable eigenfunction (see for example figures \ref{fig:db1_CoherentStructures_nwcollection}(\subref{fig:db1_CoherentStructures_nwcollection_9}), \ref{fig:db1_CoherentStructures_nwcollection}(\subref{fig:db1_CoherentStructures_nwcollection_10}) and \ref{fig:db1_CoherentStructures_nwcollection}(\subref{fig:db1_CoherentStructures_nwcollection_11})). 
This is also consistent with the observations in \S\ref{sec:wpod spectrum and modes for the Kolmogorov flow}. 
Notably, such fragmented and sheared versions of the unstable eigenfunction are modes that are typical to the burst events. 

\subsection{Tracking composite-WPOD modes in a time-series}
\label{sec:tracking composite-wpod modes}

So far, we have identified coherent structures that exist in the quiet region and the burst events. 
In this section, we track these coherent structures in a time-series obtained from the flow. 
In other words, we are interested in analysing how the contributions of these coherent structures to the flow evolve with time. 
This will pave the way for the discussion in the next section (\S\ref{sec:predicting burst events using wpod modes - method 2}), where we will use these coherent structures to introduce the WPOD-based method for predicting the burst events. 

We have two sets of composite-WPOD modes: (i) let $L_q$ represent the set of $M_q = N_e \times N_q$ number of composite-WPOD modes corresponding to the quiet region with the $j^{th}$ mode in $L_q$ having WPOD energy $\mathcal{E}_j^q$ and similarly (ii) let $L_b$ represent the set of $M_b = N_e \times N_b$ modes for the burst event with energies $\mathcal{E}_j^b$. 
Consider the modes $\bm{\phi}^{q}_j(x,y)$ ($j={1,\cdots,M_q}$) from $L_q$. 
Now consider a time series $\bm{q}(x,y,t)$ obtained from flow, where $\bm{q}=(u,v)$ is the state vector. 
Note that $\bm{q}(x,y,t)$ can be a time-series of arbitrary length with any number of burst events occurring at any point in time. 
Additionally, $\bm{q}(x,y,t)$ could lie outside the time-window used to obtain the $N_e$ realisations of the burst event for WPOD (as required for the problem of predicting the burst events in the next section). 
At each time $t$, we aim to assess the presence of the structures $\bm{\phi}^{q}_j(x,y)$ in the flow field $\bm{q}(x,y,t)$.   
For this, we first compute $E_{qj}(t)$, which is the magnitude of energy shared between a mode $\bm{\phi}^{q}_j(x,y)$ and $\bm{q}(x,y,t)$ as:
\begin{equation}
E_{qj}(t) := \int_{k_y} \int_{k_x} \left| \mbox{FT}_{xy}(q)(k_x,k_y;t) \mbox{ } \mbox{FT}_{xy}(\bm{\phi}^{q}_j)(k_x,k_y;t)^* \right|^2 \mbox{} dk_x dk_y.
\label{eqn:prediction_cross_spec}
\end{equation} 
Here $\mbox{FT}_{xy}(\cdot)$ represents the 2D Fourier transform in the spatial directions, and $k_x$ and $k_y$ are the wavenumbers in the $x$ and $y$-directions, respectively. 
We normalise $E_{qj}(t)$ using (i) $E_{jj} := \langle \bm{\phi}^{q}_j, \bm{\phi}^{q}_j \rangle_{x,y}$ which is the energy of $\bm{\phi}^{q}_j(x,y)$ and (ii) $E_{qq}(t) := \langle q, q \rangle_{x,y}$ which is the energy of $\bm{q}(x,y,t)$. 

The coherence $\gamma^q(t)$ between $\bm{q}(x,y,t)$ and the modes in $L_q$ can now be defined as:
\begin{equation}
\begin{split}
\gamma^q_j(t) = \frac{E_{qj}(t)}{E_{qq}(t) \mbox{ } E_{jj}}, \qquad
\gamma^q(t) = \sum_{j=1}^{M_q}{  \left( \frac{\mathcal{E}_i^q}{\sum_{j=0}^{M_q}\mathcal{E}_j^q} \right)  \gamma^q_j(t) }. 
\label{eqn:prediction_coherence}
\end{split}
\end{equation}
Here $\gamma^q_j(t)$ is the coherence between a point in the time-series from the flow $\bm{q}(x,y,t)$ and the $j^{th}$ composite-WPOD mode in $L_q$. 
The value of $\gamma^q_j(t)$ is bounded between $0 \leq \gamma^q_j(t)\leq 1$, with $0$ indicating no coherence between $\bm{\phi}^{q}_j(x,y)$ and $\bm{q}(x,y,t)$ and $1$ indicating perfect coherence. 
Hence, $\gamma^q(t)$ is the weighted average of the coherence across the $M_q$ different modes in $L_q$, weighted by the fraction of energy that each mode contributes to the energy of the quiet region. 
The value of $\gamma^q(t)$ therefore also lies between $0$ and $1$, where $0$ indicates that the modes in $L_q$ are not present in $\bm{q}(x,y,t)$ at that time $t$ and $1$ indicates that the modes in $L_q$ are the only structures present in $\bm{q}(x,y,t)$. 
Similarly, we can also define $\gamma^{b}(t)$ as the average coherence of $\bm{q}(x,y,t)$ with structures in $L_{b}$. 
It should be noted that, the actual value of $\gamma^q(t)$ and $\gamma^b(t)$ do not hold physical significance. 
Instead, our interest is in the trends of $\gamma^q(t)$ and $\gamma^b(t)$ over time. 

%
%
\begin{figure}[t]
\captionsetup[subfigure]{labelformat=empty,skip=-15pt}
\begin{subfigure}[b]{\textwidth}
\centering
\includegraphics[width=1\textwidth]{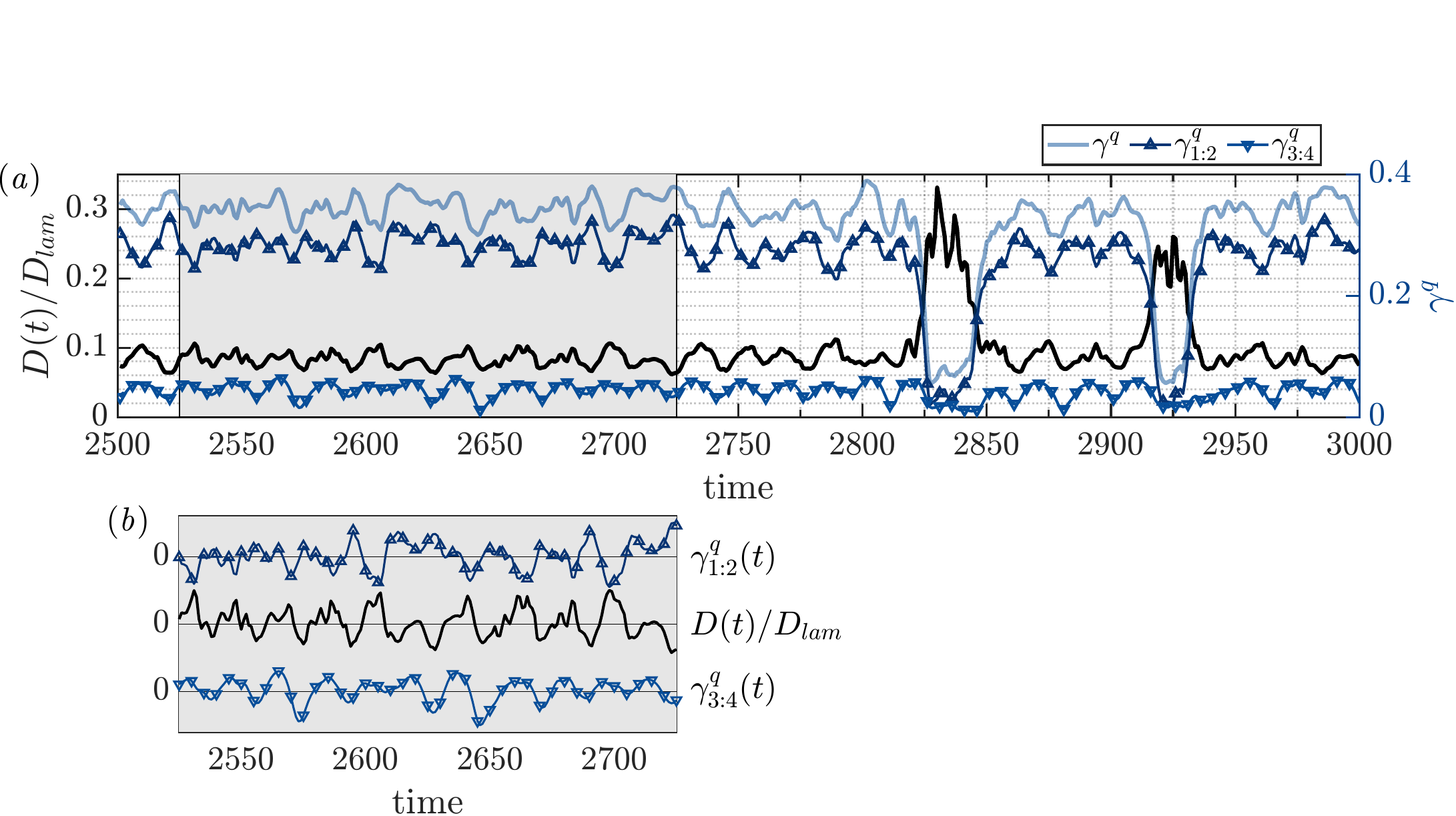}
\caption{}
\label{fig:db1_PODBasedCoherence_quiet_main}
\end{subfigure}%
\begin{subfigure}[b]{\textwidth}
\centering
\caption{}
\label{fig:db1_PODBasedCoherence_quiet_comp_zoom}
\end{subfigure}
\caption{(\textit{a}) Tracking coherent structures for the quiet region using $\gamma^q$ (blue line). 
Two components of $\gamma^q$ are also shown: $\gamma^q_{1:2}(t)$ ({\small $\triangle$}) and $\gamma^q_{3:4}(t)$ ($\triangledown$). 
(\textit{b}) Also shown are the mean-removed and normalised profiles of $D(t)/D_{lam}$, $\gamma^q_{1:2}(t)$ and $\gamma^q_{3:4}(t)$ for the time-window indicated by the grey-shaded box in (\textit{a}). }
\label{fig:db1_PODBasedCoherence_quiet}
\end{figure}
Let us first consider $\gamma^q(t)$ for a sample time-series. 
Figure \ref{fig:db1_PODBasedCoherence_quiet}(\subref{fig:db1_PODBasedCoherence_quiet_main}) displays $\gamma^q(t)$ in blue, along with the time-series of $D(t)/D_{lam}$ in black. 
To obtain a smoother curve, the average of $\gamma^{q}(t-2)$, $\gamma^{q}(t-1)$ and $\gamma^{q}(t)$ is computed at each time $t$. 
Similar averaging is also later carried out for $\gamma^b(t)$ 
(at each time $t_1$, only values $t\leq t_1$ are used for this averaging). 
Looking at $\gamma^q(t)$ we observe that: (i) within the quiet region, $\gamma^q(t)$ remains relatively high and (ii) $\gamma^q(t)$ plummets down when burst events occur. 
It is evident that $\gamma^q(t)$ exhibits distinctive changes in its trends during a burst event. 
This is consistent with the observations in the literature that shows that the occurrence of a burst event is correlated with how near or far the flow is from equilibrium solutions of the flow \citep{farazmand2016adjoint, page2021revealing}. 
Looking ahead to the problem of predicting the burst events, $\gamma^q(t)$ should, therefore, become a valuable tool. 

To further probe this, figure \ref{fig:db1_PODBasedCoherence_quiet}(\subref{fig:db1_PODBasedCoherence_quiet_main}) also shows two components of $\gamma^q(t)$: (i) $\gamma^q_{1:2}(t)$, which shows the weighted average of the coherence of just the first two modes in $L_q$, i.e.\ the unstable eigenfunction 
(modes in figures \ref{fig:db1_CoherentStructures_nwcollection}(\subref{fig:db1_CoherentStructures_nwcollection_1}, \subref{fig:db1_CoherentStructures_nwcollection_2})) 
and (ii) $\gamma^q_{3:4}(t)$, which is the weighted average coherence of modes $3$ and $4$, i.e.\ the shearing motions.  
(modes in figures \ref{fig:db1_CoherentStructures_nwcollection}(\subref{fig:db1_CoherentStructures_nwcollection_3}, \subref{fig:db1_CoherentStructures_nwcollection_4})). 
The first significant observation is that $\gamma^q_{1:2}(t)$ closely follows the trends of $\gamma^q(t)$. 
This is not surprising given that, together, modes $1$ and $2$ capture more than $70\%$ of the energy of the quiet region (as seen in figure \ref{fig:db1_CoherentStructures_nwcollection}). 
Therefore, we can say that the trends of $\gamma^q(t)$ are most significantly impacted by the flow due to modes $1$ and $2$, i.e.\ the unstable eigenfunction. 
The second noteworthy observation is that $\gamma^q_{3:4}(t)$ tends to move out of phase with $\gamma^q_{1:2}(t)$. 
This becomes more evident in figure \ref{fig:db1_PODBasedCoherence_quiet}(\subref{fig:db1_PODBasedCoherence_quiet_comp_zoom}) where mean-removed and normalised profiles of $D(t)/D_{lam}$, $\gamma^q_{1:2}(t)$ and $\gamma^q_{3:4}(t)$ are shown. 
The curves are vertically shifted in order to make the trends clearer. 
From this figure, we see that, when $D(t)/D_{lam}$ increases, the presence of the flow due to the unstable eigenfunction ($\gamma^q_{1:2}(t)$) decreases and that of the shearing structure ($\gamma^q_{3:4}(t)$) increases.  
This observation provides additional support to our initial hypothesis that the shearing motions disrupt the flow generated by the unstable eigenfunction, thereby increasing dissipation. 

%
%

\begin{figure}[t]
\captionsetup[subfigure]{labelformat=empty,skip=-15pt}
\begin{subfigure}[b]{\textwidth}
\centering
\includegraphics[width=1\textwidth]{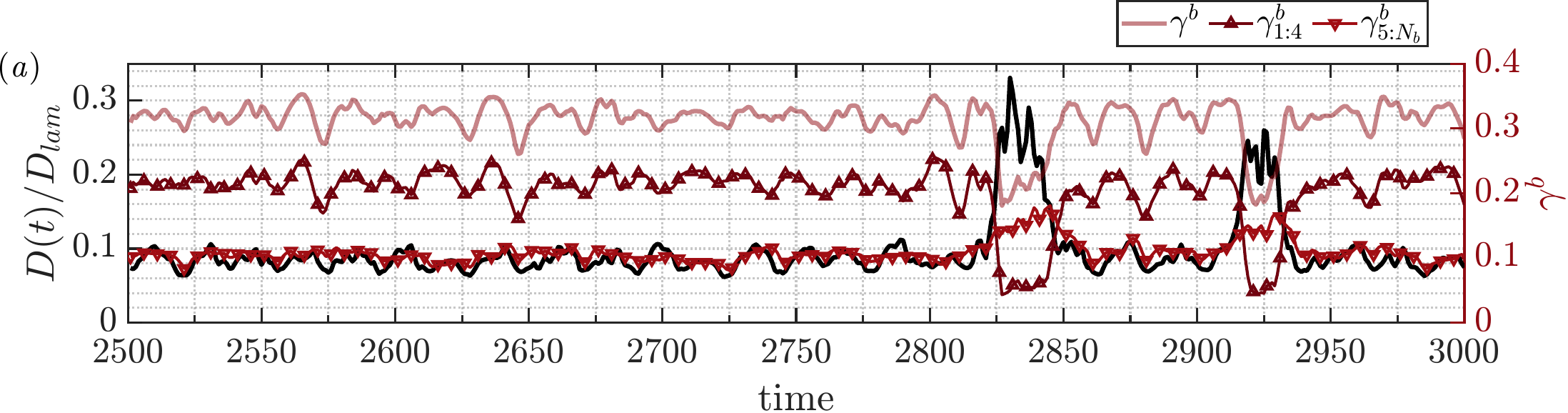}
\caption{}
\label{fig:db1_PODBasedCoherence_burst_main}
\end{subfigure}%
\begin{subfigure}[b]{\textwidth}
\centering
\caption{}
\label{fig:db1_PODBasedCoherence_burst_comp_zoom}
\end{subfigure}
\caption{(\textit{a}) Tracking coherent structures for the burst events using $\gamma^b$ (red line). 
Two components of $\gamma^b$ are also shown: $\gamma^b_{1:4}(t)$ ({\small $\triangle$}) and $\gamma^b_{5:N_b}(t)$ ($\triangledown$).  }
\label{fig:db1_PODBasedCoherence_burst}
\end{figure}

Let us now examine $\gamma^b(t)$ for the same time-series. 
Figure \ref{fig:db1_PODBasedCoherence_burst} shows $\gamma^b(t)$ in red alongside the time-series of $D(t)/D_{lam}$ in black. 
Upon comparing $\gamma^b(t)$ with $\gamma^q(t)$ in figure \ref{fig:db1_PODBasedCoherence_quiet}(\subref{fig:db1_PODBasedCoherence_quiet_main}), the first apparent observation is that the trends between them are strikingly similar. 
This observation is consistent with the earlier finding from figure \ref{fig:db1_CoherentStructures_nwcollection}, that nearly $60\%$ of the energy of the burst events is captured by modes that are present in the quiet region.  
Looking ahead to the problem of predicting the burst events, this similarity between $\gamma^q(t)$ and $\gamma^b(t)$ brings the unfortunate conclusion  that the trends of the full $\gamma^b(t)$ may not be useful for prediction.  
However, our earlier observations showed the existence of suboptimal structures that are unique to the burst events (see figures \ref{fig:db1_CoherentStructures_nwcollection}). 
Presumably, these structures exhibit trends dissimilar to $\gamma^q(t)$. 
The question, therefore, is whether we can identify a component of $\gamma^b(t)$ that has predictive value?

To explore this further, figure \ref{fig:db1_PODBasedCoherence_burst} also shows two components of $\gamma^b(t)$: (i) $\gamma^b_{1:4}(t)$ computed using just the first four modes in $L_b$, i.e.\ the modes that resemble those from the quiet region (modes in figures \ref{fig:db1_CoherentStructures_nwcollection}(\subref{fig:db1_CoherentStructures_nwcollection_5})-\ref{fig:db1_CoherentStructures_nwcollection}(\subref{fig:db1_CoherentStructures_nwcollection_8})) and (ii) $\gamma^b_{5:N_b}(t)$ computed using modes $5$ and onward (the first few of these modes are those in figures \ref{fig:db1_CoherentStructures_nwcollection}(\subref{fig:db1_CoherentStructures_nwcollection_9})-\ref{fig:db1_CoherentStructures_nwcollection}(\subref{fig:db1_CoherentStructures_nwcollection_14})). 
Two observations become immediately apparent. 
Firstly, as expected, $\gamma^b_{1:4}(t)$ follow trends similar to the full $\gamma^b(t)$ as well as $\gamma^q(t)$. 
Secondly, and also more interestingly, $\gamma^b_{5:N_b}(t)$ increases during the burst event. 
This again shows that these suboptimal modes are unique to the burst events. 
Again looking ahead to the problem of predicting the burst events, these observations show that, although the full $\gamma^b(t)$ may not be helpful in identifying the burst events, the distinctive trends of $\gamma^b_{5:N_b}(t)$ during the burst event will be. 
(Note, Appendix \ref{sec:insensitivity of prediction methods to parameter choices} shows that it is possible to predict the burst events using just $\gamma^q(t)$, ignoring the trends of $\gamma^b(t)$. However, incorporating the trends of $\gamma^b_{5:N_b}(t)$ does improve the obtained predictions.)

\subsection{Obtaining WPOD--based predictions of burst events}
\label{sec:predicting burst events using wpod modes - method 2}

\begin{figure}[t]
\captionsetup[subfigure]{labelformat=empty,skip=-15pt}
\begin{subfigure}[b]{\textwidth}
\centering
\includegraphics[width=1\textwidth]{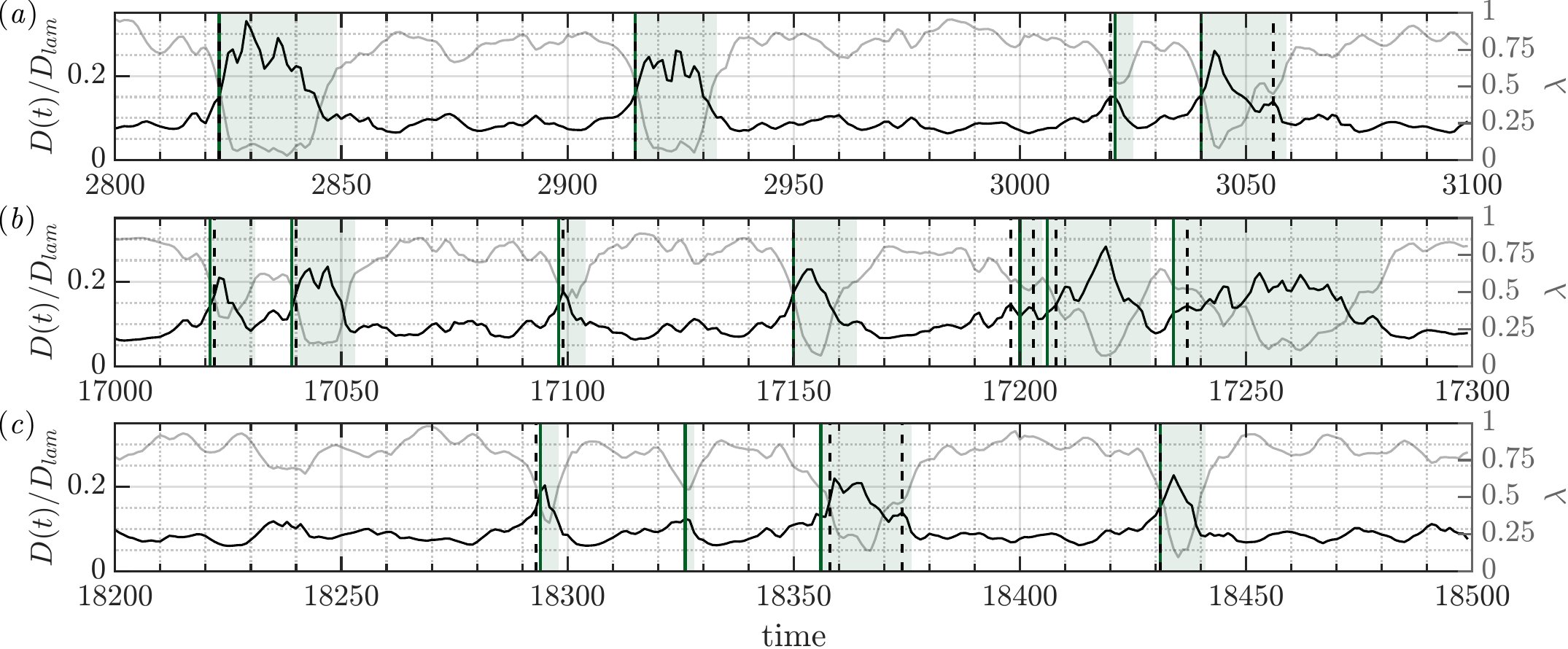}
\caption{}
\label{fig:db1_PODBasedPredictions_snap1}
\end{subfigure}%
\begin{subfigure}[b]{\textwidth}
\centering
\caption{}
\label{fig:db1_PODBasedPredictions_snap2}
\end{subfigure}%
\begin{subfigure}[b]{\textwidth}
\centering
\caption{}
\label{fig:db1_PODBasedPredictions_snap3}
\end{subfigure}
\caption{Predictions of the burst events obtained using the WPOD-based method for three separate time-series from DNS. 
The green shaded regions indicate the identified burst regions, where the predictor $\lambda$ (grey line) goes below the defined threshold value, i.e.\ $\lambda<0.95\lambda_t$. 
The vertical green lines mark the onset of these burst regions, and therefore represent the predictions of the burst event obtained from the WPOD-based method. 
Predicted times are compared to the black dashed vertical lines that indicate the predictions obtained from the energy-based method.
}
\label{fig:db1_PODBasedPredictions}
\end{figure}

In this section, to predict burst events, we use the distinctive trends of $\gamma^q(t)$ and $\gamma^b_{5:N_b}(t)$ identified in the previous section. 
For this purpose, a predictor $\lambda=\gamma^{q}(t)-\gamma_{5:N_b}^{b}(t)$ is defined. 
The average of $\lambda(t-2)$, $\lambda(t-1)$ and $\lambda(t)$ is computed at each time $t$. 
Figure \ref{fig:db1_PODBasedPredictions} shows $D(t)/D_{lam}$ along with the evolution of $\lambda$ for three different time-windows.  
The time-window in figure \ref{fig:db1_PODBasedPredictions}(\subref{fig:db1_PODBasedPredictions_snap1}) was included among the $N_e=50$ realisations of the burst events that were used for computing the WPOD modes, while the time-windows in figures \ref{fig:db1_PODBasedPredictions}(\subref{fig:db1_PODBasedPredictions_snap2}) and \ref{fig:db1_PODBasedPredictions}(\subref{fig:db1_PODBasedPredictions_snap3}) fall outside these $N_e$ realisations. 
In other words, while the data in figure \ref{fig:db1_PODBasedPredictions}(\subref{fig:db1_PODBasedPredictions_snap1}) fall within the `training dataset', figures \ref{fig:db1_PODBasedPredictions}(\subref{fig:db1_PODBasedPredictions_snap2}) and \ref{fig:db1_PODBasedPredictions}(\subref{fig:db1_PODBasedPredictions_snap3}) fall outside this training data, and will therefore serve to illustrate the generalisability of the prediction method discussed here. 
Since we are interested in the trends of $\lambda$, and not the magnitudes, we here consider normalised $\lambda$ (normalised by the minimum and maximum values obtained within a time window $t=0-15000$). 

From figure \ref{fig:db1_PODBasedPredictions}, we see that $\lambda$ decreases during burst events.  
Consequently, we designate a threshold $\lambda_t$, and values of $\lambda$ below $\lambda_t$ will be identified as burst events. 
To calculate $\lambda_t$, we begin by computing the mean of $\lambda$ minus the variance, where both these statistics are calculated for time instances that correspond to the quiet regions (i.e.\ when $D(t)/D_{lam} \leq 0.1$) between $t=0-15000$. 
The threshold $\lambda_t$ is set to be $0.95$ times this mean minus the variance. 
(The impact of varying this coefficient $0.95$ to other values will be discussed in \S\ref{sec:varying the threshold of the predictor}).  
The green shaded regions in figure \ref{fig:db1_PODBasedPredictions} mark the times when $\lambda<\lambda_t$, and the solid green vertical lines mark the beginnings of these regions. 
A majority of these green-shaded regions correspond to burst events. 
The green vertical lines are therefore the predictions of the burst events obtained using this WPOD-based method. 

To assess the prediction performance, the obtained prediction times are here compared to those obtained from a more straightforward strategy of tracking the energy of the flow. 
In this case, the predictor becomes $\lambda = -D(t)/D_{lam}$ (where the negative sign makes comparison with the wavelet-based predictors more straightforward).  
Again, the average of $\lambda(t-2)$, $\lambda(t-1)$ and $\lambda(t)$ is computed at each time $t$. 
Using the same procedure as for the WPOD-based method, a threshold is computed as $0.95$ times the mean of $\lambda$ minus the variance computed for the quiet region between $t=0-15000$. 
The vertical black dashed lines in figure \ref{fig:db1_PODBasedPredictions} indicate the beginnings of the time-windows where the predictor goes below the threshold value. 
These lines therefore represent the predictions obtained from tracking the energy of the flow, which can directly be compared to the predictions obtained from the WPOD-based method in green. 

From figure \ref{fig:db1_PODBasedPredictions}, we see that the WPOD-based method is able to predict the burst events well. 
However, for this chosen threshold, we obtain false-positives, i.e.\ predictions of oncoming burst-events in the absence of such events (e.g.\ $t \approx 18325$ in figure \ref{fig:db1_PODBasedPredictions}(\subref{fig:db1_PODBasedPredictions_snap3})). 
When compared to the energy-based method, the WPOD-based method does seem to give improved prediction times (see, for instance, $t \approx 17235$ and $t \approx 18355$ in figure \ref{fig:db1_PODBasedPredictions}). 
However, we also seem to obtain more false positives from the WPOD-based method. 

To quantify the prediction performance, we should consider predictions over long time windows. 
Here we take a time-window of length $T=20000$ and $dt=1$ outside the $N_e$ realisations used for WPOD (i.e.\ outside the `training dataset'), which contains $133$ burst events.
We compute three quantities. 
First, the average prediction time $\tau$, which is computed as the difference between the obtained prediction time and the time when the burst begins, i.e.\ when $D(t)/D_{lam}\geq0.15$. 
(Note, $\tau$ is shown in units of $\sqrt{L_y/2\pi \zeta}$.) 
Second, the percentage of predictions that are false positives, denoted by FP\%. 
A false positive is here defined as a prediction where $D(t)/D_{lam}$ of the flow does not go above $0.15$. 
Third, the percentage of false negatives, denoted by FN\%. 
These are instances when the method does not identify a region with $D(t)/D_{lam}\geq0.15$ as a burst event. 
Both FP\% and FN\% are represented as a percentage of the total number of predictions of the burst event obtained from the method. 
For the WPOD-based method $\tau=1.06$, FP\%$=20$ and FN\%$=2$, which can be compared to the values for the energy-based method $\tau=0.36$, FP\%$=13$ and FN\%$=0$. 
The WPOD-based method has difficulty in capturing burst events that persist for lesser than $1$-$2$ time units, and these contribute to the very small number of false negatives obtained. 
From these numbers, we can conclude that the WPOD-based method is capable of predicting burst events with improved prediction times compared to the more straightforward method of tracking energy, albeit with the possibility of slightly increased number of false-positives and false-negatives. 

%
%

\section{Prediction method 2: Wavelet-Resolvent Analysis (WRA) based}
\label{sec:forcing to the coherent structures}

Up to this point, our focus has been on the coherent structures in the flow. 
We now shift our attention to the forcing that generate these coherent structures. 
This will thereafter pave the way for the discussion on the WRA-based prediction method, where we will use the obtained forcing to predict the burst events. 
To 
identify this forcing, we use resolvent analysis. 
Within the resolvent analysis framework, the non-linear terms of the linearized Navier--Stokes equations are considered to be a forcing to the linear equations \citep[e.g][]{mckeon2010critical, hwang2010linear, moarref2013model, morra2021colour, towne2020resolvent}. 
Using this method, we obtain a complete basis that can be used to represent the flow. 
Additionally, the method also gives a forcing, i.e.\ the non-linear terms, that force the linearized equations to generate the coherent structures. 

Conventionally, the resolvent analysis framework is established using a Fourier transform in time. 
However, here, for precise decomposition of the flow into the quiet and burst regions, we need to use a wavelet basis (see \S\ref{sec:decomposing the flow: wavelet bases}). 
Consequently, to obtain forcing structures for the quiet and the burst regions, we need to use wavelet-based resolvent operators. 
While a substantial amount of literature considers the Fourier-based resolvent analysis, studies that concentrate on the wavelet-based resolvent analysis (WRA) are limited and very recent \citep{ballouz2023wavelet, ballouz2023transient}. 
The next section introduces the resolvent operator for a broader range of bases, which includes the Fourier and wavelet bases. 

\subsection{Resolvent operator}
\label{sec:wavelet-based resolvent operator}

To define the resolvent operator, we start with the Fourier-transformed linearized Navier--Stokes equations that were laid out in \eqref{eqn:state_space} and reproduced here:
\begin{equation}
\begin{split}
\dot{\widetilde{\omega}}(y,t;k_x) + \bm{A}\widetilde{\omega}(y,t;k_x) = \widetilde{F}(y,t;k_x). 
\label{eqn:state_space_copy1}
\end{split}
\end{equation}
Let us consider the basis vectors $\Theta_{n_w}(t)$. 
We can write $\widetilde{\omega}$ in terms of its projection onto this basis as \citep{farge1992wavelet}: 
\begin{equation}
\begin{split}
\widetilde{\omega}(y,t;k_x) &= \sum_{n_w=1}^{N_{w}} \widehat{\widetilde{\omega}}(y;k_x,n_w) \Theta_{ n_w}(t), 
\quad \mbox{ where } \\ 
\widehat{\widetilde{\omega}}(y;k_x,n_w) &= \int_0^{T} \widetilde{\omega}(y,t;k_x) \Theta_{n_w}^*(t) dt.
\end{split}
\end{equation}
We can similarly obtain the coefficients $\widehat{\widetilde{F}}(y;k_x,n_w)$ of $\widetilde{F}(y,t;k_x)$. 
Here $T$ is the length of the time-window used. 
For the case considered here, $T$ (here $T=200$) is the length of a realisation of the burst event that we used for WPOD (see \S\ref{sec:Burst centred windowing}). 
Please note that, for the rest of this manuscript, just to simplify notation, we replace the notation $(\widehat{\widetilde{\cdot}})$ with $(\widehat{\cdot})$ and also denote $n_w$ as a subscript. 
In other words, we use $\widehat{\omega}_{n_w}(y;k_x)$ and $\widehat{F}_{n_w}(y;k_x)$ to represent $\widehat{\widetilde{\omega}}(y;k_x,n_w)$ and $\widehat{\widetilde{F}}(y;k_x,n_w)$, respectively. 

The basis $\Theta_{n_w}(t)$ is chosen such that it is complete and orthonormal,
i.e. \ $\int_0^T \Theta^*_{m_w} \Theta_{n_w} dt = \delta_{n_wm_w}$, where $\delta_{n_wm_w}$ is the Dirac delta function. 
The basis could, for instance, be the Fourier basis, where $\Theta_{n_w}(t)= \exp{i(2\pi/T) n_w t}$, or as will be considered here, the wavelet basis discussed in \S\ref{sec:daubechies 1 wavelet basis and wavelet transform}.  
In terms of these transforms, the linearized equations in \eqref{eqn:state_space_copy1} become 
\begin{equation}
\begin{split}
\sum_{n_w=1}^{N_w} \widehat{\omega}_{n_w}(y;k_x) \left(\frac{d\Theta_{n_w}(t)}{dt}\right) &+ \bm{A} \sum_{n_w=1}^{N_w} \widehat{\omega}_{n_w}(y;k_x)\Theta_{n_w}(t) \\ &= \sum_{n_w=1}^{N_w} \widehat{F}_{n_w}(y;k_x)\Theta_{n_w}(t)
\label{eqn:state_space_wavelet}
\end{split}
\end{equation}
Here $N_w$ is the number of wavelet coefficients obtained from the wavelet transform (determined by the type of wavelet chosen and the number of time-instances $N_t$ chosen for the time-window). 
The bases vectors $\Theta_{n_w}(t)$ are not guaranteed to be continuous or differentiable. 
For instance, the discrete Daubechies $1$ wavelet that we use (see \S\ref{sec:daubechies 1 wavelet basis and wavelet transform}) is discontinuous and not differentiable. 
To consider the formulation in terms of the most general bases, including discontinuous ones, we re-write \eqref{eqn:state_space_wavelet} in terms of the integrals of the bases $\Theta_{n_w}(t)$, therefore giving us
\begin{equation}
\begin{split}
\sum_{n_w=1}^{N_w} \widehat{\omega}_{n_w}(y;k_x)\Theta_{n_w}(t) &+ \bm{A} \sum_{n_w=1}^{N_w} \widehat{\omega}_{n_w}(y;k_x) \left(\int_0^{t} \Theta_{n_w}(t') \mbox{ } dt'\right) \\ &= 
\sum_{n_w=1}^{N_w} \widehat{F}_{n_w}(y;k_x) \left(\int_0^{t} \Theta_{n_w}(t') \mbox{ } dt'\right). 
\label{eqn:state_space_wavelet_integral}
\end{split}
\end{equation}
Since the basis are complete, the integral of a particular basis vector can be written as a linear superposition of all basis vectors, i.e.\ $\int_0^{t} \Theta_{n_w}(t') dt' = \sum_{m_w=1}^{N_w} c_{n_wm_w}\Theta_{m_w}(t)$, where $c_{n_wm_w}$ are scalar coefficients. 
This gives,
\begin{equation}
\begin{split}
\sum_{n_w=1}^{N_w} \widehat{\omega}_{n_w}(y;k_x)\Theta_{n_w}(t) &+ \bm{A} \sum_{n_w=1}^{N_w} \widehat{\omega}_{n_w}(y;k_x) \left(\sum_{m_w=1}^{N_w} c_{n_wm_w}\Theta_{m_w}(t)\right) \\ &= 
\sum_{n_w=1}^{N_w} \widehat{F}_{n_w}(y;k_x) \left(\sum_{m_w=1}^{N_w} c_{n_wm_w}\Theta_{m_w}(t)\right). 
\label{eqn:state_space_wavelet_integral_2}
\end{split}
\end{equation} 
Using the orthogonality of the basis vectors, the equation is now rewritten as:  
\begin{equation}
\begin{split}
\widehat{\omega}_{n_w}(y;k_x) + \bm{A} \sum_{m_w=1}^{N_w} c_{n_wm_w} \widehat{\omega}_{m_w}(y;k_x)  & = \sum_{m_w=1}^{N_w} c_{n_wm_w} \widehat{F}_{m_w}(y;k_x). 
\label{eqn:state_space_wavelet_integral_3}
\end{split}
\end{equation}
If we were dealing with the Fourier-based equations $c_{n_wm_w} = in_w\delta_{n_wm_w}$. 
The summations in \eqref{eqn:state_space_wavelet_integral_3} thereby drop-off, giving independent equations for each $n_w$. 
This is not true for the case of all bases, such as a wavelet bases, where the response at one particular wavelet $\Theta_{n_w}$ cannot be isolated from the responses at the other wavelets. 
In matrix form, this equation will therefore become:
\begin{equation}
\begin{split}
\begin{bmatrix}
\widehat{\omega}_1(y;k_x) \\ 
\widehat{\omega}_2(y;k_x) \\ 
\widehat{\omega}_3(y;k_x) \\ \vdots
\end{bmatrix}
\mbox{ } + \mbox{ }
\underbrace{
\begin{bmatrix}
\bm{A} & 0 & 0 & \hdots \\
0 & \bm{A} & 0 & \hdots \\
0 & 0 & \bm{A} & \hdots \\
\vdots & \vdots & \vdots
\end{bmatrix}
\begin{bmatrix}
c_{11}\bm{I} & c_{12}\bm{I} & c_{13}\bm{I} & \hdots \\
c_{21}\bm{I} & c_{22}\bm{I} & c_{23}\bm{I} & \hdots \\
c_{31}\bm{I} & c_{32}\bm{I} & c_{33}\bm{I} & \hdots \\
\vdots & \vdots & \vdots
\end{bmatrix}
}_{\overline{\bm{A}}(y,n_w;k_x)}
\begin{bmatrix}
\widehat{\omega}_1(y;k_x) \\ 
\widehat{\omega}_2(y;k_x) \\ 
\widehat{\omega}_3(y;k_x) \\ \vdots
\end{bmatrix}
= \\
\underbrace{
\begin{bmatrix}
c_{11}\bm{I} & c_{12}\bm{I} & c_{13}\bm{I} & \hdots \\
c_{21}\bm{I} & c_{22}\bm{I} & c_{23}\bm{I} & \hdots \\
c_{31}\bm{I} & c_{32}\bm{I} & c_{33}\bm{I} & \hdots \\
\vdots & \vdots & \vdots
\end{bmatrix}}_{\overline{\bm{B}}(y,n_w;k_x)}
\begin{bmatrix}
\widehat{F}_1(y;k_x) \\ 
\widehat{F}_2(y;k_x) \\ 
\widehat{F}_3(y;k_x) \\ \vdots
\end{bmatrix}
\label{eqn:matrix_wavelet}
\end{split}
\end{equation} 
We are considering the discretised equations with $N_y$ number of grid-points in the $y$-direction and $N_w$ wavelets. 
The matrix $\bm{A}$ in \eqref{eqn:matrix_wavelet} is therefore of size $N_y\times N_y$ and $\bm{I}$ represents an identity matrix of the same size. 
The projection coefficients $c_{n_wm_w}$ are scalars. 
The matrices $\overline{\bm{A}}(y,n_w;k_x)$ and $\overline{\bm{B}}(y,n_w;k_x)$ are therefore of sizes $(N_wN_y)\times(N_wN_y)$, and the vectors $\widehat{\omega}=[\widehat{\omega}_1,\widehat{\omega}_2,\cdots \widehat{\omega}_{N_w}]$ and $\widehat{F}=[\widehat{F}_1,\widehat{F}_2,\cdots \widehat{F}_{N_w}]$ are each of size $N_y N_w$. 
Using these definitions, we can write the wavelet-based linearized equation as:
\begin{equation}
\begin{split}
\widehat{\omega}(y,n_w;k_x) + \overline{\bm{A}}(y,n_w;k_x) \widehat{\omega}(y,n_w;k_x) = \overline{\bm{B}}(y,n_w;k_x) \widehat{F}(y,n_w;k_x)
\label{eqn:state_space_general}
\end{split}
\end{equation} 
To obtain the state-vector $\widehat{\bm{q}}(y,n_w;k_x) = (\widehat{u}(y,n_w;k_x), \widehat{v}(y,n_w;k_x))$, we introduce an output-matrix $\overline{\bm{C}}$ such that
\begin{equation}
\widehat{\bm{q}}(y,n_w;k_x) = \overline{\bm{C}}(y,n_w;k_x) \widehat{\omega}(y,n_w;k_x).
\label{eqn:output_matrix}
\end{equation}
The matrix $\overline{\bm{C}}$ can also be used to `mask' the resolvent. 
For instance, in the upcoming sections, we will compute the responses at specific values of $n_w$. 
By selectively including and excluding rows of the matrix $\overline{\bm{C}}$, we can mask the response such that we obtain responses exclusively at a particular $n_w$ as in \S\ref{sec:wavelet-based resolvent and forcing modes}, or to obtain responses at specific sets of $n_w$ as in \S\ref{sec:reconstructing DNS data using wavelet-based resolvent modes}. 
It should be noted that, similar to this, the matrix $\overline{\bm{B}}$ can be used to mask the forcing to the resolvent operator. 
However, in the present study, we do not mask the forcing. 
This is because, while we are interested in responses at specific values of $n_w$ that we can separately trace back to the quiet region or the burst events, we do not have a physical argument for restricting the forcing similarly. 
In the wavelet resolvent, the response at a particular $n_w$ can be forced by all other $n_w$. 

From this point, the procedure to obtain the resolvent operator follows \citet{mckeon2010critical}. 
Rearranging \eqref{eqn:state_space_general} gives us:
\begin{equation}
\begin{split}
\widehat{\bm{q}}(y,n_w;k_x) = \underbrace{\overline{\bm{C}} \left[\bm{W}^{1/2}(I + \overline{\bm{A}})^{-1} \bm{W}^{-1/2}\right] \overline{\bm{B}}}_{\overline{\bm{H}}(y,n_w;k_x)} \mbox{ } \widehat{F}(y,n_w;k_x),
\end{split}
\label{eqn:resolvent_operator}
\end{equation}
Here $\overline{\bm{H}}(y,n_w;k_x)$ represents the resolvent operator. 
If a non-uniform grid is used to discretise $y$, the weight matrix $\bm{W}$ will contain the weights corresponding to the grid. 
In this study, we employ a Fourier grid with $N=51$ grid points, resulting in  $\bm{W}=\bm{I}$. 
As for the WPOD (see \S\ref{sec:daubechies 1 wavelet basis and wavelet transform}), here the Daubechies 1 wavelets serve as the basis vectors $\Theta_{n_w}$, with a time window of $T=200$ and a time-step of $dt=1$. 
A similar wavelet-based operator as that obtained in \eqref{eqn:resolvent_operator} has recently also been used in \citet{ballouz2023transient, ballouz2023wavelet}.

\subsection{WRA response and forcing modes} 
\label{sec:wavelet-based resolvent and forcing modes} 

SVD is used to analyse the resolvent operator, $\overline{\bm{H}}(y,n_w;k_x)$
\begin{equation}
\overline{\bm{H}}(y,n_w;k_x) = \sum\limits_{i=1}^{N_y N_w} \bm{\psi}_i(y,n_w) \mbox{ } \sigma_i \mbox{ } \phi_i(y,n_w).
\label{eqn:svd}
\end{equation}
The singular values are arranged in increasing order such that $\sigma_i \geq \sigma_{i+1}$. 
The left singular vectors $\bm{\psi}_i(y,n_w)$ represent the resolvent response modes and the right singular vectors $\phi_i(y,n_w)$ represent the resolvent forcing modes. 
Essentially, the forcing $\phi_i(y,n_w)$, when put through the resolvent operator, yields a response $\bm{\psi}_i(y,n_w)$ amplified by $\sigma_i$. 
Both $\bm{\psi}_i(y,n_w)$ and $\phi_i(y,n_w)$ are functions of $y$ and the wavelets $n_w$. 
The most amplified response is $\bm{\psi}_1(y,n_w)$ corresponding to $\sigma_1$, and the corresponding most sensitive forcing direction is $\phi_1(y,n_w)$. 

\begin{figure}[t]
\captionsetup[subfigure]{labelformat=empty,skip=-25pt}
\begin{subfigure}[b]{\textwidth}
\centering
\includegraphics[width=1\textwidth]{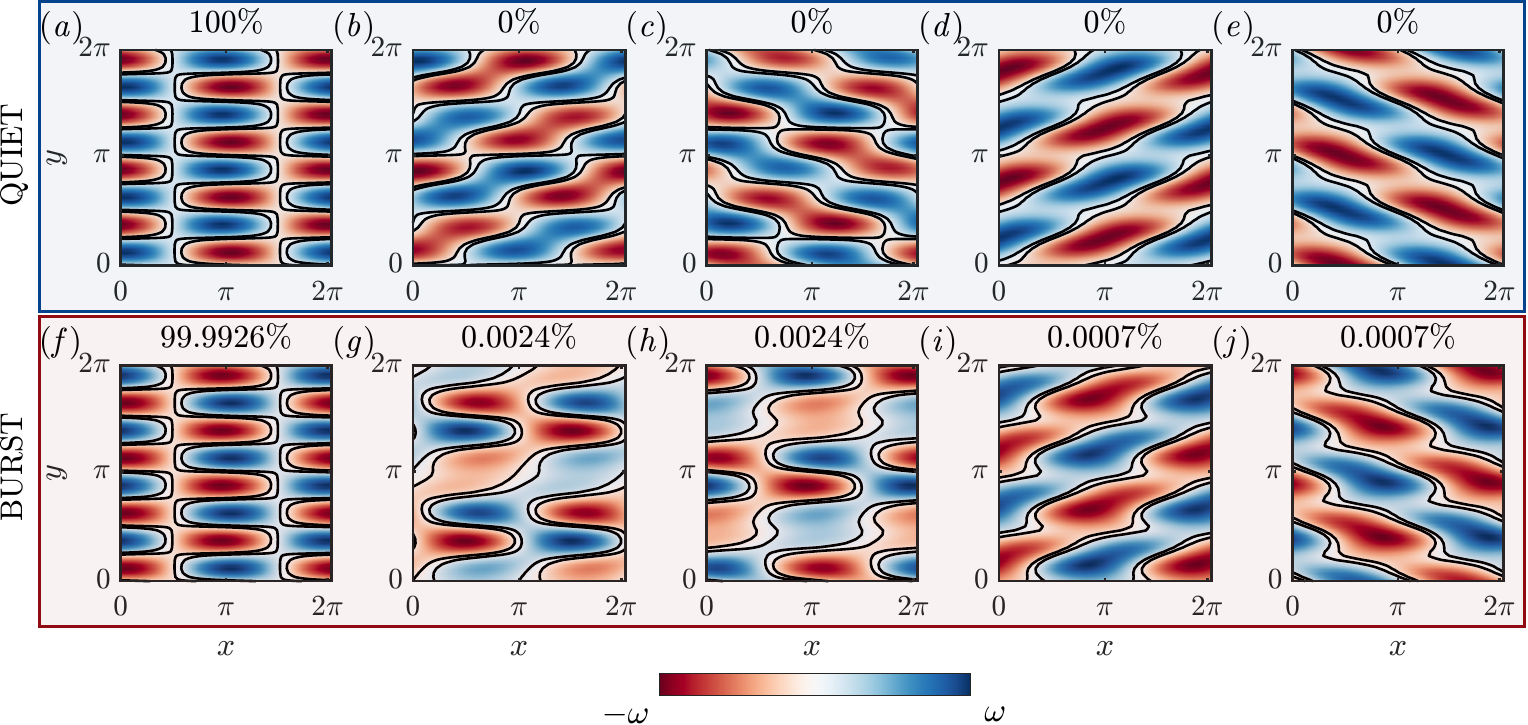}
\caption{}
\label{fig:resolvent_modes_1}
\end{subfigure}
\begin{subfigure}[b]{\textwidth}
\caption{}
\label{fig:resolvent_modes_2}
\end{subfigure}%
\begin{subfigure}[b]{\textwidth}
\caption{}
\label{fig:resolvent_modes_3}
\end{subfigure}%
\begin{subfigure}[b]{\textwidth}
\caption{}
\label{fig:resolvent_modes_4}
\end{subfigure}%
\begin{subfigure}[b]{\textwidth}
\caption{}
\label{fig:resolvent_modes_5}
\end{subfigure}%
\begin{subfigure}[b]{\textwidth}
\caption{}
\label{fig:resolvent_modes_6}
\end{subfigure}%
\begin{subfigure}[b]{\textwidth}
\caption{}
\label{fig:resolvent_modes_7}
\end{subfigure}%
\caption{First five resolvent response modes (\textit{a}-\textit{e}) at $n_w=2$ that contributes to the quiet region, and (\textit{f}-\textit{j}) at $n_w=22$ that contributes to the burst region, are shown. 
The titles show the fraction of energy (at that $n_w$) captured by the respective mode, computed as $\sigma_i^2/\sum_j{\sigma_j^2}$ (rounded off to eight decimal places for \textit{a}-\textit{e}).} 
\label{fig:resolvent_modes}
\end{figure}

First, let us consider resolvent response modes at specific values of $n_w$. 
To obtain these modes, we can selectively mask specific rows of the output matrix $\overline{\bm{C}}$ (see \eqref{eqn:output_matrix}). 
For the resolvent, we need to choose a wavenumber in the $x$-direction $k_x$. 
The dominant modes in this flow have $k_x=1$ (see figure \ref{fig:db1_CoherentStructures_nwcollection}). 
Therefore, in figure \ref{fig:resolvent_modes}, we show resolvent modes at  $k_x=1$.  
(The shearing motions in figure \ref{fig:db1_CoherentStructures_nwcollection} can be obtained as $k_x\approx0$ modes, which are not shown here for the sake of brevity). 
Figure \ref{fig:resolvent_modes} shows the resolvent response modes for two values of $n_w$: (i) $n_w=2$ that corresponds to the quiet region and (ii) $n_w=22$ that corresponds to the burst event. 
The percentages in the title of the figure represent the fraction of energy (at that $n_w$) captured by the respective mode $\bm{\psi}_i(y,n_w)$ computed as $\sigma_i^2/\sum_j{\sigma_j^2}$. 

%
%

Consider the modes from the quiet region in the first row of figure \ref{fig:resolvent_modes}. 
The first resolvent mode captures almost the full energy at this $n_w$ (the percentage is rounded off to eight decimal places).  
The mode corresponds to the unstable eigenfunction of the linearized Navier--Stokes equations (see appendix \ref{sec:unstable eigenvector of the linearized navier--stokes equations}) \citep{chandler2013invariant}. 
Crucially, when compared to the first row of figure \ref{fig:WPOD_spectrum_structures}, the first resolvent mode resembles the first WPOD mode. 
In \S\ref{sec:tracking composite-wpod modes}, we found that this first WPOD mode dominates the dynamics of the quiet region in DNS. 
Therefore, for the quiet region, we can conclude that the model is able to predict the dominant structure. 
Additionally, the model also recognises the significance of this structure, as indicated by the mode capturing almost all the energy at this $n_w$. 

Now consider the modes for the burst event in the second row of figure \ref{fig:resolvent_modes}.  
The first resolvent mode for this $n_w$ is the same as for the quiet region, representing the unstable eigenfunction and capturing the majority of the energy at this $n_w$. 
A similar trend was apparent when considering the WPOD modes in figure \ref{fig:WPOD_spectrum_structures}, where we found that the same modes as in the quiet region were also dominant for the burst event, and captured $60\%$ of the energy of the burst events. 
In the case of the resolvent modes, however, the quiet region mode captures almost the full energy. 
Examining the structure of the suboptimal resolvent modes, inclined and fragmented versions of the unstable eigenfunction are apparent, reminiscent of the WPOD modes. 
For instance, compare figure \ref{fig:db1_CoherentStructures_nwcollection}(\subref{fig:db1_CoherentStructures_nwcollection_10}) with figure \ref{fig:resolvent_modes}(\subref{fig:resolvent_modes_7}).  
Therefore, for the burst events, the suboptimal resolvent modes do capture the relevant structures. 
However, the model is not able to identify the significance of these suboptimal modes, as suggested by the percentage of energy captured by them. 

\subsection{Reconstructing DNS data using WRA modes} 
\label{sec:reconstructing DNS data using wavelet-based resolvent modes} 

%
%
\begin{figure}[t]
\captionsetup[subfigure]{labelformat=empty,skip=-15pt}
\begin{subfigure}[b]{\textwidth}
\centering
\includegraphics[width=\textwidth]{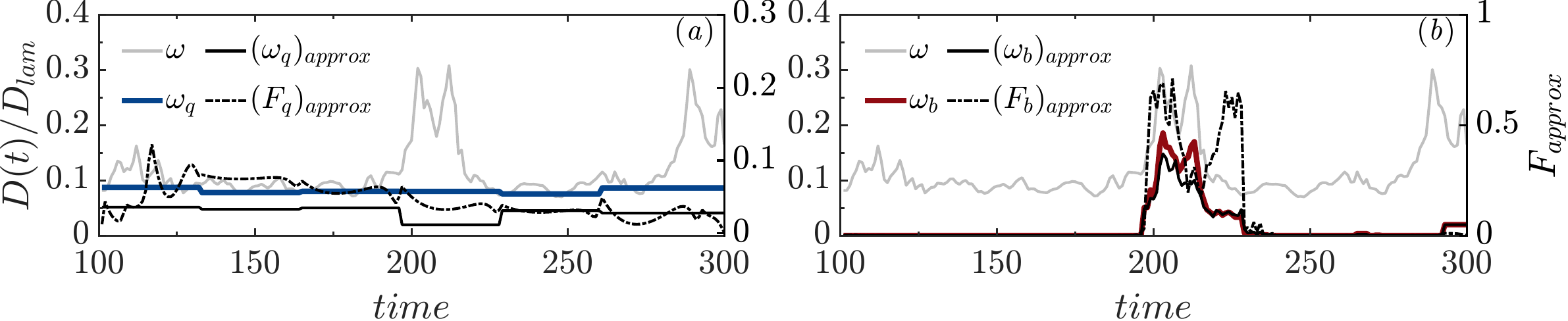}
\caption{}
\label{fig:ResRecon_quiet}
\end{subfigure}%
\begin{subfigure}[b]{\textwidth}
\centering
\caption{}
\label{fig:ResRecon_burst}
\end{subfigure}%
\caption{
$D(t)/D_{lam}$ computed from the full vorticity data for $k_x={0,1,2,3}$ (grey line) are shown alongside the wavelet-based decomposition of (\textit{a}) the quiet region in blue and (\textit{b}) the burst event in red. 
The decomposed data are compared to their respective resolvent-based reconstructions (solid black lines). 
The forcing $F_{approx}$ that generates these responses are also shown (dashed black lines). 
(The y-axis labels on the right in both (\textit{a}) and (\textit{b}) correspond to the forcing).
}
\label{fig:ResRecon}
\end{figure}
Our purpose in introducing the resolvent here is to find the forcing structures, i.e.\ the structures in the non-linear terms, that produces the quiet region and burst events. 
Studies have used the Fourier-based resolvent operators to find the underling forcing structures that are important in a flow \citep[e.g.][]{towne2015stochastic,karban2022self}. 
The right and left singular vectors of the resolvent, $\phi_i$ and $\bm{\psi}_i$, form complete bases. 
Consequently, any response $\widehat{\bm{q}}$, and any forcing $\widehat{F}$, that is obtained from data (here from DNS) can be expressed in terms of these basis vectors as
\begin{equation}
\begin{split}
\widehat{F} &= \sum_{i=1}^{N_yN_w} \chi_i \phi_i \hphantom{\sigma_i} \qquad \chi_i \hphantom{\sigma_i} = \sum_{n_w=1}^{N_w} \int_0^{L_y} \widehat{F}_{n_w}(y;k_x) (\phi_i)_{n_w}(y) \mbox{ } dy \\
\widehat{\bm{q}} &= \sum_{i=1}^{N_yN_w} \chi_i \sigma_i \bm{\psi}_i \qquad \chi_i \sigma_i = \sum_{n_w=1}^{N_w} \int_0^{L_y} \widehat{\bm{q}}_{n_w}(y;k_x) (\bm{\psi}_i)_{n_w}(y) \mbox{ } dy, 
\label{eqn:forcing_response_reconstruction}
\end{split}
\end{equation}
where $(\phi_i)_{n_w}(y)$ and $(\bm{\psi}_i)_{n_w}(y)$ are just representing the component of $\phi_i(y,n_w)$ and $\bm{\psi}_i(y,n_w)$ at a particular $n_w$. 
Therefore, given DNS data $\widetilde{\bm{q}}(y,t;k_x)$, we first obtain the wavelet coefficients $\widehat{\bm{q}}_{n_w}(y;k_x)$ through a wavelet transform, and thereafter compute $\chi_i \sigma_i$ by projecting the wavelet coefficients onto the resolvent response modes $\bm{\psi}_i(y,n_w)$. 
A reduced-order description of the flow can then be obtained by reconstructing the data using a truncated basis, 
\begin{equation}
\widehat{\bm{q}}_{approx} = \sum_{i=1}^{N_1} \chi_i \sigma_i \bm{\psi}_i, 
\end{equation}
where $N_1$ is the number of modes that are retained in the truncated data. 
The approximate data $\bm{q}_{approx}$ is then obtained from an inverse wavelet transform in time and an inverse Fourier transform in the $x$-direction. 

From the obtained product $\chi_i \sigma_i$, we then extract $\chi_i$. 
The $N_1$ is chosen such that we eliminate modes which contribute less than $0.1\%$ energy to any particular $k_x$ (i.e. the eliminated modes have $\chi_i\chi_i^* \sigma_i^2$ less that $0.1\%$ of the energy at that $k_x$). 
This $\chi_i$ is then used to compute the forcing $\widehat{F}_{approx} = \sum_{i=1}^{N_1} \chi_i \phi_i$. 
The obtained forcing $\widehat{F}_{approx}$ when put through the resolvent operator will give a response $\widehat{\bm{q}}_{approx}$. 
In other words, we can find the component of the full forcing $\widehat{F}$ from DNS (i.e.\ the non-linear term) that is specifically responsible for amplifying the response $\widehat{\bm{q}}_{approx}$. 
The forcing $F_{approx}$ (in time) can then be obtained from an inverse wavelet transform in time and an inverse Fourier transform in the $x$-direction. 
(It should be noted that, unlike in the case of the Fourier resolvent, the forcing obtained from the wavelet resolvent, $\widehat{F}_{approx}$, does not represent a particular spatial mode shape. 
Instead, $\widehat{F}_{approx}$ represents mode shapes at different wavelets $n_w$. 
This is because, unlike the Fourier resolvent, for the wavelet resolvent, response at a particular $n_w$ can be forced by multiple separate wavelets simultaneously.) 

To obtain $F_{approx}$ specifically for the quiet region, we obtain resolvent modes for the set of wavelets $\bm{n}_w^q$ corresponding to the quiet region. 
The matrix $\overline{\bm{C}}$ can be used to mask the resolvent so as to obtain a response solely for this particular set of wavelets. 
The wavelet coefficients obtained from DNS for $\bm{n}_w^q$ are then projected onto the obtained resolvent modes. 
Thereafter, following the procedure above, we can obtain the forcing for the quiet region. 
Similarly, we can also obtain the forcing for the burst region. 
An illustration of this procedure is shown in figure \ref{fig:ResRecon}, where figures \ref{fig:ResRecon}(\subref{fig:ResRecon_quiet}) and \ref{fig:ResRecon}(\subref{fig:ResRecon_burst}) show the vorticity $\omega$ from DNS decomposed into the quiet (blue) and the burst (red) regions, respectively.
The corresponding resolvent based reconstructions $\omega_{approx}$ (black solid lines) are also shown, where $(\omega_q)_{approx}$ and $(\omega_b)_{approx}$ represent the reconstructions of the quiet region and the burst events, respectively. 
Additionally, the forcing $F_{approx}$ that generate these responses are shown (black dashed lines) in these figures. 
Using WRA, we have therefore constructed forcing $(F_q)_{approx}$ and $(F_b)_{approx}$ that separately generate the quiet region and the burst events, respectively. 

\subsection{Coherent structures of the forcing} 
\label{sec:coherent structures of the forcing} 

%
%
\begin{figure}[t]
\captionsetup[subfigure]{labelformat=empty,skip=-15pt}
\begin{subfigure}[b]{\textwidth}
\centering
\includegraphics[width=\textwidth]{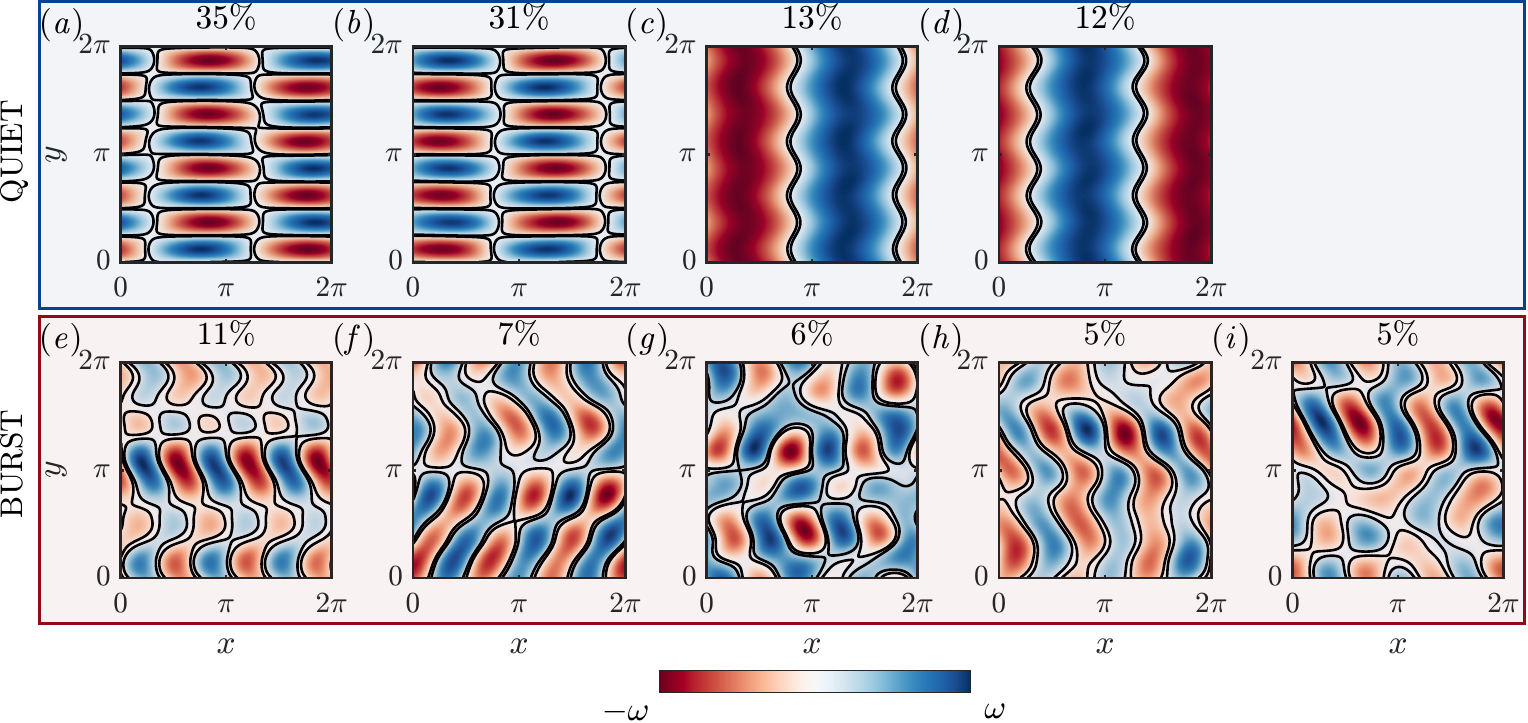}
\caption{}
\label{fig:forcing_structure_1}
\end{subfigure}%
\begin{subfigure}[b]{\textwidth}
\centering
\caption{}
\label{fig:forcing_structure_2}
\end{subfigure}%
\begin{subfigure}[b]{\textwidth}
\centering
\caption{}
\label{fig:forcing_structure_3}
\end{subfigure}%
\begin{subfigure}[b]{\textwidth}
\centering
\caption{}
\label{fig:forcing_structure_4}
\end{subfigure}%
\begin{subfigure}[b]{\textwidth}
\centering
\caption{}
\label{fig:forcing_structure_5}
\end{subfigure}%
\begin{subfigure}[b]{\textwidth}
\centering
\caption{}
\label{fig:forcing_structure_6}
\end{subfigure}%
\begin{subfigure}[b]{\textwidth}
\centering
\caption{}
\label{fig:forcing_structure_7}
\end{subfigure}%
\begin{subfigure}[b]{\textwidth}
\centering
\caption{}
\label{fig:forcing_structure_8}
\end{subfigure}%
\begin{subfigure}[b]{\textwidth}
\centering
\caption{}
\label{fig:forcing_structure_9}
\end{subfigure}%
\begin{subfigure}[b]{\textwidth}
\centering
\caption{}
\label{fig:forcing_structure_10}
\end{subfigure}
\caption{The first few POD modes obtained from the forcing corresponding to the resolvent reconstruction of (\textit{a}-\textit{d}) the quiet region and (\textit{e}-\textit{i}) the burst events. 
The titles of these plots show the percentage energy captured by the mode computed as $100\times\mathcal{E}_i^q/\sum_{j=0}^M \mathcal{E}_j^q$ for the quiet region, and similarly for the burst events. }
\label{fig:forcing_structure}
\end{figure}

Our aim is to identify the coherent structures in the forcing that generates the quiet region and the burst events. 
In order to find these coherent structures, we first start by using the resolvent to reconstruct the flow field as illustrated in figure \ref{fig:ResRecon}, and thereby obtain the corresponding forcing. 
This reconstruction is performed for $M$ different realisations of the quiet region and the burst events (here $M=50$). 
Consequently, we obtain $M$ realisations of resolvent-based vorticity reconstructions $(\omega_{approx,1}(x,y,t),\omega_{approx,2}(x,y,t),\cdots,\omega_{approx,M}(x,y,t))$ and the corresponding forcing $(F_{approx,1}(x,y,t),F_{approx,2}(x,y,t),\cdots,F_{approx,M}(x,y,t))$. 
It is important to note that, for the reconstructions done here, $k_x=(0,1,2,3)$ wavenumbers in the $x$-direction are utilised. 
Subsequently, to extract the coherent structures, we perform a POD on the $M$ realisations of the forcing. 
This involves constructing a data matrix $P$:
\begin{equation}
\begin{split}
P = [ &F_{approx,1}(x,y,t_1), F_{approx,1}(x,y,t_2), \cdots, F_{approx,1}(x,y,T), \\ 
&F_{approx,2}(x,y,t_1), F_{approx,2}(x,y,t_2), \cdots, F_{approx,2}(x,y,T), \cdots ]. 
\end{split}
\end{equation}
A SVD of the data matrix $P$ gives us the POD modes. 
We obtain such POD modes for the quiet region and the burst events. 

The top row of figure \ref{fig:forcing_structure} shows the first few forcing POD modes obtained for the quiet region. 
The titles of these figures show the percentage of energy that is captured by the mode.
Notably, the first two modes capture a majority of the energy, and the modes resemble the unstable eigenfunction itself. 
This again suggests that the dominant amplification mechanism is a normal-mode mechanism, where the forcing and response modes can be expected to be structurally similar. 
Interestingly, modes 3 and 4 resemble the dominant resolvent forcing to the unstable eigenfunction (see appendix \ref{sec:unstable eigenvector of the linearized navier--stokes equations}), which suggests that there is a second route through which the unstable eigenfunction can be amplified.  
We can also obtain the forcing POD modes for the burst events, and the first five of these modes are shown in the bottom row of figure \ref{fig:forcing_structure}. 
Notably, inclined structures dominate the forcing to the burst events. 

\subsection{Obtaining WRA--based predictions of burst events}
\label{sec:predicting burst events using resolvent forcing structures - method 3}

Now that we have identified the coherent structures in the forcing that generates the quiet regions and the burst events, we will track them in a time-series obtained from the flow. 
More specifically, our interest is in analysing how the contributions of these structures to the non-linear terms of the flow evolve with time.  
We have two sets of forcing coherent structures: (i) let $F_q$ represent the modes corresponding to the quiet region, with each mode $\phi_i^q(x,y)$ having POD energy $\mathcal{E}_i^q$, and (ii) $F_b$ the modes for the burst event, with each mode $\phi_i^b(x,y)$ having POD energy $\mathcal{E}_i^b$. 
Both $F_q$ and $F_b$ contain $M \times N_t$ modes, where $M$ is the number of resolvent reconstructions used (here $M=50$) and $N_t$ is the number of time-snapshots of DNS data in each realisation (here $N_t=200$). 
To simplify computation, here we ignore modes that cumulatively contribute less than $1\%$ energy (i.e. POD modes $k$ with  $\sum_{i=k}^M \mathcal{E}_i^q/\sum_{j=0}^M \mathcal{E}_j^q < 0.01$). 
At each time $t$, we aim to assess the presence of the structures in $F_q$ and $F_b$ in the non-linear terms $F(x,y,t)$ obtained from the flow. 

\begin{figure}[t]
\captionsetup[subfigure]{labelformat=empty,skip=-20pt}
\begin{subfigure}[b]{\textwidth}
\centering
\includegraphics[width=1\textwidth]{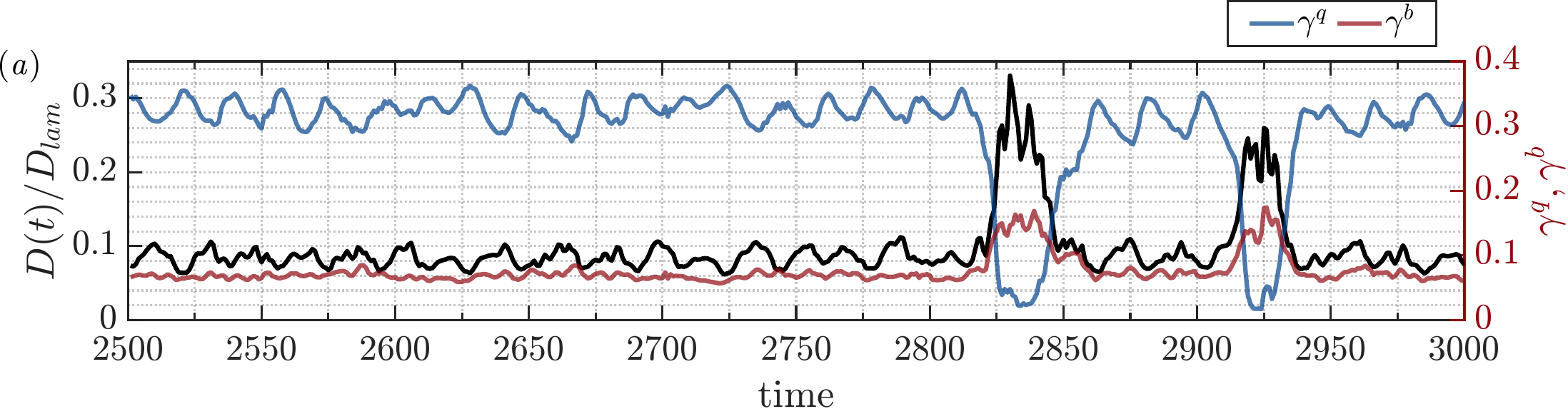}
\caption{}
\label{fig:CoherenceOfStructures_forcing_quiet}
\end{subfigure}%
\begin{subfigure}[b]{\textwidth}
\centering
\caption{}
\label{fig:CoherenceOfStructures_forcing_quietUD}
\end{subfigure}%
\begin{subfigure}[b]{\textwidth}
\centering
\caption{}
\label{fig:CoherenceOfStructures_forcing_burst}
\end{subfigure}
\caption{
Tracking forcing structures in a time series. 
The evolution of $\gamma^q(t)$ (blue) and $\gamma^b(t)$ (red) obtained by tracking forcing structures for the quiet region and the burst events, respectively, are shown. 
$D(t)/D_{lam}$ is also shown (in black). }
\label{fig:CoherenceOfStructures_forcing}
\end{figure}

For this purpose, we follow the methodology laid out in \S\ref{sec:tracking composite-wpod modes}, with $\bm{q}(x,y,t)$ now taken to be the non-linear terms $F(x,y,t)$ obtained from DNS. 
As explained in \S\ref{sec:tracking composite-wpod modes}, we obtain $\gamma^q(t)$, which is the weighted average of the coherence of the modes in $F_q$ with each point in the time-series $F(x,y,t)$. 
Equivalently, we also obtain $\gamma^b(t)$ for the modes in $F_b$. 
(As in \S\ref{sec:tracking composite-wpod modes}, we are here interested in the trends of $\gamma^q(t)$ and $\gamma^b(t)$, and not the actual values.) 
Figure \ref{fig:CoherenceOfStructures_forcing} shows the profiles of $\gamma^q(t)$ and $\gamma^b(t)$. 
We see that $\gamma^q(t)$ and $\gamma^b(t)$ follow distinctive trends during a burst event, with $\gamma^q(t)$ (blue line) decreasing and $\gamma^b(t)$ (red line) increasing. 
The trends of $\gamma^q(t)$ and $\gamma^b(t)$ can therefore be used for predicting the burst event. 

%
%

\begin{figure}[t]
\captionsetup[subfigure]{labelformat=empty,skip=-15pt}
\begin{subfigure}[b]{\textwidth}
\centering
\includegraphics[width=1\textwidth]{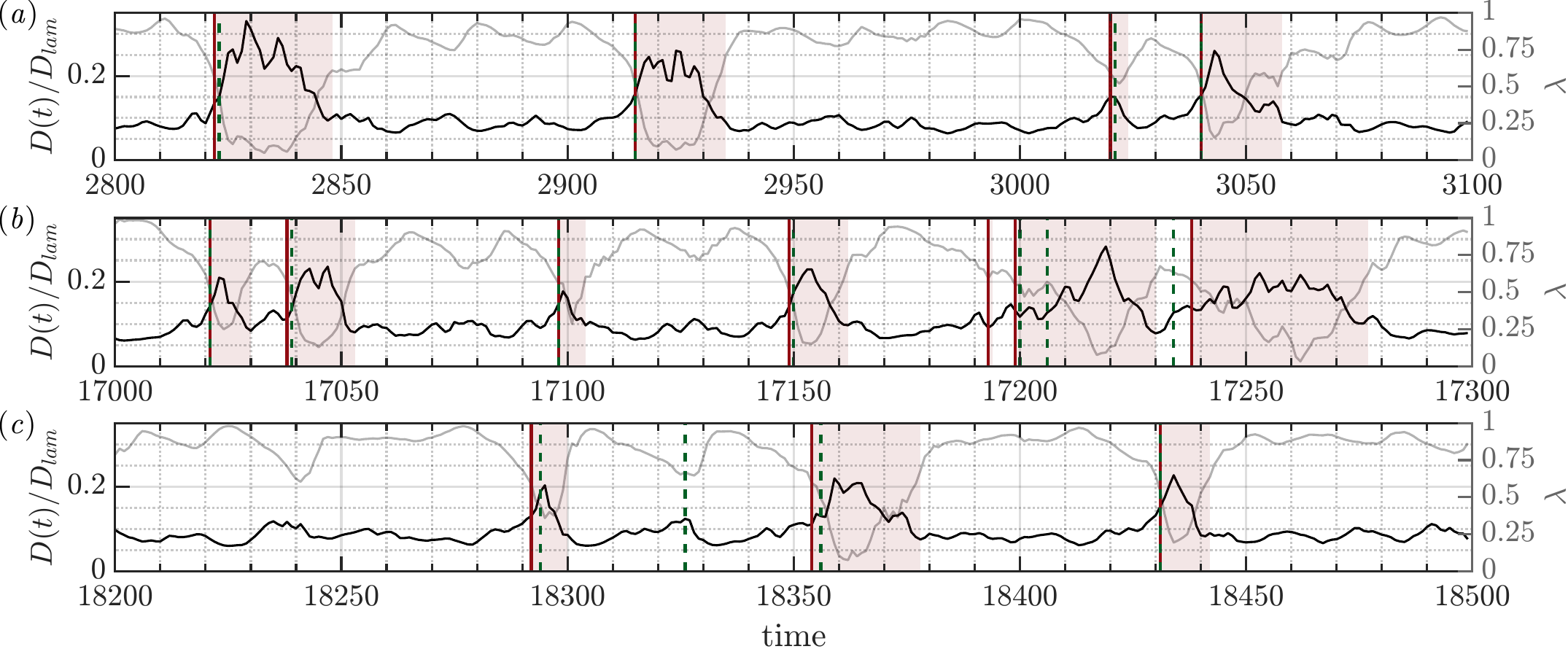}
\caption{}
\label{fig:predictions_forcing_snap1}
\end{subfigure}%
\begin{subfigure}[b]{\textwidth}
\centering
\caption{}
\label{fig:predictions_forcing_snap2}
\end{subfigure}%
\begin{subfigure}[b]{\textwidth}
\centering
\caption{}
\label{fig:predictions_forcing_snap3}
\end{subfigure} 
\caption{Predictions of the burst events obtained using the WRA-based method for three separate time-series from DNS. 
The red shaded regions indicate the identified burst regions, where the predictor $\lambda$ (grey line) goes below the defined threshold value, i.e.\ $\lambda<0.95\lambda_t$. 
The vertical red lines mark the onset of these burst regions, and therefore represent the predictions of the burst event obtained from the WRA-based method. 
Predicted times are compared to the green dashed vertical lines that indicate the predictions obtained from the WPOD-based method. 
}
\label{fig:predictions_forcing}
\end{figure}

Like in \S\ref{sec:predicting burst events using wpod modes - method 2}, here we define a predictor $\lambda = \gamma^q(t) - \gamma^b(t)$. 
Figure \ref{fig:predictions_forcing} shows the evolution of this predictor for three different time-windows. 
(As in \S\ref{sec:predicting burst events using wpod modes - method 2}, here $\lambda$ is normalised by the minimum and the maximum values obtained within the time window $t=0-15000$.) 
Using the same method as in \S\ref{sec:predicting burst events using wpod modes - method 2}, we again choose a threshold $\lambda_t$.  
The times when $\lambda<\lambda_t$ are identified as predictions of the burst regions, and the red shaded regions in figure \ref{fig:predictions_forcing} denotes these predictions. 
These prediction times are compared to the green dashed lines, which denote the predictions obtained from the WPOD-based method in \S\ref{sec:predicting burst events using wpod modes - method 2}. 
First, we note that, by tracking the forcing structures, we are indeed able to obtain predictions of the burst event. 
These prediction times are improvements over the WPOD method.  
We also note that false positives are obtained using this method as well (e.g.\ $t\approx17195$ in figure \ref{fig:predictions_forcing}). 
As in \S\ref{sec:predicting burst events using wpod modes - method 2}, for a more accurate comparison, we compute the average prediction time $\tau$, the percentage of false positives FP\% and the percentage of false negatives FN\%. 
For the WRA-based method $\tau=1.88$, FP\%$=25$ and FN\%$=4$ (compared to $\tau=1.06$, FP\%$=20$ and FN\%$=2$ for the WPOD-based method and $\tau=0.36$, FP\%$=13$ and FN\%$=0$ for the energy-based method). 
Therefore, the WRA-based method gives improved prediction times in comparison to the WPOD-based method, albeit with slight increases in the false positives and false negatives obtained.

%
%


\section{A discussion of the predictions obtained}
\label{sec:a discussion of the predictions obtained}

\subsection{A comparison of the three prediction of methods}
\label{sec:varying the threshold of the predictor}

\begin{figure}[t]
\captionsetup[subfigure]{labelformat=empty,skip=-15pt}
\begin{subfigure}[b]{\textwidth}
\centering
\includegraphics[width=\textwidth]{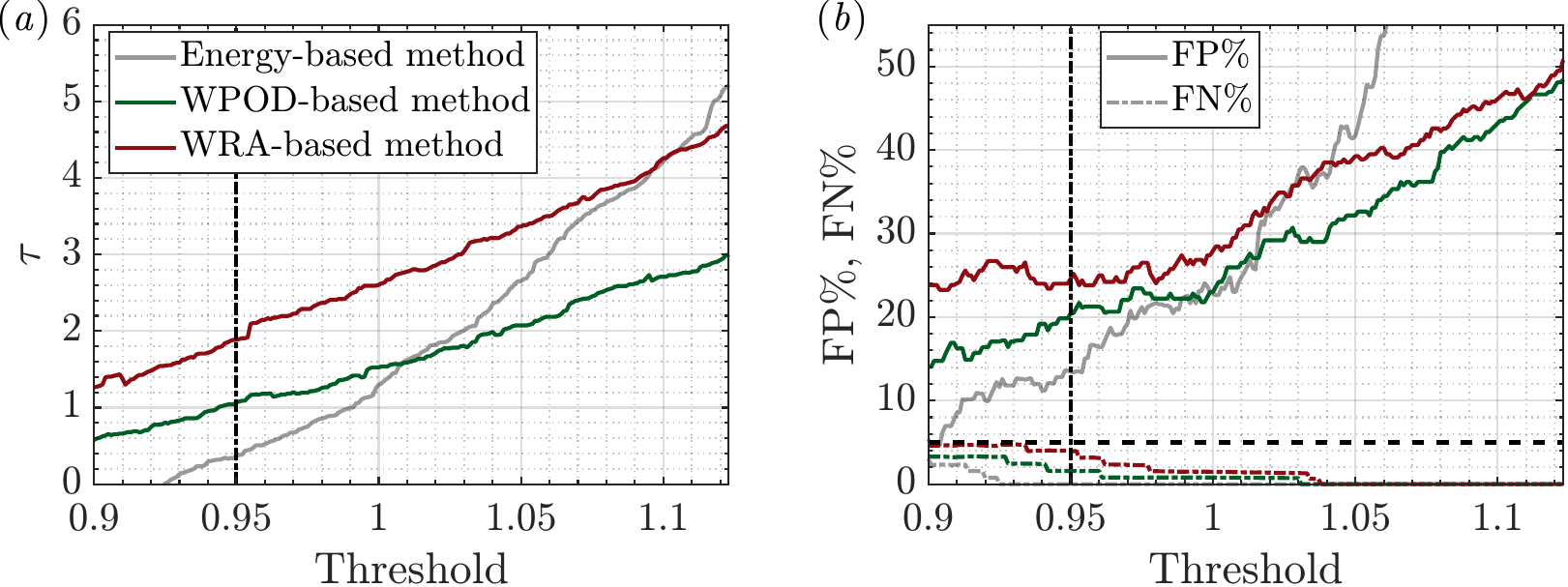}
\caption{}
\label{fig:QuantifyPredictions_pod}
\end{subfigure}%
\begin{subfigure}[b]{\textwidth}
\centering
\caption{}
\label{fig:QuantifyPredictions_res}
\end{subfigure}%
\caption{(\textit{a}) The average prediction times $\tau$ are shown as a function of the threshold of the predictors for the energy-based method (grey), the WPOD-based method (green) and the WRA-based method (red). 
(\textit{b}) The percentage of the obtained predictions that are false positives FP\% (solid lines) and false negatives FN\% (dashed-dot lines) are also shown for the three methods. }
\label{fig:QuantifyPredictions}
\end{figure}

The prediction methods discussed in this manuscript rely on a predictor going below a threshold value. 
Therefore, the predictions will vary with this choice of the threshold. 
The thresholds were chosen as $0.95$ times the mean of the predictor $\lambda$ minus the variance, where these statistics were computed specifically for time instances corresponding to the quiet regions (see \S\ref{sec:predicting burst events using wpod modes - method 2}).  
Figure \ref{fig:QuantifyPredictions} illustrates how varying this coefficient $0.95$ impacts predictions obtained from the energy-based method in grey, WPOD-based method in green and WRA-based method in red. 
The comparison between the methods is carried out using the three time averaged quantities defined in \S\ref{sec:predicting burst events using wpod modes - method 2}: (i) average prediction time $\tau$ in figure \ref{fig:QuantifyPredictions}(\subref{fig:QuantifyPredictions_pod}), (ii) the percentage of false positives FP\% and (iii) the percentage of false negatives FN\% in figure \ref{fig:QuantifyPredictions}(\subref{fig:QuantifyPredictions_res}). 

From figure \ref{fig:QuantifyPredictions}(\textit{a}), we see that, below a threshold value of approximately $1$, both the WPOD-based and WRA-based methods give better prediction times than the energy-based method. 
From figure \ref{fig:QuantifyPredictions}(\textit{b}), we see that this improvement comes at the cost of a slight increase in the number of false positives. 
Moving back to figure \ref{fig:QuantifyPredictions}(\textit{a}), we see that above this threshold of approximately $1$, the energy-based method starts performing better than the WPOD-based method, but still worse than the WRA-based method. 
For very high values of the threshold, the energy-based method seems to perform better than both the wavelet-based methods. 
However, from figure \ref{fig:QuantifyPredictions}(\textit{b}) we note that, this improvement in the performance of the energy-based method comes with a jump in the number of false positives obtained. 
For the very high thresholds, more than $50\%$ of the predictions given by the energy-based method are false, i.e.\ one in two predictions are false. 
We can therefore conclude that, both the wavelet-based methods give a more robust prediction performance when compared to the energy-based method. 
Additionally, from figure \ref{fig:QuantifyPredictions}(\textit{a}), we see that the WRA-based method always outperforms the WPOD-based method. 
However, this does come with the cost of a slight increase in the false positives. 
Finally, considering the false negatives obtained, all three methods give a small number of false negatives. 
Both the wavelet based techniques have difficulty in identifying short-lived burst events that exist for $1$-$2$ time units. 
These events contribute to the false negatives obtained. 

(Appendix \ref{sec:prediction using daubechies 2 wavelets} shows how this prediction performance changes for a different choice of wavelets, Daubechies 2 instead of the Daubechies 1 used here.)

%
%

\subsection{Outlook}
\label{sec:outlook}

So far, we have looked at two wavelet-based methods to predict intermittent events in a flow: (i) tracking the coherent structures obtained from WPOD and (ii) tracking the forcing structures that generate these coherent structures through the resolvent operator. 
We illustrated these methods using the 2-D Kolmogorov flow, which is governed by the unstable eigenfunction of the linearized Navier--Stokes equations. 
The quiet region of this flow is dominated by this unstable eigenfunction, and the burst events seem to happen because of a disruption of this structure. 
In other words, there is a single eigenfunction of the linearized Navier--Stokes equations that seems to govern the flow. 
The energy amplification mechanism in this flow therefore is likely a normal-mode amplification mechanism. 
In this scenario, the resolvent analysis shouldn't actually be expected to give a significant time-delay between the forcing and the response. 

Contrary to this, we could analyse a flow where the energy amplification occurs due to non-normal mechanisms, such as for instance the streak-generation in channel flows.
In this case, the forcing will have a time-delay from the response. 
This is due to the transient growth mechanisms that cause energy amplification in non-normal flows \citep[e.g.][]{trefethen1993hydrodynamic, schmid2007nonmodal}. 
In this case, the WRA-based method will likely provide better prediction times in comparison to the WPOD-based method. 
Comparing these methods for non-normal flows is therefore an important future direction of work. 

Applying these methods in practise in experiments would first involve finding the dominant energetic structures or forcing structures. 
For instance, a calibration experiment with good enough resolution to capture these structures or an LES simulation that captures the relevant dynamics can aid in this purpose. 
The obtained structures can then be tracked in an experiment at a lower resolution. 
How sparse these measurements can be, and what spatial regions of the flow the measurements need to be concentrated in, are important questions to be addressed in future work. 
These factors will also closely depend on the flow considered. 
For instance, for the flow considered here, although $256$ wavenumbers are present in the $x$-direction, only $7$ ($k_x=[-3,-2,-1,0,1,2,3]$) are required for the POD-based prediction.

\section{Conclusions}
\label{sec:conclusions}

In this study, we used wavelet-based methods to understand, and therefore predict, high-energy intermittent bursting events in the 2D Kolmogorov flow at Reynolds number $Re=40$ forced by a sinusoidal body forcing with wavenumber $n=4$. 
In this regime, for a majority of the time, the flow remains quiet with minor oscillations in energy. 
This quiet region is occasionally interrupted by intermittent high-energy burst events. 
The focus was on finding distinctive flow patterns for the quiet regions and the burst events, and thereafter track these structures to obtain predictions of oncoming burst events.

Due to the time-localised nature of the burst events, Fourier-based methods proved inefficient at capturing them (figure \ref{fig:DNSSplit_fourier}). 
Consequently, we used wavelet-based methods (figure \ref{fig:DNSSplit_wavelet}). 
Two wavelet-based techniques were employed:  
(i) wavelet-based Proper Orthogonal Decomposition (WPOD) to distinguish the dominant flow patterns of the quiet region and burst events (figure \ref{fig:db1_CoherentStructures_nwcollection}), and  
(ii) wavelet-based resolvent analysis (WRA) to identify  forcing structures that
generate the quiet region and burst events (figure \ref{fig:forcing_structure}). 
Subsequently, coherence-based methods were used to track these structures in an evolving time-series obtained from the flow. 
This approach yielded two effective strategies to predict oncoming burst events: (i) the WPOD-based method involved tracking the flow patterns obtained from WPOD (figure \ref{fig:db1_PODBasedPredictions}) and (ii) the WRA-based method involved tracking the forcing-patterns obtained from WRA (figure \ref{fig:predictions_forcing}). 
These predictions were then compared to those obtained from the more straightforward energy-based method, which focused on tracking the energy of the flow. 

The WPOD analysis revealed three dominant flow patterns (figure \ref{fig:db1_CoherentStructures_nwcollection}): 
(i) the unstable eigenfunction of the linearized Navier--Stokes equations that dominates the quiet region, 
(ii) shearing structures, also present in the quiet region, and 
(iii) fragmented or distorted versions of the unstable eigenfunction, crucial during burst events. 
Tracking these modes revealed that the shearing motions move out of phase with the flow due to the unstable eigenfunction. 
Moreover, the presence of the unstable eigenfunction decreases during the burst events, while the presence of the fragmented versions of the unstable eigenfunction increases. 
Based on these observations, we hypothesise that in the 2-D Kolmogorov flow, shearing motions distort the flow due to the unstable eigenfunction, leading intermittently to burst events.  
Identifying regions where the flow due to the fragmented or distorted eigenfunctions dominates over the flow due to the eigenfunction itself allowed us to develop the WPOD-based method for predicting oncoming burst events. 

Thereafter, using resolvent analysis, we identified the forcing structures that generate the unstable eigenfunction, dominant in the quiet region, as well as inclined forcing structures, dominant during the burst events. 
Identifying instances when the burst event forcing structures dominate over the quiet region structures led to the development of the WRA-based method of predicting oncoming burst events. 
Notably, both the WPOD-based and the WRA-based methods were able to predict oncoming burst events and, on average, demonstrated improved prediction times over the energy-based method. 
However, false positives, where a burst event is predicted while none occurs in the flow, were observed for both methods, with the WRA-based method more prone than the WPOD-based method (figure \ref{fig:QuantifyPredictions}). 
On the other hand, the WRA-based method gives improved prediction times over the WPOD-based method (figure \ref{fig:QuantifyPredictions}), thereby suggesting that tracking forcing structures might be a more efficient prediction strategy.  

The initial expectation was that the WRA-based method would greatly outperform the WPOD-based method. 
However, for the 2D Kolmogorov flow, while the WRA-based method does yield improved predictions, the extent of these improvements is not as substantial as initially expected. 
To understand the underlying reason for this, we can look at the linear mechanisms active in the flow. 
The dynamics of the 2D Kolmogorov flow are predominantly governed by the unstable eigenvector of the linearized Navier--Stokes equations. 
Consequently, the energy amplification mechanism that dominates this flow is likely a normal-mode mechanism. 
As a result, the appearance of the coherent structures in the flow, and the forcing that generates them, may occur without significant time-delay.  
In order to evaluate the prediction methods, there is therefore a requirement to compare the methods in flows where non-normal mechanisms are active. 
A natural progression to this work, therefore, is to evaluate burst events in wall-bounded flows, that arise from the breakdown of streaks generated by non-normal mechanisms \citep[e.g.][]{jimenez2018coherent}. 
We hope to report on findings from this line of research in the near future. 

\mbox{}

\noindent
\textbf{Acknowledgements}: The authors would like to thank the Isaac Newton Institute for Mathematical Sciences for support and hospitality during the programme `Mathematical aspects of turbulence: where do we stand?' where part of the work on this paper was undertaken. This was supported by: EPSRC grant number EP/R014604/1. 
We would also like to thank Dr. Sean Symon for helpful suggestions regarding this work. 

\mbox{}

\noindent
\textbf{Declaration of interests}: The authors report no conflicts of interest.

\appendix

\section{Unstable eigenvector of the linearized Navier--Stokes equations}
\label{sec:unstable eigenvector of the linearized navier--stokes equations}

\begin{figure}[t]
\captionsetup[subfigure]{labelformat=empty,skip=-15pt}
\begin{subfigure}[b]{\textwidth}
\centering
\includegraphics[width=0.9\textwidth]{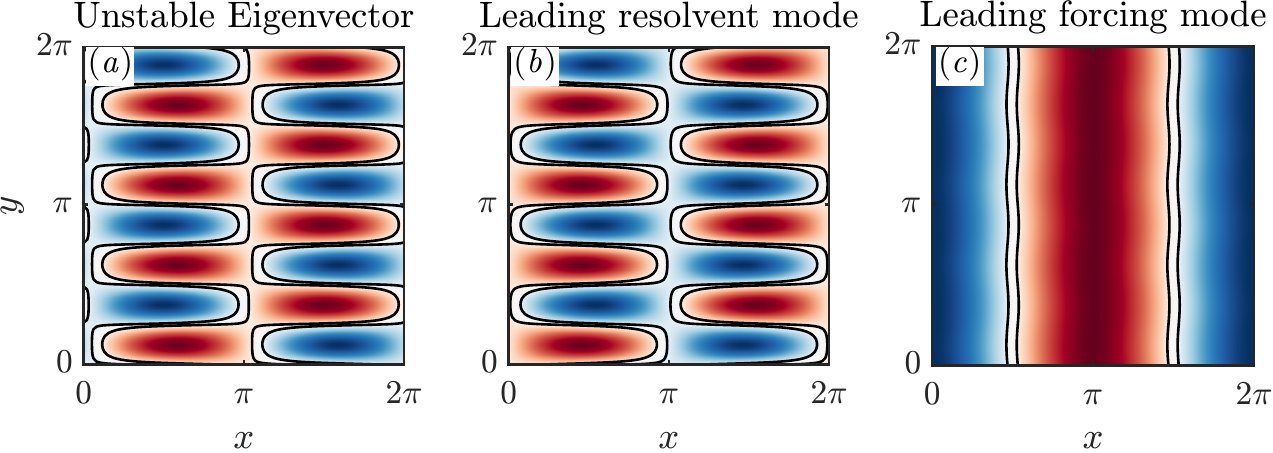}
\caption{}
\label{fig:UnstEigVect_eig}
\end{subfigure}%
\begin{subfigure}[b]{\textwidth}
\centering
\caption{}
\label{fig:UnstEigVect_res}
\end{subfigure}%
\begin{subfigure}[b]{\textwidth}
\centering
\caption{}
\label{fig:UnstEigVect_frc}
\end{subfigure}%
\caption{(\textit{a}) The unstable eigenvector of the linearized Navier--Stokes equations is shown for $k_x=1$. 
Also shown are the Fourier-based (\textit{b}) resolvent response mode and (\textit{c}) the corresponding forcing mode for a temporal frequency of $\Omega=0$. }
\label{fig:UnstEigVect}
\end{figure}

Figure \ref{fig:UnstEigVect}(\subref{fig:UnstEigVect_eig}) shows the unstable eigenvector of the linearized Navier--Stokes equations for $k_x=1$ obtained from matrix $\bm{A}$ in \eqref{eqn:state_space}. 
The resemblance between the leading WPOD modes obtained in figures \ref{fig:WPOD_spectrum_structures} and \ref{fig:db1_CoherentStructures_nwcollection} and the unstable eigenvector is evident. 
Additionally, figures \ref{fig:UnstEigVect}(\subref{fig:UnstEigVect_res}-\subref{fig:UnstEigVect_frc}) presents the Fourier-based resolvent mode for $k_x=1$ at temporal frequency $\Omega=0$. 
The leading resolvent response mode is shown in \ref{fig:UnstEigVect}(\subref{fig:UnstEigVect_res}), and this mode is similar to the unstable eigenvector. 
The corresponding leading forcing mode in \ref{fig:UnstEigVect}(\subref{fig:UnstEigVect_frc}) therefore represents the forcing that captures the unstable eigenvector. 
The similarity between this forcing mode, and the suboptimal forcing modes obtained from the WRA method for the quiet region in figure \ref{fig:forcing_structure} is apparent. 

\section{Prediction using Daubechies 2 wavelets}
\label{sec:prediction using daubechies 2 wavelets}

\begin{figure}[t]
\captionsetup[subfigure]{labelformat=empty,skip=-15pt}
\begin{subfigure}[b]{\textwidth}
\centering
\includegraphics[width=\textwidth]{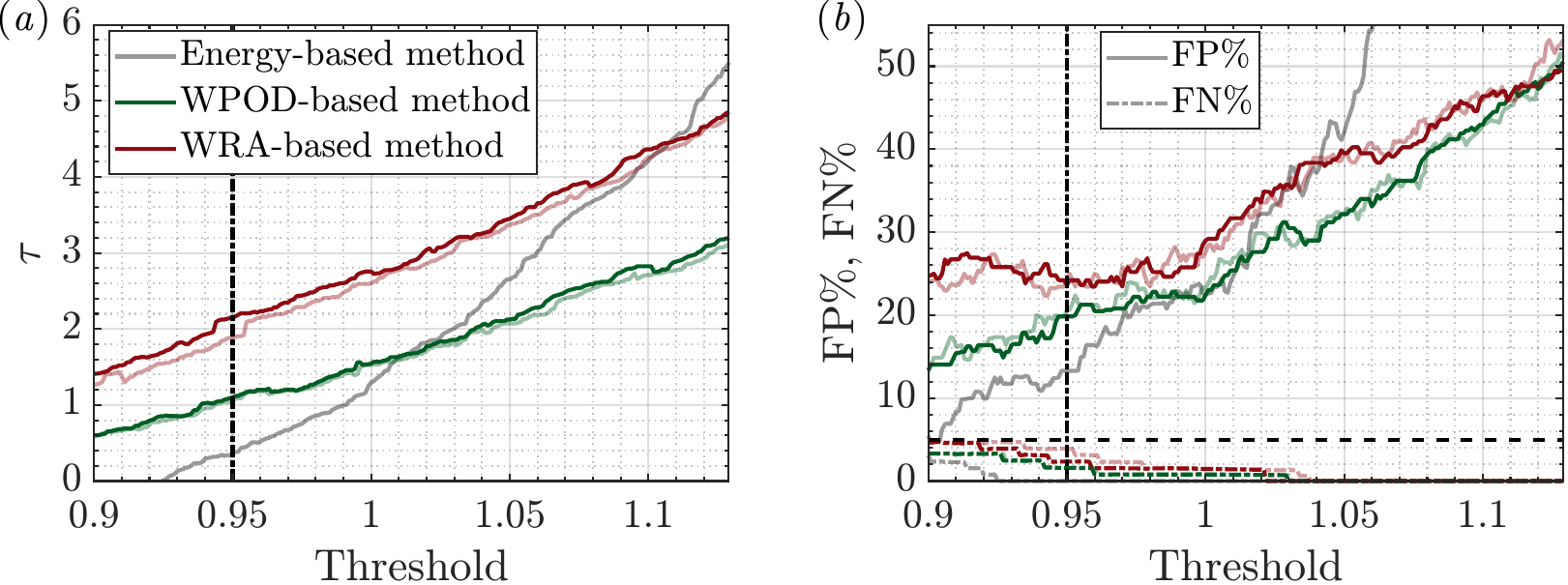}
\caption{}
\label{fig:db2VSdb1_QuantifyPredictions_thresh}
\end{subfigure}%
\begin{subfigure}[b]{\textwidth}
\centering
\caption{}
\label{fig:db2VSdb1_QuantifyPredictions_FP}
\end{subfigure}%
\caption{WPOD and WRA-based predictions obtained using the Daubechies 2 wavelets (dark-coloured lines) are compared to the predictions obtained using the Daubechies 1 wavelets (lines from figure \ref{fig:QuantifyPredictions} reproduced here as the lighter coloured lines).
(\textit{a}) The average prediction times $\tau$ are shown as a function of the threshold of the predictors for the energy-based method (grey), the WPOD-based method (green) and the WRA-based method (red). 
(\textit{b}) The percentage of the obtained predictions that are false positives FP\% (solid lines) and false negatives FN\% (dashed-dot lines) are also shown for the three methods. }
\label{fig:db2VSdb1_QuantifyPredictions}
\end{figure}

In this section, we reproduce the WPOD-based and the WRA-based prediction performance results, but this time using the Daubechies 2 (DB2) wavelets, instead of the Daubechies 1 (DB1) wavelets used in the rest of the manuscript. 
Figure \ref{fig:db2VSdb1_QuantifyPredictions} shows the comparison in figure \ref{fig:QuantifyPredictions} for predictions using DB2. 
As in figure \ref{fig:QuantifyPredictions}, two quantities are shown in figure \ref{fig:db2VSdb1_QuantifyPredictions}. 
The obtained average prediction times (denoted by $\tau$) in figure \ref{fig:db2VSdb1_QuantifyPredictions}(\subref{fig:db2VSdb1_QuantifyPredictions_thresh}), and the percentage of predictions that are false positives FP\% and false negatives FN\% in figure \ref{fig:db2VSdb1_QuantifyPredictions}(\subref{fig:db2VSdb1_QuantifyPredictions_FP}). 
The lighter coloured lines are reproductions of the corresponding lines in figure \ref{fig:QuantifyPredictions} (i.e.\ the predictions using DB1) shown here again for comparison. 
Using DB2 instead of DB1, we obtain marginally improved (earlier) predictions from the WRA-based method. 
Apart from that, we note that the obtained prediction performances are similar for both DB1 and DB2. 
This shows that the results are fairly insensitive to the choice of wavelets between DB1 and DB2. 

\section{Insensitivity of prediction methods to parameter choices}
\label{sec:insensitivity of prediction methods to parameter choices}

\subsection{Insensitivity to the exact choice of burst wavelets}

\begin{figure}[t]
\captionsetup[subfigure]{labelformat=empty,skip=-15pt}
\begin{subfigure}[b]{\textwidth}
\centering
\includegraphics[width=\textwidth]{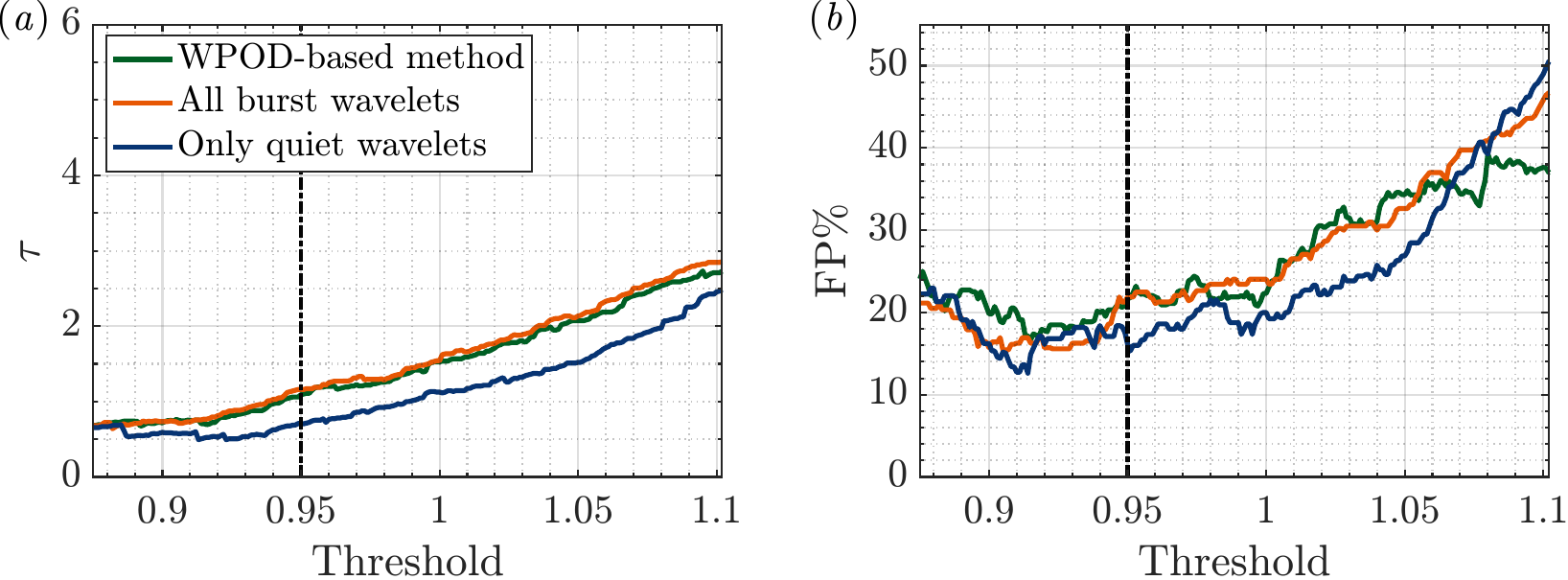}
\caption{}
\label{fig:AllBurst_QuantifyPredictions_pod}
\end{subfigure}%
\begin{subfigure}[b]{\textwidth}
\centering
\caption{}
\label{fig:AllBurst_QuantifyPredictions_res}
\end{subfigure}%
\caption{(\textit{a}) The average prediction time $\tau$ is shown for two modifications of the WPOD-based method.  
To obtain the first modified WPOD-based method (orange line), instead of choosing only the most energetic $20\%$ of the wavelets in the levels corresponding to burst events, all wavelets in those levels are considered. 
The second modification of the WPOD-based method (blue line) is obtained by defining a predictor using $\gamma^q$ alone (not including $\gamma^b$). 
The original prediction results from the WPOD-based method, shown in figure \ref{fig:QuantifyPredictions}, are reproduced here for comparison (green line).
(\textit{b}) The percentage of the obtained predictions that are false positives are also shown.}
\label{fig:AllBurst_QuantifyPredictions}
\end{figure}
One of the aims of the current section is to show that the prediction results presented here are not sensitive to the selection of wavelets. 
For a majority of this study, the first three levels of the wavelet transform correspond to the quiet region and $20\%$ of the most energetic wavelets in the remaining levels represent the burst events. 
The easiest way to show the insensitivity of the results obtained to this selection, is by performing the same prediction using all the wavelets in the quiet region as well as the burst events. 
In other words, the wavelets in the first three levels still represent the quiet region, while all the wavelets in levels $4-8$ represent the burst events. 
This, of course, makes the problem computationally a lot more challenging. 

The orange curve in figure \ref{fig:AllBurst_QuantifyPredictions} shows the WPOD-based prediction results when using all wavelets, which can be compared to the original WPOD-based prediction in figure \ref{fig:QuantifyPredictions}, reproduced here as the green line. 
Two quantities are shown here, the average prediction time $\tau$ and the percentage of false positives FP$\%$. 
We can see that, as per both metrics, the performance of both the methods are similar. 
Therefore, choosing just the most energetic wavelets for the burst events, is the computationally more efficient method to choose. 
In other words, the method is not sensitive to the choice of wavelets, as long as the most energetic wavelets are still chosen. 

\subsection{WPOD-based method using only the quiet region}

Additionally, here we show how the predictions obtained would vary when considering wavelets in the quiet region alone. 
We therefore compare predictions obtained using just the quiet region structures, to those obtained using the original WPOD-based method. 
For this purpose, we consider a modified WPOD-based method, where the predictor simply is $\gamma^q$ (and does not involve $\gamma^b$). 
The dark blue line in figure \ref{fig:AllBurst_QuantifyPredictions} shows the performance of this predictor, which can be compared to the green line obtained for the original WPOD-based method. 
We note that the prediction performance of this modified method, based only on the quiet region, is worse than the performance obtained from the original WPOD-based method. 

\subsection{Insensitivity to the definition of the burst event}

\begin{figure}[t]
\captionsetup[subfigure]{labelformat=empty,skip=-15pt}
\begin{subfigure}[b]{\textwidth}
\centering
\includegraphics[width=\textwidth]{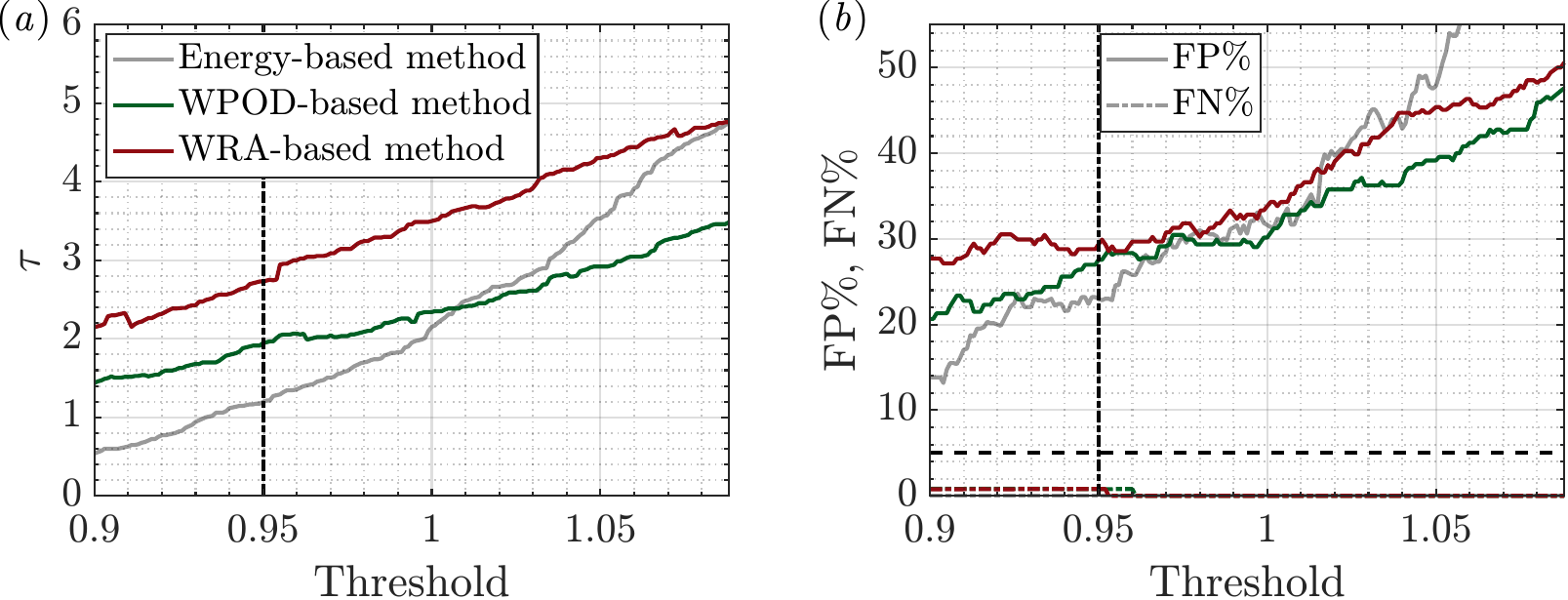}
\caption{}
\label{fig:modifiedBurstDef_QuantifyPredictions_pod}
\end{subfigure}%
\begin{subfigure}[b]{\textwidth}
\centering
\caption{}
\label{fig:modifiedBurstDef_QuantifyPredictions_res}
\end{subfigure}%
\caption{Figure \ref{fig:QuantifyPredictions} is recreated here with a modified definition of the burst events as times when $D(t)/D_{lam}\geq0.17$. }
\label{fig:modifiedBurstDef_QuantifyPredictions}
\end{figure}
Finally, the insensitivity of the results to the definition chosen for the burst event is shown here in figure \ref{fig:modifiedBurstDef_QuantifyPredictions}. 
In \S\ref{sec:direct numerical simulations} we choose the definition of a burst event in the 2D Kolmogorov flow as times when $D(t)/D_{lam} \geq 0.15$. 
In figure \ref{fig:QuantifyPredictions}, the comparison of the performances of the three prediction methods was done using this definition of a burst event. 
Figure \ref{fig:modifiedBurstDef_QuantifyPredictions} shows the same comparison as in figure \ref{fig:QuantifyPredictions}, but with the burst events defined as $D(t)/D_{lam} \geq 0.17$. 
As expected, although the quantitative values change, the relative trends between the three methods remain the same between figures \ref{fig:QuantifyPredictions} and \ref{fig:modifiedBurstDef_QuantifyPredictions}. 
Since the conclusions drawn in this manuscript are based on these relative trends between the methods, we can be confident that these conclusions are insensitive to the exact definition of the burst event. 

\bibliographystyle{jfm}
\bibliography{ref}

\begin{thebibliography}{58}
\expandafter\ifx\csname natexlab\endcsname\relax\def\natexlab#1{#1}\fi
\def\au#1{#1} \def\ed#1{#1} \def\yr#1{#1}\def\at#1{#1}\def\jt#1{\textit{#1}}
  \def\bt#1{#1}\def\bvol#1{\textbf{#1}} \def\vol#1{#1} \def\pg#1{#1}
  \def\publ#1{#1}\def\arxiv#1{#1}\def\org#1{#1}\def\st#1{\textit{#1}}

\bibitem[Babaee \& Sapsis(2016)]{babaee2016minimization}
{\sc \au{Babaee, H.} \& \au{Sapsis, T.~P.}} \yr{2016}  \at{A minimization
  principle for the description of modes associated with finite-time
  instabilities}.  \jt{Proc. R. Soc. Lond. A}  \bvol{472}~(2186),
  \pg{20150779}.

\bibitem[Ballouz {\em et~al.\/}(2023{\natexlab{{\em a\/}}})Ballouz, Dawson \&
  Bae]{ballouz2023transient}
{\sc \au{Ballouz, E.}, \au{Dawson, S.} \& \au{Bae, H.~J.}}
  \yr{2023{\natexlab{{\em a\/}}}}  \at{Transient growth of wavelet-based
  resolvent modes in the buffer layer of wall-bounded turbulence}.  \jt{arXiv
  preprint arXiv:2312.15465} .

\bibitem[Ballouz {\em et~al.\/}(2023{\natexlab{{\em b\/}}})Ballouz,
  Lopez-Doriga, Dawson \& Bae]{ballouz2023wavelet}
{\sc \au{Ballouz, E.}, \au{Lopez-Doriga, B.}, \au{Dawson, S.~T.} \& \au{Bae,
  H.~J.}} \yr{2023{\natexlab{{\em b\/}}}} Wavelet-based resolvent analysis for
  statistically-stationary and temporally-evolving flows.  \bt{In {\em AIAA
  SCITECH 2023 Forum\/}},  \pg{p. 0676}.

\bibitem[Barthel \& Sapsis(2023)]{barthel2023harnessing}
{\sc \au{Barthel, B.} \& \au{Sapsis, T.}} \yr{2023}  \at{Harnessing the
  instability mechanisms in airfoil flow for the data-driven forecasting of
  extreme events}.  \jt{arXiv preprint arXiv:2303.07056} .

\bibitem[Bay{\i}nd{\i}r(2016)]{bayindir2016early}
{\sc \au{Bay{\i}nd{\i}r, C.}} \yr{2016}  \at{Early detection of rogue waves by
  the wavelet transforms}.  \jt{Phys. Lett. A.}  \bvol{380}~(1-2),
  \pg{156--161}.

\bibitem[Blonigan {\em et~al.\/}(2019)Blonigan, Farazmand \&
  Sapsis]{blonigan2019extreme}
{\sc \au{Blonigan, P.~J.}, \au{Farazmand, M.} \& \au{Sapsis, T.~P.}} \yr{2019}
  \at{Are extreme dissipation events predictable in turbulent fluid flows?}
  \jt{Phys. Rev. Fluids}  \bvol{4}~(4),  \pg{044606}.

\bibitem[Chandler \& Kerswell(2013)]{chandler2013invariant}
{\sc \au{Chandler, G.~J.} \& \au{Kerswell, R.~R.}} \yr{2013}  \at{Invariant
  recurrent solutions embedded in a turbulent two-dimensional {K}olmogorov
  flow}.  \jt{J. Fluid Mech.}  \bvol{722},  \pg{554--595}.

\bibitem[Daubechies(1988)]{daubechies1988orthonormal}
{\sc \au{Daubechies, I.}} \yr{1988}  \at{Orthonormal bases of compactly
  supported wavelets}.  \jt{Commun. Pure Appl. Math}  \bvol{41}~(7),
  \pg{909--996}.

\bibitem[Doan {\em et~al.\/}(2021)Doan, Polifke \& Magri]{doan2021short}
{\sc \au{Doan, N. A.~K.}, \au{Polifke, W.} \& \au{Magri, L.}} \yr{2021}
  \at{Short-and long-term predictions of chaotic flows and extreme events: a
  physics-constrained reservoir computing approach}.  \jt{Proc. R. Soc. A}
  \bvol{477}~(2253),  \pg{20210135}.

\bibitem[Donzis \& Sreenivasan(2010)]{donzis2010short}
{\sc \au{Donzis, D.~A.} \& \au{Sreenivasan, K.~R.}} \yr{2010}  \at{Short-term
  forecasts and scaling of intense events in turbulence}.  \jt{J. Fluid Mech.}
  \bvol{647},  \pg{13--26}.

\bibitem[Dysthe {\em et~al.\/}(2008)Dysthe, Krogstad \&
  M{\"u}ller]{dysthe2008oceanic}
{\sc \au{Dysthe, K.}, \au{Krogstad, H.~E.} \& \au{M{\"u}ller, P.}} \yr{2008}
  \at{Oceanic rogue waves}.  \jt{Annu. Rev. Fluid Mech.}  \bvol{40},
  \pg{287--310}.

\bibitem[Farazmand(2016)]{farazmand2016adjoint}
{\sc \au{Farazmand, M.}} \yr{2016}  \at{An adjoint-based approach for finding
  invariant solutions of navier--stokes equations}.  \jt{J. Fluid Mech.}
  \bvol{795},  \pg{278--312}.

\bibitem[Farazmand \& Sapsis(2016)]{farazmand2016dynamical}
{\sc \au{Farazmand, M.} \& \au{Sapsis, T.~P.}} \yr{2016}  \at{Dynamical
  indicators for the prediction of bursting phenomena in high-dimensional
  systems}.  \jt{Phys. Rev. E}  \bvol{94}~(3),  \pg{032212}.

\bibitem[Farazmand \& Sapsis(2017)]{farazmand2017variational}
{\sc \au{Farazmand, M.} \& \au{Sapsis, T.~P.}} \yr{2017}  \at{A variational
  approach to probing extreme events in turbulent dynamical systems}.  \jt{Sci.
  Adv.}  \bvol{3}~(9),  \pg{e1701533}.

\bibitem[Farazmand \& Sapsis(2019{\natexlab{{\em a\/}}})]{farazmand2019closed}
{\sc \au{Farazmand, M.} \& \au{Sapsis, T.~P.}} \yr{2019{\natexlab{{\em a\/}}}}
  \at{Closed-loop adaptive control of extreme events in a turbulent flow}.
  \jt{Phys. Rev. E}  \bvol{100}~(3),  \pg{033110}.

\bibitem[Farazmand \& Sapsis(2019{\natexlab{{\em b\/}}})]{farazmand2019extreme}
{\sc \au{Farazmand, M.} \& \au{Sapsis, T.~P.}} \yr{2019{\natexlab{{\em b\/}}}}
  \at{Extreme events: {M}echanisms and prediction}.  \jt{Appl. Mech. Rev.}
  \bvol{71}~(5),  \pg{050801}.

\bibitem[Farge(1992)]{farge1992wavelet}
{\sc \au{Farge, Marie}} \yr{1992}  \at{Wavelet transforms and their
  applications to turbulence}.  \jt{Annu. Rev. Fluid Mech.}  \bvol{24}~(1),
  \pg{395--458}.

\bibitem[Farge {\em et~al.\/}(2001)Farge, Pellegrino \&
  Schneider]{farge2001coherent}
{\sc \au{Farge, M.}, \au{Pellegrino, G.} \& \au{Schneider, K.}} \yr{2001}
  \at{Coherent vortex extraction in {3D} turbulent flows using orthogonal
  wavelets}.  \jt{Phys. Rev. Lett.}  \bvol{87}~(5),  \pg{054501}.

\bibitem[Farge \& Rabreau(1988)]{farge1988transformee}
{\sc \au{Farge, M.} \& \au{Rabreau, G.}} \yr{1988}  \at{Transform{\'e}e en
  ondelettes pour d{\'e}tecter et analyser les structures coh{\'e}rentes dans
  les {\'e}coulements turbulents bidimensionnels}.  \jt{C. R. Acad. Sci. Paris}
   \bvol{307},  \pg{1479--1486}.

\bibitem[Farge {\em et~al.\/}(2006)Farge, Schneider \&
  Devynck]{farge2006extraction}
{\sc \au{Farge, M.}, \au{Schneider, K.} \& \au{Devynck, P.}} \yr{2006}
  \at{Extraction of coherent bursts from turbulent edge plasma in magnetic
  fusion devices using orthogonal wavelets}.  \jt{Phys. Plasmas}
  \bvol{13}~(4).

\bibitem[Farge {\em et~al.\/}(2003)Farge, Schneider, Pellegrino, Wray \&
  Rogallo]{farge2003coherent}
{\sc \au{Farge, M.}, \au{Schneider, K.}, \au{Pellegrino, G.}, \au{Wray, A.~A.}
  \& \au{Rogallo, R.~S.}} \yr{2003}  \at{Coherent vortex extraction in
  three-dimensional homogeneous turbulence: {C}omparison between {CVS}-wavelet
  and {POD-F}ourier decompositions}.  \jt{Phys. Fluids}  \bvol{15}~(10),
  \pg{2886--2896}.

\bibitem[Floryan \& Graham(2021)]{floryan2021discovering}
{\sc \au{Floryan, D.} \& \au{Graham, M.~D.}} \yr{2021}  \at{Discovering
  multiscale and self-similar structure with data-driven wavelets}.  \jt{Proc.
  Natl. Acad. Sci.}  \bvol{118}~(1),  \pg{e2021299118}.

\bibitem[Fox {\em et~al.\/}(2023)Fox, Constante-Amores \&
  Graham]{fox2023predicting}
{\sc \au{Fox, A.~J.}, \au{Constante-Amores, C.~R.} \& \au{Graham, M.~D.}}
  \yr{2023}  \at{Predicting extreme events in a data-driven model of turbulent
  shear flow using an atlas of charts}.  \jt{Phys. Rev. Fluids}  \bvol{8}~(9),
  \pg{094401}.

\bibitem[Gupta {\em et~al.\/}(2022)Gupta, Shanbhogue, Shimura, Ghoniem \&
  Hemchandra]{gupta2022impact}
{\sc \au{Gupta, S.}, \au{Shanbhogue, S.}, \au{Shimura, M.}, \au{Ghoniem, A.} \&
  \au{Hemchandra, S.}} \yr{2022}  \at{Impact of a centerbody on the unsteady
  flow dynamics of a swirl nozzle: Intermittency of precessing vortex core
  oscillations}.  \jt{J. Eng. Gas Turbines Power}  \bvol{144}~(2),
  \pg{021014}.

\bibitem[Guth \& Sapsis(2019)]{guth2019machine}
{\sc \au{Guth, S.} \& \au{Sapsis, T.~P.}} \yr{2019}  \at{Machine learning
  predictors of extreme events occurring in complex dynamical systems}.
  \jt{Entropy}  \bvol{21}~(10),  \pg{925}.

\bibitem[Hwang \& Cossu(2010)]{hwang2010linear}
{\sc \au{Hwang, Y.} \& \au{Cossu, C.}} \yr{2010}  \at{Linear non-normal energy
  amplification of harmonic and stochastic forcing in the turbulent channel
  flow}.  \jt{J. Fluid Mech.}  \bvol{664},  \pg{51--73}.

\bibitem[Jim{\'e}nez(2018)]{jimenez2018coherent}
{\sc \au{Jim{\'e}nez, J.}} \yr{2018}  \at{Coherent structures in wall-bounded
  turbulence}.  \jt{J. Fluid Mech.}  \bvol{842},  \pg{P1}.

\bibitem[Karban {\em et~al.\/}(2022)Karban, Martini, Cavalieri, Lesshafft \&
  Jordan]{karban2022self}
{\sc \au{Karban, U.}, \au{Martini, E.}, \au{Cavalieri, A. V.~G.},
  \au{Lesshafft, L.} \& \au{Jordan, P.}} \yr{2022}  \at{Self-similar mechanisms
  in wall turbulence studied using resolvent analysis}.  \jt{J. Fluid Mech.}
  \bvol{939},  \pg{A36}.

\bibitem[Krah {\em et~al.\/}(2022)Krah, Engels, Schneider \&
  Reiss]{krah2022wavelet}
{\sc \au{Krah, P.}, \au{Engels, T.}, \au{Schneider, K.} \& \au{Reiss, J.}}
  \yr{2022}  \at{Wavelet adaptive proper orthogonal decomposition for
  large-scale flow data}.  \jt{Adv. Comput. Math.}  \bvol{48}~(2),  \pg{10}.

\bibitem[Lumley(1967)]{lumley1967structure}
{\sc \au{Lumley, J.~L.}} \yr{1967}  \at{The structure of inhomogeneous
  turbulent flows}.  \jt{Atmospheric turbulence and radio wave propagation}
  \pg{pp. 166--178}.

\bibitem[Lumley(1970)]{lumley1970stochastic}
{\sc \au{Lumley, J.~L.}} \yr{1970} {\em Stochastic tools in turbulence\/}.
  \publ{Academic Press}.

\bibitem[McKeon \& Sharma(2010)]{mckeon2010critical}
{\sc \au{McKeon, B.~J.} \& \au{Sharma, A.~S.}} \yr{2010}  \at{A critical-layer
  framework for turbulent pipe flow}.  \jt{J. Fluid Mech.}  \bvol{658},
  \pg{336--382}.

\bibitem[Meneveau(1991)]{meneveau1991analysis}
{\sc \au{Meneveau, C.}} \yr{1991}  \at{Analysis of turbulence in the
  orthonormal wavelet representation}.  \jt{J. Fluid Mech.}  \bvol{232},
  \pg{469--520}.

\bibitem[Moarref {\em et~al.\/}(2013)Moarref, Sharma, Tropp \&
  McKeon]{moarref2013model}
{\sc \au{Moarref, R.}, \au{Sharma, A.~S.}, \au{Tropp, J.~A.} \& \au{McKeon,
  B.~J.}} \yr{2013}  \at{Model-based scaling of the streamwise energy density
  in high-{R}eynolds number turbulent channels}.  \jt{J. Fluid Mech.}
  \bvol{734},  \pg{275--316}.

\bibitem[Morra {\em et~al.\/}(2021)Morra, Nogueira, Cavalieri \&
  Henningson]{morra2021colour}
{\sc \au{Morra, P.}, \au{Nogueira, P. A.~S.}, \au{Cavalieri, A. V.~G.} \&
  \au{Henningson, D.~S.}} \yr{2021}  \at{The colour of forcing statistics in
  resolvent analyses of turbulent channel flows}.  \jt{Journal of Fluid
  Mechanics}  \bvol{907}.

\bibitem[Neelin {\em et~al.\/}(1998)Neelin, Battisti, Hirst, Jin, Wakata,
  Yamagata \& Zebiak]{neelin1998enso}
{\sc \au{Neelin, J.~D.}, \au{Battisti, D.~S.}, \au{Hirst, A.~C.}, \au{Jin,
  F.-F.}, \au{Wakata, Y.}, \au{Yamagata, T.} \& \au{Zebiak, S.~E.}} \yr{1998}
  \at{{ENSO} theory}.  \jt{J. Geophys. Res.}  \bvol{103}~(C7),
  \pg{14261--14290}.

\bibitem[Page {\em et~al.\/}(2021)Page, Brenner \& Kerswell]{page2021revealing}
{\sc \au{Page, J.}, \au{Brenner, M.~P.} \& \au{Kerswell, R.~R.}} \yr{2021}
  \at{Revealing the state space of turbulence using machine learning}.
  \jt{Phys. Rev. Fluids}  \bvol{6}~(3),  \pg{034402}.

\bibitem[Pyragas \& Pyragas(2020)]{pyragas2020using}
{\sc \au{Pyragas, V.} \& \au{Pyragas, K.}} \yr{2020}  \at{Using reservoir
  computer to predict and prevent extreme events}.  \jt{Phys. Lett. A}
  \bvol{384}~(24),  \pg{126591}.

\bibitem[Qi \& Majda(2020)]{qi2020using}
{\sc \au{Qi, D.} \& \au{Majda, A.~J.}} \yr{2020}  \at{Using machine learning to
  predict extreme events in complex systems}.  \jt{Proc. Natl. Acad. Sci.}
  \bvol{117}~(1),  \pg{52--59}.

\bibitem[Racca \& Magri(2022)]{racca2022data}
{\sc \au{Racca, A.} \& \au{Magri, L.}} \yr{2022}  \at{Data-driven prediction
  and control of extreme events in a chaotic flow}.  \jt{Phys. Rev. Fluids}
  \bvol{7}~(10),  \pg{104402}.

\bibitem[Ren {\em et~al.\/}(2021)Ren, Mao \& Fu]{ren2021image}
{\sc \au{Ren, J.}, \au{Mao, X.} \& \au{Fu, S.}} \yr{2021}  \at{Image-based flow
  decomposition using empirical wavelet transform}.  \jt{J. Fluid Mech.}
  \bvol{906},  \pg{A22}.

\bibitem[Robinson(1991)]{robinson1991coherent}
{\sc \au{Robinson, S.~K.}} \yr{1991}  \at{Coherent motions in the turbulent
  boundary layer}.  \jt{Annu. Rev. Fluid Mech.}  \bvol{23}~(1),  \pg{601--639}.

\bibitem[Rowley \& Dawson(2017)]{rowley2017model}
{\sc \au{Rowley, C.~W.} \& \au{Dawson, S. T.~M.}} \yr{2017}  \at{Model
  reduction for flow analysis and control}.  \jt{Annu. Rev. Fluid Mech}
  \bvol{49},  \pg{387--417}.

\bibitem[Rudy \& Sapsis(2022)]{rudy2022prediction}
{\sc \au{Rudy, S.~H.} \& \au{Sapsis, T.~P.}} \yr{2022}  \at{Prediction of
  intermittent fluctuations from surface pressure measurements on a turbulent
  airfoil}.  \jt{AIAA J.}  \bvol{60}~(7),  \pg{4174--4190}.

\bibitem[Sapsis(2021)]{sapsis2021statistics}
{\sc \au{Sapsis, T.~P.}} \yr{2021}  \at{Statistics of extreme events in fluid
  flows and waves}.  \jt{Annu. Rev. Fluid Mech.}  \bvol{53},  \pg{85--111}.

\bibitem[Schmid(2007)]{schmid2007nonmodal}
{\sc \au{Schmid, P.~J.}} \yr{2007}  \at{Nonmodal stability theory}.  \jt{Annu.
  Rev. Fluid Mech.}  \bvol{39},  \pg{129--162}.

\bibitem[Schmid(2022)]{schmid2022dynamic}
{\sc \au{Schmid, P.~J.}} \yr{2022}  \at{Dynamic mode decomposition and its
  variants}.  \jt{Annu. Rev. Fluid Mech.}  \bvol{54},  \pg{225--254}.

\bibitem[Schmid {\em et~al.\/}(2018)Schmid, Garc{\'\i}a-Gutierrez \&
  Jim{\'e}nez]{schmid2018description}
{\sc \au{Schmid, P.~J.}, \au{Garc{\'\i}a-Gutierrez, A.} \& \au{Jim{\'e}nez,
  J.}} \yr{2018} Description and detection of burst events in turbulent flows.
  \bt{In {\em J. Phys.: Conf. Ser.\/}}, ,  \vol{vol. 1001},  \pg{p. 012015}.
  IOP Publishing.

\bibitem[Schmidt \& Schmid(2019)]{schmidt2019conditional}
{\sc \au{Schmidt, O.~T.} \& \au{Schmid, P.~J.}} \yr{2019}  \at{A conditional
  space--time {POD} formalism for intermittent and rare events: example of
  acoustic bursts in turbulent jets}.  \jt{J. Fluid Mech.}  \bvol{867},
  \pg{R2}.

\bibitem[Schneider \& Vasilyev(2010)]{schneider2010wavelet}
{\sc \au{Schneider, K.} \& \au{Vasilyev, O.~V.}} \yr{2010}  \at{Wavelet methods
  in computational fluid dynamics}.  \jt{Annu. Rev. Fluid Mech.}  \bvol{42},
  \pg{473--503}.

\bibitem[Srirangarajan {\em et~al.\/}(2013)Srirangarajan, Allen, Preis, Iqbal,
  Lim \& Whittle]{srirangarajan2013wavelet}
{\sc \au{Srirangarajan, S.}, \au{Allen, M.}, \au{Preis, A.}, \au{Iqbal, M.},
  \au{Lim, H.~B.} \& \au{Whittle, A.~J.}} \yr{2013}  \at{Wavelet-based burst
  event detection and localization in water distribution systems}.  \jt{J.
  Signal Process. Syst.}  \bvol{72},  \pg{1--16}.

\bibitem[Taira {\em et~al.\/}(2017)Taira, Brunton, Dawson, Rowley, Colonius,
  McKeon, Schmidt, Gordeyev, Theofilis \& Ukeiley]{taira2017modal}
{\sc \au{Taira, K.}, \au{Brunton, S.~L.}, \au{Dawson, S. T.~M.}, \au{Rowley,
  C.~W.}, \au{Colonius, T.}, \au{McKeon, B.~J.}, \au{Schmidt, O.~T.},
  \au{Gordeyev, S.}, \au{Theofilis, V.} \& \au{Ukeiley, L.~S.}} \yr{2017}
  \at{Modal analysis of fluid flows: {A}n overview}.  \jt{AIAA J.}
  \bvol{55}~(12),  \pg{4013--4041}.

\bibitem[Towne {\em et~al.\/}(2015)Towne, Colonius, Jordan, Cavalieri \&
  Bres]{towne2015stochastic}
{\sc \au{Towne, A.}, \au{Colonius, T.}, \au{Jordan, P.}, \au{Cavalieri, A.~V.}
  \& \au{Bres, G.~A.}} \yr{2015} Stochastic and nonlinear forcing of
  wavepackets in a {M}ach 0.9 jet.  \bt{In {\em AIAA Paper\/}},  \pg{p. 2217}.

\bibitem[Towne {\em et~al.\/}(2020)Towne, Lozano-Dur{\'a}n \&
  Yang]{towne2020resolvent}
{\sc \au{Towne, A.}, \au{Lozano-Dur{\'a}n, A.} \& \au{Yang, X.}} \yr{2020}
  \at{Resolvent-based estimation of space--time flow statistics}.  \jt{J. Fluid
  Mech.}  \bvol{883}.

\bibitem[Towne {\em et~al.\/}(2018)Towne, Schmidt \&
  Colonius]{towne2018spectral}
{\sc \au{Towne, A.}, \au{Schmidt, O.~T.} \& \au{Colonius, T.}} \yr{2018}
  \at{Spectral proper orthogonal decomposition and its relationship to dynamic
  mode decomposition and resolvent analysis}.  \jt{J. Fluid Mech.}  \bvol{847},
   \pg{821--867}.

\bibitem[Trefethen {\em et~al.\/}(1993)Trefethen, Trefethen, Reddy \&
  Driscoll]{trefethen1993hydrodynamic}
{\sc \au{Trefethen, L.~N.}, \au{Trefethen, A.~E.}, \au{Reddy, S.} \&
  \au{Driscoll, T.~A.}} \yr{1993}  \at{Hydrodynamic stability without
  eigenvalues}.  \jt{Science}  \bvol{261},  \pg{578--584}.

\bibitem[Wan {\em et~al.\/}(2018)Wan, Vlachas, Koumoutsakos \&
  Sapsis]{wan2018data}
{\sc \au{Wan, Z.~Y.}, \au{Vlachas, P.}, \au{Koumoutsakos, P.} \& \au{Sapsis,
  T.}} \yr{2018}  \at{Data-assisted reduced-order modeling of extreme events in
  complex dynamical systems}.  \jt{PLoS ONE}  \bvol{13}~(5),  \pg{e0197704}.

\bibitem[Yeung {\em et~al.\/}(2015)Yeung, Zhai \&
  Sreenivasan]{yeung2015extreme}
{\sc \au{Yeung, P.~K.}, \au{Zhai, X.~M.} \& \au{Sreenivasan, K.~R.}} \yr{2015}
  \at{Extreme events in computational turbulence}.  \jt{Proc. Natl. Acad. Sci.}
   \bvol{112}~(41),  \pg{12633--12638}.

\end{thebibliography}

\end{document}